\documentclass[showpacs,preprintnumbers,nofootinbib,amsmath,amssymb]{revtex4}
\usepackage{hyperref,slashed,color}
\usepackage{graphicx,subfig}
\allowdisplaybreaks
 \def\nx{\overline{\nabla}_x}
  \def\pps{\frac{\partial}{\partial s}}
   \def\sk{\Sigma_k}
 \def\na{{\nabla_a}}
 \def\gamf{\gamma_5}
  \def\fm{f_-}
   \def\fp{f_+}

   \def\chip{\chi_+}
    \def\chim{\chi_-}
     \def\la{\langle}
      \def\ra{\rangle}
 \def\skp{\Sigma_k'}
 \def\skpp{\Sigma_k''}
 \def\skppp{\Sigma_k'''}
 \def\skpppp{\Sigma_k''''}
 \def\skppppp{\Sigma_k'''''}
 
\begin{document}
\preprint{TUHEP-TH-09169}
\title{Computation of the $p^6$ order chiral Lagrangian coefficients
\\from the underlying theory of QCD}

\bigskip
\author{Shao-Zhou
Jiang$^{1,2}$\footnote{Email:\href{mailto:jsz@mails.tsinghua.edu.cn}{jsz@mails.tsinghua.edu.cn}.},
Ying Zhang$^{3}$\footnote{Email:
\href{mailto:hepzhy@mail.xjtu.edu.cn}{hepzhy@mail.xjtu.edu.cn}.},
Chuan
Li$^{1,2}$\footnote{Email:\href{mailto:lcsyhshy2008@yahoo.com.cn}{lcsyhshy2008@yahoo.com.cn}.},
 and Qing Wang$^{1,2}$\footnote{Email:
\href{mailto:wangq@mail.tsinghua.edu.cn}{wangq@mail.tsinghua.edu.cn}.}\footnote{corresponding
author}\\~}

\bigskip
\affiliation{$^1$Center for High Energy Physics, Tsinghua University, Beijing 100084, P.R. China\\
$^2$Department of Physics, Tsinghua University, Beijing 100084,
P.R. China\footnote{mailing address}\\
$^3$School of Science, Xi'an Jiaotong University, Xi'an, 710049,
P.R. China}

\begin{abstract}
We present results of computing the $p^6$ order low energy constants
in the  normal part of chiral Lagrangian  both for two and three
flavor pseudo-scalar mesons. This is a generalization of our
previous work on calculating the $p^4$ order
 coefficients of the chiral Lagrangian in
 terms of the quark self energy
$\Sigma(p^2)$ approximately from QCD. We show that most of our
results are consistent with those we can find in the literature.
\end{abstract}
 \maketitle
%\tableofcontents
%%%%%%%%%%%%%%%%%%%%%%%%%%%%%%%%%%%%%%%%%%%%%%%%%%%%%%
\section{Introduction}

Chiral Lagrangian for low lying pseudoscalar
mesons\cite{weinberg}\cite{GS} as the most successful effective
field theory is now widely used in various strong, weak and
electromagnetic processes. To match the increasing demand for higher
precision in low energy description of QCD,  the applications of the
low energy expansion of the chiral Lagrangian is extended from early
time discussions on the leading $p^2$ and next to leading $p^4$
orders to present $p^6$ order. For the latest review, see
Ref.\cite{Review}. In the chiral Lagrangian, there are many unknown
phenomenological low energy constants (LECs) which appear in front
of each Goldstone field dependent operators and the number of the
LECs increases rapidly when we go to the higher orders of the low
energy expansion. For example for the three flavor case, the $p^2$
and $p^4$ order chiral Lagrangian have 2 and 10 LECs respectively,
while the normal part of $p^6$ order chiral Lagrangian have 90 LECs.
Such a large number of LECs is very difficult to fix from the
experiment data. This badly reduces the predictive power of the
chiral Lagrangian and blur the check of its convergence. The area of
estimating $p^6$ order LECs is where most improvement is needed in
the future of higher order chiral Lagrangian calculations.

A way to increase the precision of the low energy expansion and
improve the present embarrassed situation is studying the relation
between the chiral Lagrangian and the fundamental principles of QCD.
We expect that this relation will be helpful for understanding the
origin of these LECs and further offer us their values. In previous
paper \cite{WQ1}, based on a more earlier study of deriving the
chiral Lagrangian from the first principles of QCD \cite{WQ0} in
which LECs are defined in terms of certain Green's functions in QCD,
we have developed techniques and calculated the $p^2$ and $p^4$
order LECs approximately from QCD. Our simple approach involves the
approximations of taking the large-$N_c$ limit, the leading order in
dynamical perturbation theory, and the improved ladder
approximation, thereby the relevant Green's functions relate to LECs
are expressed in terms of the quark self energy $\Sigma(p^2)$. The
result chiral Lagrangian in terms of the quark self energy is proved
equivalent to a gauge invariant, nonlocal, dynamical (GND) quark
model\cite{WQ2}. By solving the Schwinger-Dyson equation (SDE) for
$\Sigma(p^2)$, we obtain the approximate QCD predicted LECs  which
are consistent with the experimental values. With these results,
generalization of the calculations to $p^6$ order LECs becomes the
next natural step. Considering that the algebraic derivations for
those formulae to express LECs in terms of the quark self energy at
$p^4$ order are lengthy (they need at least several months of
handwork), it is almost impossible to achieve the similar works for
the $p^6$ order calculations just by hand. Therefore, to realize the
calculations for the $p^6$ order LECs, we need to computerize the
original calculations and this is a very hard task. The key
difficulty comes from that the formulation developed in
Ref.\cite{det0} and exploited in Ref.\cite{WQ1} not automatically
keeps the local chiral covariance of the theory and one has to
adjust the calculation procedure by hand to realize the covariance
of the results. To match with the computer program, we need to
change the original formulation to a chiral covariant one. In
Ref.\cite{covariant,covariant1,covariant2}, we have built and
developed such a formulation, followed by next several year's
efforts, we now successfully encode the formulation into computer
programs. With the help of these computer codes we can reproduce
analytical results on the computer originally derived by hand in
Ref.\cite{WQ1} within 15 minutes now. This not only confirms the
reliability of the program itself, but also checks the correctness
of our original formulae. Based on these progresses, in this paper,
we generalize our previous works on calculating the $p^4$ order LECs
to computing the $p^6$ order LECs of chiral Lagrangian both for two
and three flavor pseudo-scalar mesons. This generalization not only
produces new numerical predictions for the $p^6$ order LECs, but
also forces us to reexamine our original formulation from a new
angle in dealing with $p^2$ and $p^4$ order LECs.

This paper is organized as follows: In Sec.II, we review our
previous calculations on the $p^2$ and $p^4$ order LECs. Then, in
Sec.III, based on the technique developed in Ref.\cite{covariant},
we reformulate the original low energy expansion used in
Ref.\cite{WQ1} into a chiral covariant one suitable for computer
derivation. In Sec.IV, from present $p^6$ order viewpoint, we
reexamine the formulation we taken before and show that if we sum
all higher order  anomaly part contributions terms together, their
total contributions to the normal part of the chiral Lagrangian
vanish. This leads a change the role of finite $p^4$ order anomaly
part contributions which originally are subtracted in the chiral
Lagrangian in Ref.\cite{WQ1} and now must be used to cancel
divergent higher order anomaly part contributions. We reexhibit the
numerical result of the $p^4$ order LECs without subtraction of
$p^4$ order anomaly part contributions. In Sec.V, we present general
$p^6$ order chiral Lagrangian in terms of rotated sources and
express the $p^6$ order LECs in terms of the quark self energy.
Sec.VI is a part where we give numerical results for $p^6$ order
LECs in the normal part of chiral Lagrangian both for two and three
flavor pseudo scalar mesons.  In Sec. VII, we apply and compare with
our results to some individuals and combinations of LECs  proposed
and estimated in the literature, checking the correctness of our
numerical predictions. Sec.VIII is a summary. In Appendices, we list
some necessary formulae and relations.

%%%%%%%%%%%%%%%%%%%%%%%%%%%%%%%%%%%%%%%%%%%%%%%%%%%%%%%%%%%%%%%%%%%%
\section{Review of the Calculations on the $p^2$ and $p^4$ Order LECs}

Theoretically, the action of the chiral Lagrangian at large $N_c$
limit derived from the first principle of QCD takes form \cite{WQ0}
\begin{eqnarray}
S_\mathrm{eff}&=&-iN_c\mathrm{Tr}\ln[i\slashed{\partial}+J_{\Omega}-\Pi_{\Omega
c}]
+iN_c\mathrm{Tr}\ln[i\slashed{\partial}+J_{\Omega}]-iN_c\mathrm{Tr}\ln[i\slashed{\partial}+J]
+N_c\mathrm{Tr}[\Phi_{\Omega c}\Pi^T_{\Omega c}]\label{Seff0}\\
&&+N_c\sum^{\infty}_{n=2}{\int}d^{4}x_1\cdots
d^4x_{n}'\frac{(-i)^{n}(N_c
g_s^2)^{n-1}}{n!}\bar{G}^{\sigma_1\cdots\sigma_n}_{\rho_1
\cdots\rho_n}(x_1,x'_1,\cdots,x_n,x'_n)\Phi^{\sigma_1\rho_1}_{\Omega
c}(x_1 ,x'_1)\cdots \Phi^{\sigma_n\rho_n}_{\Omega c}(x_n
,x'_n)+O(\frac{1}{N_c})\,\nonumber
\end{eqnarray}
in which $J_{\Omega}$ is external source $J$  including currents and
densities after Goldstone field dependent chiral rotation $\Omega$
\begin{eqnarray}
J_{\Omega}=[\Omega
P_R+\Omega^{\dag}P_L][J+i\slashed{\partial}][\Omega P_R+\Omega^\dag
P_L]=\slashed{v}_\Omega+\slashed{a}_\Omega\gamma_5-s_\Omega+ip_\Omega\gamma_5\hspace{1cm}
J=\slashed{v}+\slashed{a}\gamma_5-s+ip\gamma_5\hspace{1cm}U=\Omega^2\;.~~~~\label{JOmega}
\end{eqnarray}
$\Phi_{\Omega c}$ and $\Pi_{\Omega c}$ are two point rotated quark
Green's function and interaction part of two point rotated quark
vertex in presence of external sources respectively,  $\Phi_{\Omega
c}$ is defined by
\begin{eqnarray}
\Phi_{\Omega c}^{\sigma\rho}(x,y)\equiv
\frac{1}{N_c}\langle\overline{\psi}_{\Omega}^{\sigma}(x)\psi_{\Omega}^{\rho}(y)\rangle
=-i[(i\slashed{\partial}+J_{\Omega}-\Pi_{\Omega
c})^{-1}]^{\rho\sigma}(y,x)\hspace{2cm}
\psi_{\Omega}(x)\equiv[\Omega(x)P_L+\Omega^\dag(x)P_R]\psi(x)\label{PhiPi}
\end{eqnarray}
with subscript $_c$ denoting the classical field and $\psi(x)$ being
light quark fields. $\bar{G}^{\sigma_1\cdots\sigma_n}_{\rho_1
\cdots\rho_n}(x_1,x'_1,\cdots,x_n,x'_n)$ is effective gluon n-point
Green's function and $g_s$ is coupling constant of QCD. It can be
shown that the last term in the first line and the term in the
second line of the r.h.s. of Eq.(\ref{Seff0}) are independent of
pseudo scalar meson field $U$ or $\Omega$ and therefore are just
irrelevant constants in the effective action. While the second and
third terms in the first line of the r.h.s. of Eq.(\ref{Seff0}) are
anomaly part contributions, since they represent the variations of
the path integral measure for light quark fields $\psi$. The
remaining first term is called normal part contributions which
relies on $\Pi_{\Omega c}$. The $\Phi_{\Omega c}$ and $\Pi_{\Omega
c}$ are related by the first equation of (\ref{PhiPi}) and
determined by
\begin{eqnarray}
&&[\Phi_{\Omega
c}+\tilde{\Xi}]^{\sigma\rho}+\sum^{\infty}_{n=1}{\int}d^4x_1d^4x'_1\cdots{d^4}x_n
d^4x'_n\frac{(-i)^{n+1}(N_c
g_s^2)^n}{n!}\overline{G}^{\sigma\sigma_1\cdots\sigma_n}_{\rho\rho_1
\cdots\rho_n}(x,y,x_1,x'_1,\cdots,x_n,x'_n)\nonumber\\
&&\times \Phi_{\Omega c}^{\sigma_1\rho_1}(x_1 ,x'_1)\cdots
\Phi_{\Omega c}^{\sigma_n\rho_n}(x_n
,x'_n)=O(\frac{1}{N_c}),\label{SDE} \label{fineqNc}
\end{eqnarray}
where $\tilde{\Xi}$ is a Lagrangian multiplier which insures the
constraint $\mathrm{tr}_l[\gamma_5\Phi_{\Omega c}^T(x,x)]=0$.
Eq.(\ref{fineqNc}) is the SDE in presence of the rotated external
source. In Ref.\cite{WQ1}, we have assumed the solution of
(\ref{fineqNc}) approximately by
\begin{eqnarray}
\Pi^{\sigma\rho}_{\Omega c}(x,y)=
[\Sigma(\overline{\nabla}^2_x)]^{\sigma\rho}\delta^4(x-y)\hspace{3cm}
\overline{\nabla}^{\mu}_x=\partial^{\mu}_x-iv_{\Omega}^{\mu}(x)\;,
\end{eqnarray}
where $\Sigma$ is the quark self energy which satisfy SDE
(\ref{SDE}) with vanishing rotated external source. Under the ladder
approximation, this SDE in Euclidean space-time is reduced to the
standard form of
\begin{eqnarray}
\Sigma(p^2)-3C_2(R)\int\frac{d^4q}{4\pi^3}
\frac{\alpha_s[(p-q)^2]}{(p-q)^2}
\frac{\Sigma(q^2)}{q^2+\Sigma^2(q^2)}=0\;, \label{eq0}
\end{eqnarray}
where $C_2(R)$ is the second order Casimir operator of the quark
representation R, in our case, quark is belong to $SU(N_c)$
fundamental representation, therefore $C_2(R)=(N_c^2-1)/2N_c$ and in
the large $N_c$ limit, we will neglect the second term of it.
$\alpha_s(p^2)$ is the running coupling constant of QCD which
depends on $N_c$ and quark flavor.
 With
these approximations, the result action (\ref{Seff0}) of the chiral
Lagrangian becomes the GND model introduced in Ref.\cite{WQ2},
\begin{eqnarray}
S_\mathrm{eff}\approx S_\mathrm{GND}+O(\frac{1}{N_c})\hspace{1cm}
S_\mathrm{GND}\equiv-iN_c\mathrm{Tr}\ln[i\slashed{\partial}+J_{\Omega}-\Sigma(\overline{\nabla}^2)]
+iN_c\mathrm{Tr}\ln[i\slashed{\partial}+J_{\Omega}]-iN_c\mathrm{Tr}\ln[i\slashed{\partial}+J]\;.~~~~~~\label{Seff1}
\end{eqnarray}
In which the third term at r.h.s. of (\ref{Seff1}) is independent of
pseudo scalar field $U$, therefore it only affects contact term of
the chiral Lagrangian. In fact, for the contact term part, we can
take $\Omega=1$ in (\ref{Seff1}), then
\begin{eqnarray}
S_{\mathrm{eff}}\bigg|_{\mathrm{contact}}
\approx-iN_c\mathrm{Tr}\ln\{i\slashed{\partial}+J-\Sigma[(\partial-iv)^2)]\}
+O(\frac{1}{N_c})\;.~~~~~~ \label{Seff2}
\end{eqnarray}
For the non-contact terms concerned in this paper, we can ignore the
third term at r.h.s. of (\ref{Seff1}) and the next key element is to
compute term
$\mathrm{Tr}\ln[i\slashed{\partial}+J_{\Omega}-\Sigma(\overline{\nabla}^2)]$.
The remaining term $\mathrm{Tr}\ln[i\slashed{\partial}+J_{\Omega}]$
in our previous work is obtained by further taking limit
$\Sigma\rightarrow 0$ in
$\mathrm{Tr}\ln[i\slashed{\partial}+J_{\Omega}-\Sigma(\overline{\nabla}^2)]$
\footnote{This will cause some confusions and we are going to
discuss them in section IV.}. Since anomaly terms are at least the
$p^4$ order and at this order, anomaly is the well known Wess-Zumino
terms which have no unknown LECs (In Ref.\cite{WQma}, we have
derived such terms from QCD). All unknown LECs at $p^2$ and $p^4$
orders are in the normal part of chiral Lagrangian, so to calculate
the $p^2$ and $p^4$ orders LECs, we only need to discuss the normal
part of chiral Lagrangian which is in fact the real part of
$\mathrm{Tr}\ln(\cdots)$. With the help of
 Schwinger proper time method \cite{det0}, this real part in Euclidean space-time\footnote{Our extension from Minkovski space to Euclidean space takes
  $x^0|_M\rightarrow-ix^4|_E$,~$x^i|_M\rightarrow x^i|_M$,~$\gamma^0|_M\rightarrow\gamma^4|_E$,
  ~$\gamma^i|_M\rightarrow i\gamma^i|_E$,~with $i=1,2,3$ being space
  indices and there $\gamma_E^\mu$ are hermitian.  $v^\mu_\Omega, a^\mu_\Omega$ transform as $x^\mu$.
  $\gamma_5|_M\rightarrow\gamma_5|_E$,~$s|_M\rightarrow-s|_E$, ~$p|_M\rightarrow-p|_E$. }
 with metric tensor $g^{\mu\nu}=\mbox{diag}(1,1,1,1)$,
 can be written as
 \begin{eqnarray}
 &&\mathrm{ReTr}\ln[\slashed{\partial}-i\slashed{v}_\Omega-i\slashed{a}_\Omega\gamma_5-s_\Omega+ip_\Omega\gamma_5
 +\Sigma(-\bar{\nabla}^2)] \nonumber\\
 &&=\frac{1}{2}\mathrm{Tr}\ln\Big[[\slashed{\partial}-i\slashed{v}_\Omega-i\slashed{a}_\Omega\gamma_5-s_\Omega
 +ip_\Omega\gamma_5+\Sigma(-\bar{\nabla}^2)]^\dag
 [\slashed{\partial}-i\slashed{v}_\Omega-i\slashed{a}_\Omega\gamma_5-s_\Omega+ip_\Omega\gamma_5+\Sigma(-\bar{\nabla}^2)]\Big]\nonumber\\
 &&=-\frac{1}{2}\lim_{\Lambda \to \infty}
 \int^\infty_{\frac{1}{\Lambda^2}}\frac{d\tau}{\tau}\int{d^4x}
 ~\mathrm{tr}\langle x|\exp\bigg[-\tau[\bar{E}-(\bar{\nabla}-ia_\Omega)^2+\Sigma^2(-\bar{\nabla}^2)
 +\hat{I}_{\Omega}\Sigma(-\bar{\nabla}^2)+\Sigma(-\bar{\nabla}^2)\tilde{I}_{\Omega}-d\!\!\!
 /\;\Sigma(-\bar{\nabla}^2)]\bigg]|x\rangle\nonumber\\
 &&=-\frac{1}{2}\lim_{\Lambda \to \infty}
 \int^\infty_{\frac{1}{\Lambda^2}}\frac{d\tau}{\tau}\int{d^4x}\int\frac{d^4k}{(2\pi)^4}
 ~\mathrm{tr}\exp\bigg[-\tau[\bar{E}+(k+i\bar{\nabla}_x+a_\Omega)^2+\Sigma^2((k+i\bar{\nabla}_x)^2)
 +\hat{I}_{\Omega}\Sigma((k+i\bar{\nabla}_x)^2)\notag\\
 &&\hspace{0.5cm}+\Sigma((k+i\bar{\nabla}_x)^2)\tilde{I}_{\Omega}-d\!\!\!
 /\;\Sigma((k+i\bar{\nabla}_x)^2)]\bigg]\;,\label{Trln}~~~~~~
 \end{eqnarray}
 in which
 \begin{eqnarray}
     &&\hspace{-0.5cm}\bar{E}=\frac{i}{4}[\gamma_\mu,\gamma_\nu]R^{\mu\nu}+\gamma_\mu
 d^\mu(s_\Omega-ip_\Omega\gamma_5)+i\gamma_\mu[a^\mu_\Omega\gamma_5(s_\Omega-ip_\Omega\gamma_5)
 +(s_\Omega-ip_\Omega\gamma_5)a_\Omega^\mu\gamma_5]+s_\Omega^2+p_\Omega^2-[s_\Omega,p_\Omega]i\gamma_5
 \nonumber\\
     &&\hspace{-0.5cm}d^\mu {\cal O}\equiv \partial^\mu{\cal O}-i[v^\mu_\Omega,{\cal O}]\hspace{0.3cm}
     ({\cal O}=\mbox{any operator})\hspace{0.7cm}
     R^{\mu\nu}\!\!=V_\Omega^{\mu\nu}\!-i[a_\Omega^\mu,a_\Omega^\nu]
     +(d^\mu a_\Omega^\nu-d^\nu a_\Omega^\mu)\gamma_5 \hspace{0.7cm}
         V_{\Omega,\mu\nu}\!\!=i[\bar{\nabla}_\mu,\bar{\nabla}_\nu]\hspace{0.7cm}\nonumber\\
         &&\hspace{-0.5cm}\bar{\nabla}^\mu_x
         =\partial^\mu\!\!-iv_\Omega^\mu(x)
     \hspace{0.7cm}\hat{I}_{\Omega}=-ia\!\!\!/_\Omega\gamma_5-s_\Omega-ip_\Omega\gamma_5
     \hspace{0.7cm}
     \tilde{I}_{\Omega}=-ia\!\!\!/_\Omega\gamma_5-s_\Omega+ip_\Omega\gamma_5\hspace{0.7cm}
     d\!\!\!/\;\Sigma(-\bar{\nabla}^2)=\gamma_\mu d^\mu\Sigma(-\bar{\nabla}^2)\;.\nonumber
 \end{eqnarray}
In (\ref{Trln}), a cutoff $\Lambda$ is introduced into the theory to
regularize the possible ultraviolet divergences. In practical
calculations,  we treat it as the physical cutoff of the theory.
Taking the low energy expansion for (\ref{Trln}), we can finally
express
$\mathrm{Tr}\ln[i\slashed{\partial}+J_{\Omega}-\Sigma(\overline{\nabla}^2)]$
in terms of power expansion of external sources with coefficients
being $\Sigma$ dependent functions. Further vanishing $\Sigma$, we
obtain $\mathrm{Tr}\ln[i\slashed{\partial}+J_{\Omega}]$. Then the
r.h.s. of (\ref{Seff1}) is expressed in terms of power expansion of
rotated external sources, compare the result with the
parametrization of the effective action without applying the
equations of motion for pseudo scalar mesons,
\begin{eqnarray}
S_\mathrm{eff}&=&\int
d^4x~\mathrm{tr}_f\bigg[F_0^2a_{\Omega}^2+F_0^2B_0s_{\Omega}-{\cal
K}_1^{(\mathrm{norm},\Pi_{\Omega
c}\neq0)}[d_{\mu}a_{\Omega}^{\mu}]^2 -{\cal
K}_2^{(\mathrm{norm},\Pi_{\Omega
c}\neq0)}(d^{\mu}a_{\Omega}^{\nu}-d^{\nu}a_{\Omega}^{\mu})
(d_{\mu}a_{\Omega,\nu}-d_{\nu}a_{\Omega,\mu})\nonumber\\
&&+{\cal K}_3^{(\mathrm{norm},\Pi_{\Omega
c}\neq0)}[a_{\Omega}^2]^2+{\cal K}_4^{(\mathrm{norm},\Pi_{\Omega
c}\neq0)}a_{\Omega}^{\mu}a_{\Omega}^{\nu}
a_{\Omega,\mu}a_{\Omega,\nu}+{\cal K}_5^{(\mathrm{norm},\Pi_{\Omega
c}\neq0)}a_\Omega^2\mathrm{tr}_f[a_{\Omega}^2]\nonumber\\
&&+{\cal K}_6^{(\mathrm{norm},\Pi_{\Omega
c}\neq0)}a_{\Omega}^{\mu}a_{\Omega}^{\nu}\mathrm{tr}_f[a_{\Omega,\mu}a_{\Omega,\nu}]+{\cal
K}_7^{(\mathrm{norm},\Pi_{\Omega c}\neq0)}s_{\Omega}^2 +{\cal
K}_8^{(\mathrm{norm},\Pi_{\Omega
c}\neq0)}s_{\Omega}\mathrm{tr}_f[s_{\Omega}] +{\cal
K}_9^{(\mathrm{norm},\Pi_{\Omega c}\neq0)}p_{\Omega}^2
\nonumber\\
&&+{\cal K}_{10}^{(\mathrm{norm},\Pi_{\Omega
c}\neq0)}p_{\Omega}\mathrm{tr}_f[p_{\Omega}] +{\cal
K}_{11}^{(\mathrm{norm},\Pi_{\Omega c}\neq0)}s_{\Omega}a_{\Omega}^2
+{\cal K}_{12}^{(\mathrm{norm},\Pi_{\Omega
c}\neq0)}s_{\Omega}\mathrm{tr}_f[a_{\Omega}^2]-{\cal
K}_{13}^{(\mathrm{norm},\Pi_{\Omega
c}\neq0)}V_{\Omega}^{\mu\nu}V_{\Omega,\mu\nu} \nonumber\\
&&+i{\cal K}_{14}^{(\mathrm{norm},\Pi_{\Omega
c}\neq0)}V_{\Omega}^{\mu\nu}a_{\Omega,\mu}a_{\Omega,\nu} +{\cal
K}_{15}^{(\mathrm{norm},\Pi_{\Omega
c}\neq0)}p_{\Omega}d_{\mu}a^{\mu}_{\Omega}\bigg]+O(p^6)+U\mbox{-independent~source~terms}.
\label{p4}
\end{eqnarray}
We can read out $F_0^2,B_0$ and
$\mathcal{K}_i^{(\mathrm{norm},\Pi_{\Omega c}\neq0)}$ for
$i=1,\ldots,15$ as functions of $\Sigma$.
$\mathcal{K}_i^{(\mathrm{norm},\Pi_{\Omega c}\neq0)}$s relate to the
 conventional $p^4$ order LECs through (25) of Ref.\cite{WQ1}.
 A superscript
$^{(\mathrm{norm},\Pi_{\Omega c}\neq0)}$ on each of $\mathcal{K}_i$
denotes the property that when $\Pi_{\Omega c}=\Sigma=0$, all
$\mathcal{K}_i$ vanish, i.e.
\begin{eqnarray}
\mathcal{K}_i^{(\mathrm{norm},\Pi_{\Omega
c}\neq0)}=\mathcal{K}_i^{(\mathrm{norm})}-\mathcal{K}_i^{(\mathrm{norm},\Pi_{\Omega
c}=0)}
\hspace{2cm}\mathcal{K}_i^{(\mathrm{norm})}\stackrel{\Sigma=0}{====}\mathcal{K}_i^{(\mathrm{norm},\Pi_{\Omega
c}=0)}\hspace{2cm}i=1,\ldots,15\;,
\end{eqnarray}
where $\mathcal{K}_i^{(\mathrm{norm})}$ and
$-\mathcal{K}_i^{(\mathrm{norm},\Pi_{\Omega c}=0)}$ are the
contributions to the effective action from the first, second and
third terms in the r.h.s. of (\ref{Seff1}) respectively. Replacing
superscript $^{(\mathrm{norm},\Pi_{\Omega c}\neq0)}$ with
$^{(\mathrm{norm})}$ in the r.h.s. of (\ref{p4}), we obtain term
$-iN_c\mathrm{Tr}\ln[i\slashed{\partial}+J_{\Omega}-\Sigma(\overline{\nabla}^2)]$.
And replacing superscript $^{(\mathrm{norm},\Pi_{\Omega}\neq0)}$
with $^{(\mathrm{norm},\Pi_{\Omega}=0)}$ and vanishing $F_0^2$ in
the r.h.s. of (\ref{p4}) , we obtain term
$-iN_c\mathrm{Tr}\ln[i\slashed{\partial}+J_{\Omega}]+iN_c\mathrm{Tr}\ln[i\slashed{\partial}+J]$.
The result formulae for $F_0^2B_0, F_0^2$ and
$\mathcal{K}_i^{(\mathrm{norm})}$ expressed in terms of $\Sigma$ are
explicitly given in (34), (35) and (36) in Ref.\cite{WQ1}.

With the analytical formulae for LECs of $F_0^2,B_0$ and
$\mathcal{K}_i^{(\mathrm{norm},\Pi_{\Omega c}\neq0)}$ for
$i=1,\ldots,15$ as functions of $\Sigma$, we can suitably choose
running coupling constant $\alpha_s(p^2)$, solve SDE (\ref{eq0})
numerically obtaining quark self energy $\Sigma$, then calculate the
numerical values of all $p^2$ and $p^4$ order LECs. To obtain the
final numerical result in Ref.\cite{WQ1}, we have assumed
$F_0=f_\pi=93$MeV as input\footnote{Later we will use a changed
value of $F_0=87$MeV for two-flavour case. For detail, see the
discussion of Eq.(58) } to fix the dimensional parameter
$\Lambda_\mathrm{QCD}$ appear in running coupling constant
$\alpha_s(p^2)$ and taken cutoff parameter $\Lambda$ appear in
(\ref{Trln}) equal to infinity and 1GeV respectively. The final
obtained values are consistent with those fixed phenomenologically.

%%%%%%%%%%%%%%%%%%%%%%%%%%%%%%%%%%%%%%%%%%%%%%%%%%%%%%%%%%%%%%%%%%%%%
\section{Chiral Covariant Low Energy Expansion}

Eq.(\ref{Trln}) is the starting point of our reformulation in this
section. In Ref.\cite{WQ1}, we expand (\ref{Trln}) up to the $p^4$
order and obtain analytical result. This expansion is not explicitly
chiral covariant, since the operator $\bar{\nabla}_x^\mu$ appears in
the formula is not always covariant under the local chiral symmetry
transformations. For example, when $\bar{\nabla}_x^\mu$ acts on a
constant number 1, it gives $\bar{\nabla}_x^\mu~1=-iv_\Omega^\mu(x)$
which is not covariant since $v_\Omega^\mu(x)$ itself behaves as the
gauge field in the local chiral symmetry transformations. Only when
they combined into commutators, such as
$[\bar{\nabla}_x^\mu,\bar{\nabla}_x^\nu]$ or
$[\bar{\nabla}_x^\mu,a_\Omega^\nu(x)]$, the covariance recovers
back. Therefore in the detail calculation, we need to confirm that
all $\bar{\nabla}_x^\mu$s appear in the result do can be arranged
into some commutators. This is a conjecture. In the original work of
Ref.\cite{WQ1}, we have found that this conjecture is valid up to
some terms with coefficients being expressed as integration over
some total derivatives, i.e. form of $\int
d^4k\frac{\partial}{\partial k^\mu}g(k)$. If we ignore these total
derivative terms, up to order of $p^4$, we can explicitly prove the
conjecture. At the stage of our earlier works, we do not question
the reason that why we can drop out those  total derivative terms
(In fact, in Eq.(74) of Ref.\cite{WQ0}, we have shown that in order
to obtain the well-known Pagels-Stokar formula, a total derivative
term must be dropped out). This leads the further discussions on the
role of total derivative terms in the quantum field theory
\cite{covariant2}. Later in this section, we will give the correct
reason of dropping out those  total derivative terms. Arranging
various $\bar{\nabla}_x^\mu$ into commutators is a very tricky and
complex task which is very hard to be achieved by computer. In order
to computerize the calculation, we need to find a way which can
automatically arrange all $\bar{\nabla}_x^\mu$s into some
commutators. This leads the developments given in
Ref.\cite{covariant,covariant1,covariant2}, where we have introduced
 \begin{eqnarray}
 k^\mu+i\bar{\nabla}_x^\mu=e^{i\bar{\nabla}_{x}\cdot\frac{\partial}{\partial k}}
 \Big(k^\mu+\tilde{F}^\mu(\bar{\nabla},\frac{\partial}{\partial k})\Big)
 e^{-i\bar{\nabla}_{x}\cdot\frac{\partial}{\partial k}}\;,\label{differential}
 \end{eqnarray}
 in which
 \begin{eqnarray}
\tilde{F}^\mu(\bar{\nabla},\frac{\partial}{\partial k})
 &\equiv&-e^{Ad(-i\bar{\nabla}_x\!\!\cdot\!\frac{\partial}{\partial k})}
 \bigg(F\Big[Ad(i\bar{\nabla}_x\cdot\frac{\partial}{\partial k})\Big]
 (i\bar{\nabla}_x^\mu)\bigg)\nonumber\\
 &=&\frac{1}{2}(\nu\mu)\frac{\partial}{\partial
 k^{\nu}}
 -\frac{i}{3}(\lambda\nu\mu)\frac{\partial^2}{\partial k^{\lambda}\partial
 k^{\nu}}
 -\frac{1}{8} (\rho\lambda\nu\mu)
 \frac{\partial^3}{\partial k^{\rho}\partial k^{\lambda}\partial k^{\nu}}
 +\frac{i}{30}(\sigma\rho\lambda\nu\mu)
 \frac{\partial^4}{\partial k^\sigma\partial k^{\rho}\partial k^{\lambda}\partial k^{\nu}}\notag\\
 &&+\frac{1}{144}(\delta\sigma\rho\lambda\nu\mu)
 \frac{\partial^5}{\partial k^\delta\partial k^\sigma\partial k^{\rho}\partial k^{\lambda}\partial k^{\nu}}+O(p^7)\;,\\
   &&\hspace{-2.3cm}F(z)=\sum^\infty_{n=2}\frac{z^{n-1}}{n!}\hspace{3cm}
  [Ad(B)]^n(C)\equiv[\underbrace{B,[B,\cdots,[B}_{\mbox{n
 times}},C]\cdots]]\;,\notag\\
 &&\hspace{-2.3cm}(\mu_n\mu_{n-1}\cdots\mu_2\mu_1)
 \equiv[\nx^{\mu_n},[\nx^{\mu_{n-1}},\cdots,[\nx^{\mu_2},\nx^{\mu_1}]\cdots]]\;,\notag
 \end{eqnarray}
 where the default set of Lorentz indices for
 $(\mu_n\mu_{n-1}\cdots\mu_2\mu_1)$ is the supperscripts, in
 some cases, we need subscript, we will use $\underline{\mu}$ to
 denote the corresponding subscript for $\mu$.
Note that in present notation for
$(\mu_n\mu_{n-1}\cdots\mu_2\mu_1)$, we don't explicitly write
$\overline{\nabla}_x$s, but only their Greek superscripts for short.
If we use other symbols, such as $s_{\Omega}$ appeared in $(\mu
s_\Omega)$ and $a_\Omega^\nu$ in $(\mu a_\Omega^\nu)$, then we take
definition that $(\mu s_\Omega)\equiv[\nx^\mu,s_\Omega]$ and $(\mu
 a_\Omega^\nu)\equiv[\nx^\mu,a_\Omega^\nu]$.

 Substitute
(\ref{differential}) into (\ref{Trln}), we change (\ref{Trln}) to
\begin{eqnarray}
 &&\mathrm{ReTr}\ln[\slashed{\partial}-i\slashed{v}_\Omega-i\slashed{a}_\Omega\gamma_5-s_\Omega+ip_\Omega\gamma_5
 +\Sigma(-\bar{\nabla}^2)] \nonumber\\
 &&=-\frac{1}{2}\lim_{\Lambda \to \infty}
 \int^\infty_{\frac{1}{\Lambda^2}}\frac{d\tau}{\tau}\int{d^4x}\int\frac{d^4k}{(2\pi)^4}\mathrm{tr}~e^{i\bar{\nabla}_{x}^{\mu}\cdot\frac{\partial}{\partial
 k}}\exp\bigg\{-\tau\bigg[\tilde{E}+(k+\tilde{F})^2
 +\tilde{a}_\mu\gamma_5(k^\mu+\tilde{F}^\mu)+(k^\mu+\tilde{F}^\mu)\tilde{a}_\mu\gamma_5+\tilde{a}^2
 \notag\\
 &&\hspace{0.4cm}+\Sigma^2\big((k+\tilde{F})^2\big)+\tilde{J}\Sigma\big((k+\tilde{F})^2\big)
 +\Sigma\big((k+\tilde{F})^2\big)\tilde{K}
 -\gamma_\mu\Big[\tilde{\bar{\nabla}}^\mu_x,\Sigma\big((k+\tilde{F})^2\big)\Big]\bigg]\bigg\}\cdot 1\label{dB0}
 \end{eqnarray}
 with tilde operation defined as
 \begin{eqnarray}
 \tilde{\mathcal{O}}
 &\equiv& \mathcal{O}-i (\nu\mathcal{O})\frac{\partial}{\partial k^{\nu}}
 -\frac{1}{2} (\lambda\nu\mathcal{O})\frac{\partial^2}{\partial k^{\lambda}\partial
 k^{\nu}}
 +\frac{i}{6} (\rho\lambda\nu\mathcal{O})\frac{\partial^3}{\partial k^\rho\partial k^{\lambda}\partial
 k^{\nu}}+\frac{1}{24}(\sigma\rho\lambda\nu\mathcal{O})
\frac{\partial^4}{\partial k^\sigma\partial k^\rho\partial
k^{\lambda}\partial
k^{\nu}}\\
 &&\hspace{0cm}-\frac{i}{120}(\delta\sigma\rho\lambda\nu\mathcal{O})
\frac{\partial^5}{\partial k^\delta\partial k^{\sigma}\partial
k^{\rho}\partial k^{\lambda}\partial k^{\nu}}+O(p^7)\;,\notag
\end{eqnarray}
 where $\tilde{\mathcal{O}}\equiv(\tilde{E},\tilde{J},\tilde{K},\tilde{a}^\mu,\tilde{a}^2,\tilde{\bar{\nabla}}^\mu_x)^T$
 and $\mathcal{O}\equiv(E,\hat{I},\tilde{I},a_\Omega^\mu,a_\Omega^2,\bar{\nabla}^\mu_x)^T$.
Note that for finite cutoff $\Lambda$, the value of parameter $\tau$
must be real and larger than zero, the term $e^{-\tau k^2}$ in
(\ref{dB0}) then provides a natural suppression factor for the
momentum integration and this leads the convergence of the
integration. For a converged integration, we can replace the term
$e^{i\bar{\nabla}_{x}^{\mu}\cdot\frac{\partial}{\partial
 k}}$ in front of the integration kernel in (\ref{dB0}) by $1$, since the difference $(e^{i\bar{\nabla}_{x}^{\mu}\cdot\frac{\partial}{\partial
 k}}-1)\cdots$ are some momentum total derivative terms which vanish as long as we have nontrivial suppression factor
$e^{-\tau k^2}$. With these considerations, (\ref{dB0}) becomes
\begin{eqnarray}
 &&\hspace{-0.5cm}\mathrm{ReTr}\ln[\slashed{\partial}-i\slashed{v}_\Omega-i\slashed{a}_\Omega\gamma_5
 -s_\Omega+ip_\Omega\gamma_5+\Sigma(-\bar{\nabla}^2)]
  =-\frac{1}{2}\lim_{\Lambda \to \infty}
 \int^\infty_{\frac{1}{\Lambda^2}}\frac{d\tau}{\tau}\int{d^4x}\int\frac{d^4k}{(2\pi)^4}\mathrm{tr}~e^B\cdot
 1\;,
 ~~~~~\label{dB1}\\
 &&\hspace{-0.5cm}B\equiv-\tau\bigg[\tilde{E}+(k+\tilde{F})^2
 +\tilde{a}_\mu\gamma_5(k^\mu+\tilde{F}^\mu)+(k^\mu+\tilde{F}^\mu)\tilde{a}_\mu\gamma_5
 +\tilde{a}^2
 +\Sigma^2\big((k+\tilde{F})^2\big)+\tilde{J}\Sigma\big((k+\tilde{F})^2\big)
 +\Sigma\big((k+\tilde{F})^2\big)\tilde{K}
 \notag\\
 &&\hspace{0.3cm}-\gamma_\mu\Big[\tilde{\bar{\nabla}}^\mu_x,\Sigma\big((k+\tilde{F})^2\big)\Big]\bigg]\;.\label{Bdef}
 \end{eqnarray}
 From (\ref{Bdef}), we see that all $\bar{\nabla}^\mu$ in (\ref{dB1}) appear as commutators,
 therefore (\ref{dB1}) and (\ref{Bdef}) offer a covariant formulation
 which matches the general result that the real part of
 $\mathrm{Trln}\cdots$ should be invariant under local chiral
 transformations. The price is that we
 need to handle many momentum derivatives on the exponential and the result computations become extremely
 lengthy. But as long as our reformulation is suitable to computerize,
 it is worth to pay such a price. To deal the next problem of derivatives on the exponential, we
 first take the low energy expansion on $B$
 \begin{eqnarray}
 B=B_0+B_1+\frac{1}{2}B_2+\frac{1}{3!}B_3+\frac{1}{4!}B_4
 +\frac{1}{5!}B_5+\frac{1}{6!}B_6+\cdots\label{Bexp}
 \end{eqnarray}
with $\frac{1}{n!}B_n$ is the $p^n$ order part of $B$. Further
introduce a parameter $t$ dependent $B(t)$ as
 \begin{eqnarray}
 B(t)=B_0+tB_1+\frac{t^2}{2}B_2+\frac{t^3}{3!}B_3+\frac{t^4}{4!}B_4
 +\frac{t^5}{5!}B_5+\frac{t^6}{6!}B_6+\cdots\hspace{2cm}B=B(t)\bigg|_{t=1}\;.
 \end{eqnarray}
 Then take Taylor expansion of $e^{B(t)}$ at point $t=0$,
 \begin{eqnarray}
 e^B=e^{B(t)}\bigg|_{t=1}=e^{B_0}+[\frac{d}{dt}e^{B(t)}]_{t=0}
 +\frac{1}{2!}[\frac{d^2}{dt^2}e^{B(t)}]_{t=0}
 +\frac{1}{3!}[\frac{d^3}{dt^3}e^{B(t)}]_{t=0}
 +\frac{1}{4!}[\frac{d^4}{dt^4}e^{B(t)}]_{t=0}
 +\ldots\;.~~~~~\label{EBexp0}
 \end{eqnarray}
 With the help of identities
 \begin{eqnarray}
 [\frac{d}{d\tau}e^B]e^{-B}=f(Ad(B))(\frac{dB}{d\tau})\hspace{3cm}
f(z)=\frac{e^z-1}{z}=1+\frac{z}{2!}+\frac{z^2}{3!}+\cdots\;.
 \end{eqnarray}
One can explicitly work out
$\frac{1}{n!}[\frac{d^n}{dt^n}e^{B(t)}]_{t=0}$, for several lowest
orders
\begin{eqnarray}
 \frac{d}{d t}e^{B(t)}\bigg|_{t=0}&=&e^{B_0}f[Ad(-B_0)](B_1)\;,\label{dEBt}\\
  \frac{d^2}{d t^2}e^{B(t)}\bigg|_{t=0}&=&\frac{d}{d t}e^{B(t)}\bigg|_{t=0}~f[Ad(-B_0)](B_1)
 +e^{B_0}\frac{d f}{d t}[Ad(-B(t))]\bigg|_{t=0}(B_1)+e^{B_0}f[Ad(-B_0)](B_2)\;,\label{d2EBt}
 \end{eqnarray}
 where
  \begin{eqnarray}
 \frac{d^mf}{d t^m}[Ad(-B(t))]={\displaystyle\sum_{n=0}^{\infty}}
 \frac{1}{(n+1)!}\frac{d^m}{dt^m}[Ad(-B(t))]^n\;.
 \end{eqnarray}
For more higher orders needed in our computations, we list the
results of $\frac{d^3}{d t^3}e^{B(t)}\bigg|_{t=0}$, $\frac{d^4}{d
t^4}e^{B(t)}\bigg|_{t=0}$, $\frac{d^5}{d t^5}e^{B(t)}\bigg|_{t=0}$
and $\frac{d^6}{d t^6}e^{B(t)}\bigg|_{t=0}$ in Appendix \ref{EBexp}.

With the help of (\ref{dEBt}), (\ref{d2EBt}) and
(\ref{d3EBt})-(\ref{d6EBt}) , as long as the
$B_0,B_1,B_2,B_3,B_4,B_5,B_6$ are known, (\ref{EBexp0}) is known and
we can substitute it back  into (\ref{dB1}) to calculate the real
part of
$-iN_c\mathrm{Tr}\ln[i\slashed{\partial}+J_{\Omega}-\Sigma(\overline{\nabla}^2)]$
order by orders up to the $p^6$ order in the low energy expansion.
To obtain $B_i$, (\ref{Bdef}) tells us that the difficulty is the
low energy expansion for $\Sigma\big((k+\tilde{F})^2\big)$. To
achieve it, we expand the argument of
$\Sigma\big((k+\tilde{F})^2\big)$ as
\begin{eqnarray}
(k+\tilde{F})^2&=&k^2+\frac{1}{2}A_2+\frac{1}{6}A_3+\frac{1}{24}A_4
 +\frac{1}{120}A_5+\frac{1}{720}A_6+O(p^{5,6})|_\mathrm{traceless}+O(p^7)\;,
 \end{eqnarray}
 in which
  \begin{eqnarray}
 A_2&=&-2 (\mu\nu)k_{\mu}\frac{\partial}{\partial k^{\nu}}\;,\\
 A_3&=&4i (\mu \nu \lambda )k_{\nu}\frac{\partial^2}{\partial k^{\mu}\partial k^{\lambda}}
 +2i (\mu \underline{\mu} \nu )\frac{\partial}{\partial k^{\nu}}\;,\\
 A_4&=&6 (\mu \nu \lambda \rho )k_{\lambda}\frac{\partial^3}{\partial k^{\mu}\partial k^{\nu}\partial k^{\rho}}
 +6 (\mu \nu )(\underline{\mu} \lambda )\frac{\partial^2}{\partial k^{\nu}\partial k^{\lambda}}
 +3 (\mu \nu \underline{\nu} \lambda )\frac{\partial^2}{\partial k^{\mu}\partial k^{\lambda}}
 +3 (\mu \nu \underline{\mu} \lambda )\frac{\partial^2}{\partial k^{\nu}\partial k^{\lambda}}\;,\\
 A_5&=&0\;,\\
 A_6&=&-90 (\mu \nu \lambda \rho )(\underline{\lambda} \sigma )\frac{\partial^4}{\partial k^{\mu}\partial k^{\nu}\partial k^{\rho}\partial k^{\sigma}}
 -80 (\mu \nu \lambda )(\rho \underline{\nu} \sigma )\frac{\partial^4}{\partial k^{\mu}\partial k^{\lambda}\partial k^{\rho}\partial
 k^{\sigma}}\;.
 \end{eqnarray}
Since we are only interested in the terms not higher than $p^6$, we
find that those traceless terms of $p^5$ and $p^6$ orders will not
make contributions to the final result. So to save space and
simplify the computations, we do not explicitly write down the
detail structure of them, just represent these terms with symbol
$O(p^{5,6})|_\mathrm{traceless}$ and remove traceless term in $A_5$
and $A_6$. Further introduce $A(t)$ as,
\begin{eqnarray}
A(t)\equiv k^2+\frac{t^2}{2}A_2+\frac{t^3}{6}A_3+\frac{t^4}{24}A_4
 +\frac{t^5}{120}A_5+\frac{t^6}{720}A_6+O(p^{5,6})|_\mathrm{traceless}+O(p^7)\hspace{1cm}A(t)=\left\{
 \begin{array}{lll}k^2&&t=0\\&&\\(k+\tilde{F})^2&&t=1\end{array}.\right.~~~~
 \end{eqnarray}
 Then
\begin{eqnarray}
\Sigma\big((k+\tilde{F})^2\big)=\Sigma(k^2) +\bigg[\frac{d}{d
t}\Sigma[A(t)]\bigg]_{t=0}
 +\frac{1}{2!}\bigg[\frac{d^2}{d t^2}\Sigma[A(t)]\bigg]_{t=0}
 +\frac{1}{3!}\bigg[\frac{d^3}{d t^3}\Sigma[A(t)]\bigg]_{t=0}
+\frac{1}{4!}\bigg[\frac{d^4}{d
 t^4}\Sigma[A(t)]\bigg]_{t=0}+\ldots\;.\label{Sigmaexp}
\end{eqnarray}
Now, we need to know  $\bigg[\frac{d^m}{d
t^m}\Sigma[A(t)]\bigg]_{t=0}$, using the following formula
 \begin{eqnarray}
 \Sigma[A(t)]&=&\Sigma[s+A(t)]\bigg|_{s=0}
 =e^{A(t)\frac{\partial}{\partial s}}\Sigma(s)
 e^{-A(t)\frac{\partial}{\partial s}}\bigg|_{s=0}
 =e^{A(t)\frac{\partial}{\partial s}}\Sigma(s)\bigg|_{s=0}\;,
 \end{eqnarray}
 then
 \begin{eqnarray}
 \bigg[\frac{d^m}{d t^m}\Sigma[A(t)]\bigg]_{t=0} =\bigg[\frac{d^m}{d
 t^m}e^{A(t)\frac{\partial}{\partial s}}
 \bigg]_{t=0}\Sigma(s)\bigg|_{s=0}\;.
 \end{eqnarray}
 Therefore to compute $\bigg[\frac{d^m}{d
t^m}\Sigma[A(t)]\bigg]_{t=0}$,  we only need to calculate
$\bigg[\frac{d^m}{d
 t^m}e^{A(t)\frac{\partial}{\partial s}}
 \bigg]_{t=0}\Sigma(s)\bigg|_{s=0}$ which is just equivalent to
 replace $B_l\rightarrow
 A_l\frac{\partial}{\partial s}$ in (\ref{dEBt}),(\ref{d2EBt}) and
(\ref{d3EBt})-(\ref{d6EBt}),  followed by multiplying an extra
factor $\Sigma(s)$ at the r.h.s. and vanishing parameter $s$ after
finishing all differential operations. Follow this calculation road
map, the detail calculation gives
 \begin{eqnarray}
 \bigg[\frac{d}{d t}\Sigma[A(t)]\bigg]_{t=0} &=&
 e^{Ad(A_0\frac{\partial}{\partial s})}(f[Ad(-A_0\frac{\partial}{\partial s})]
 A_1)\Sigma'(s+A_0)\bigg|_{s=0}=0\;,\\
 \frac{1}{2}\bigg[\frac{d^2}{d t^2}\Sigma[A(t)]\bigg]_{t=0}
 &=&\frac{1}{2}e^{Ad(A_0\frac{\partial}{\partial s})}\bigg[\bigg(e^{-A_0\frac{\partial}{\partial s}}
 \frac{d}{d t}e^{A(t)\frac{\partial}{\partial s}}\bigg|_{t=0}
 f[Ad(-A_0\frac{\partial}{\partial s})](A_1\frac{\partial}{\partial s})
 +\frac{d f}{d t}[Ad(-A(t)\frac{\partial}{\partial s})]\bigg|_{t=0}(A_1\frac{\partial}{\partial s})\nonumber\\
 &&+f[Ad(-A_0\frac{\partial}{\partial s})]
 (A_2\frac{\partial}{\partial
 s})\bigg)\bigg]\Sigma(s+A(t))\bigg|_{s=0}=- (\mu,\nu)k_{\mu}\Sigma'_k
 \frac{\partial}{\partial k^{\nu}}\;,
 \end{eqnarray}
where $\Sigma_k\equiv\Sigma(k^2)$. For more higher orders, we list
the results of $\bigg[\frac{d^3}{d t^3}\Sigma[A(t)]\bigg]_{t=0}$,
$\bigg[\frac{d^4}{d t^4}\Sigma[A(t)]\bigg]_{t=0}$,
$\bigg[\frac{d^5}{d t^5}\Sigma[A(t)]\bigg]_{t=0}$ and
$\bigg[\frac{d^6}{d t^6}\Sigma[A(t)]\bigg]_{t=0}$ in Appendix
\ref{EBexp}. With these results, we finally obtain the low energy
expansion of $B$,
\begin{eqnarray}
 B_0&=&-\tau (k^2+\sk^2)\;,\\
 B_1&=&2\tau(-a^{\mu}_{\Omega}k_{\mu}
 +ia_\Omega^{\mu}\gamma_\mu\sk)\gamma_5\;, \\
 B_2&=&-2 a_\Omega^2\tau
 + (\mu \nu )\gamma_{\mu}\gamma_{\nu}\tau
 - a_\Omega^{\mu}a_\Omega^{\nu}[\gamma_{\mu},\gamma_{\nu}]\tau
  -i (d^{\mu}a_\Omega^\nu-d^{\nu}a_\Omega^{\mu})\gamma_{\mu}\gamma_{\nu}\gamf \tau
 +4 s_\Omega \tau \sk+2 (\mu \nu )\tau k_{\mu}\frac{\partial}{\partial k^{\nu}}
 +4 (\mu a_\Omega^{\nu} )\gamma_{\nu}\gamf \tau k_{\mu}\skp\notag\\
 &&
 +4 (\mu a_\Omega^{\nu} )\gamma_{\nu}\gamf \tau \sk \frac{\partial}{\partial k^{\mu}}
 +2i (\mu a_{\Omega\mu} )\gamf \tau
 +4i (\mu a_\Omega^{\nu} )\gamf \tau k_{\nu}\frac{\partial}{\partial k^{\mu}}+4i (\mu \nu )\gamma_{\mu}\tau k_{\nu}\skp
 +4 (\mu \nu )\tau k_{\mu}\sk \skp \frac{\partial}{\partial
 k^{\nu}}\;.
\end{eqnarray}
We list $B_3,B_4, B_5, B_6$ in Appendix.\ref{EBexp}.  With these
explicit expressions for $B_0,B_1,B_2,B_3,B_4,B_5,B_6$, using
(\ref{dEBt}), (\ref{d2EBt}) and (\ref{d3EBt})-(\ref{d6EBt}) , we get
(\ref{EBexp0}) and further substitute (\ref{EBexp0})  back into
(\ref{dB1}), we can obtain the real part of
$-iN_c\mathrm{Tr}\ln[i\slashed{\partial}+J_{\Omega}-\Sigma(\overline{\nabla}^2)]$
order by orders up to the $p^6$ order in the low energy expansion.
The analytical results of $p^2$ and $p^4$ orders  are the same as
those given by (34),(35) and (36) in Ref.\cite{WQ1}, except some
total derivative terms which, as we mentioned before, can be ignored
as long as we take finite cutoff $\Lambda$.
%%%%%%%%%%%%%%%%%%%%%%%%%%%%%%%%%%%%%%%%%%%%%%%%%%%%%%%%%%%%%%%%%
\section{Ambiguities in the Anomaly Part Contributions to the Chiral Lagrangian}

In the last section, we have introduced a  chiral covariant method
to calculate
$\mathrm{Tr}\ln[i\slashed{\partial}+J_{\Omega}-\Sigma(\overline{\nabla}^2)]$
which is already computerized now. With the help of computer, for
the $p^2$ and $p^4$ order analytical formulae in the low energy
expansion, we can get results within 15 minutes, while for the $p^6$
order terms, we need roughly 13 hours to output all expansion
results. From our general result (\ref{Seff1}), the term
$-iN_c\mathrm{Tr}\ln[i\slashed{\partial}+J_{\Omega}-\Sigma(\overline{\nabla}^2)]$
 is the normal part. To get the full result of the chiral
 Lagrangian, we need to calculate the remaining anomaly part
 contributions
$iN_c\mathrm{Tr}\ln[i\slashed{\partial}+J_{\Omega}]-iN_c\mathrm{Tr}\ln[i\slashed{\partial}+J]$.
 As the discussion of Ref.\cite{WQ3}, in 1980s there is a class of
 works (see references given in \cite{WQ3}) identifying this part as the
 full chiral Lagrangian, and in Ref.\cite{WQ3} we refer them as the anomaly approach of calculating LECs.
 In our previous work \cite{WQ1}, we pointed out that this anomaly part contributions are completely
 canceled by the normal part contribution, left nontrivial pure
 $\Sigma$ dependent terms contribute to the chiral Lagrangian.

For the anomaly part contributions,  the key is to calculate the $U$
field dependent term
$\mathrm{Tr}\ln[i\slashed{\partial}+J_{\Omega}]$ which, as we
mentioned before, can be obtained by vanishing $\Sigma$ in
$\mathrm{Tr}\ln[i\slashed{\partial}+J_{\Omega}-\Sigma(\overline{\nabla}^2)]$.
In practice, the limit was taken by first assuming $\Sigma$ being a
constant mass $m$ and then letting $m\rightarrow 0$. For $p^2$
order, this operation gives null result, while for $p^4$ order, it
gives the result originally presented in anomaly approach. Now in
this work, naively what we need to do is to generalize the
calculation to $p^6$ order. But to our surprise, we get many terms
with divergent coefficients. Checking the calculation carefully, we
find that the reason of appearance of infinities is due to the fact
that most of the coefficients in front of the $p^6$ order operators
have dimension of $1/m^2$ which goes to infinity when we take limit
$\Sigma=m\rightarrow 0$. Note that the $p^6$ terms may also have
coefficients of $1/\Lambda^2$ which are finite in the limit of
$m\rightarrow 0$ , although they vanish when we take
$\Lambda\rightarrow\infty$. These terms are irrelevant to our
discussion on the divergence of $p^6$ order terms and therefore we
do not need to care about them. Applying the argument on $1/m^2$
dependence of the $p^6$ order coefficients back to the $p^2$ and
$p^4$ order results we discussed before, coefficients in front of
$p^2$ order operators have dimension of $m^2$ which goes to zero,
this explains the phenomena that anomaly approach can not produce
$p^2$ order terms. For $p^4$ order, the coefficients in front of
operators are dimensionless and therefore the $m$ dependence is at
most logarithmic of form $\ln m/\Lambda$ which implies existence of
a logarithmic ultraviolet divergence. Since we know that in the
large $N_c$ limit, the $p^4$ order LECs (non-contact coefficients)
are not divergent, the $\ln m/\Lambda$ term then can not appear in
the final expression of these LECs, therefore in $p^4$ order,
anomaly approach leads finite result LECs. In general for a $p^{2n}$
order operator, the corresponding coefficient should has dimension
$1/m^{2(n-2)}$. This implies that the infinity in the anomaly part
contributions will be a general phenomena, when we go to the higher
orders of the low energy expansion, since the more higher the order
is, the more negative powers of $m$ dependence the coefficient will
have and these negative powers of $m$ will result in infinities as
we take limit $\Sigma=m\rightarrow 0$.

  The appearance of these high order infinities provides another
evidence that the anomaly approach is not a correct formulation in
calculating LECs, at least not for the $p^6$ and more higher order
LECs. Since high order divergence term is as an addition part of our
general result (\ref{Seff1}), we can not avoid them in our
computations. How to deal with these high order infinities from
negative powers of $m$? There exists an alternative way, not relying
on the low energy expansion, to examine this anomaly part
contributions in which we must exploit the first equation of
 (\ref{JOmega}) and we find
\begin{eqnarray}
\mathrm{Tr}\ln[i\slashed{\partial}+J_{\Omega}]-\mathrm{Tr}\ln[i\slashed{\partial}+J]
&=&\ln\mathrm{Det}[i\slashed{\partial}+J_{\Omega}]-\ln\mathrm{Det}[i\slashed{\partial}+J]\nonumber\\
&=&\ln\mathrm{Det}\bigg[[\Omega
P_R+\Omega^{\dag}P_L][J+i\slashed{\partial}][\Omega P_R+\Omega^\dag
P_L]\bigg]-\ln\mathrm{Det}[i\slashed{\partial}+J]\nonumber\\
&=&\ln\mathrm{Det}\bigg[[\Omega P_R+\Omega^{\dag}P_L][\Omega
P_R+\Omega^\dag P_L]\bigg]=\mathrm{Tr}\ln\bigg[[\Omega
P_R+\Omega^{\dag}P_L][\Omega P_R+\Omega^\dag P_L]\bigg]\;.
\end{eqnarray}
For our interests, we are only interested in the real part of it,
then
\begin{eqnarray}
\mathrm{ReTr}\ln[i\slashed{\partial}+J_{\Omega}]-\mathrm{ReTr}\ln[i\slashed{\partial}+J]
&=&\frac{1}{2}\mathrm{Tr}\ln\bigg[[\Omega
P_R+\Omega^{\dag}P_L][\Omega P_R+\Omega^\dag
P_L]\bigg]+\frac{1}{2}\mathrm{Tr}\ln\bigg[[\Omega
P_R+\Omega^{\dag}P_L]^\dag[\Omega P_R+\Omega^\dag
P_L]^\dag\bigg]\nonumber\\
&=&\frac{1}{2}\mathrm{Tr}\ln\bigg[[\Omega
P_R+\Omega^{\dag}P_L][\Omega P_R+\Omega^\dag
P_L]\bigg]+\frac{1}{2}\mathrm{Tr}\ln\bigg[[\Omega^\dag P_R+\Omega
P_L][\Omega^\dag P_R+\Omega P_L]\bigg]\nonumber\\
&=&\frac{1}{2}\mathrm{Tr}\ln\bigg[[P_R+P_L][P_R+P_L]\bigg]=\frac{1}{2}\mathrm{Tr}\ln
1=0\;.
\end{eqnarray}
Which shows that the compact form of anomaly part contributions to
normal part of the chiral Lagrangian is zero !

How can this null result be consistent with another divergent result
obtained from the low energy expansion ? The only possible
explanation is that the $p^4$ order finite term plus all those
higher order infinities results a zero ! Is it possible ? A
well-known positive example is the expansion
$1/(1+x)=1-x+x^2-x^3+x^4-x^5+x^6-\cdots$ goes to zero when $x$ is
very large, which implies that the summation of series
$x-x^2+x^3-x^4+x^5-x^6+\cdots$ converges to $1$ when $x$ is very
large and each individual term in the series diverges. In the
following we take a more realistic but simplified example to show
that this really happens in our formulation. Our example starts from
(\ref{Trln}) for the case of $\Sigma$ equal to a constant mass $m$
 \begin{eqnarray}
 &&\mathrm{ReTr}\ln[\slashed{\partial}-i\slashed{v}_\Omega-i\slashed{a}_\Omega\gamma_5-s_\Omega+ip_\Omega\gamma_5
 +m]=-\frac{1}{2}\lim_{\Lambda \to \infty}
 \int^\infty_{\frac{1}{\Lambda^2}}\frac{d\tau}{\tau}\int{d^4x}\int\frac{d^4k}{(2\pi)^4}
 ~\mathrm{tr}e^{-\tau(k^2+k\cdot b'+m^2+bm+C)}\;,~~~~~~\\
 &&{b'}^\mu\equiv2i\bar{\nabla}_x^\mu+2a_\Omega^\mu\hspace{2cm}
 b\equiv\hat{I}_\Omega+\tilde{I}_\Omega\hspace{2cm}C\equiv\bar{E}+(i\bar{\nabla}_x+a_\Omega)^2\;.
\end{eqnarray}
For simplicity, we ignore the contributions from $b'$ which does not
change the key result of our discussion. Then our example becomes to
investigate following integration
\begin{eqnarray}
I\equiv\int_{\frac{1}{\Lambda^2}}^\infty\frac{d\tau}{\tau}\int_0^\infty
k^2 dk^2 e^{-\tau k^2-\tau m^2-\tau bm-\tau C}
\end{eqnarray}
with $b$ and $C$ not commuting each other. We will show that high
order terms in the low energy expansion of the above integration $I$
will go to infinity when we take $m\rightarrow 0$, but if summing
all the expansion terms together, we get finite result which
corresponds to previous null result of summing all higher order
terms of anomaly part contributions into a compact form. We use
three different methods to finish above integration and explain our
point. The first method is to vanish $m$ firstly and then to finish
the integration, i.e.
 \begin{eqnarray}
 I\bigg|_{m=0}=\int_{\frac{1}{\Lambda^2}}^\infty\frac{d\tau}{\tau}\int_0^\infty
k^2 dk^2 e^{-\tau k^2-\tau C}
 =\int_{\frac{1}{\Lambda^2}}^\infty\frac{d\tau}{\tau^3} e^{-\tau C}
=\frac{\Lambda^4}{2}e^{-\frac{C}{\Lambda^2}}
 -\frac{\Lambda^2}{2}Ce^{-\frac{C}{\Lambda^2}}-\frac{1}{2}C^2\mathbf{Ei}(-\frac{C}{\Lambda^2})\;,~~~~~~\label{I0}
 \end{eqnarray}
where
\begin{eqnarray}
\mathbf{Ei}(-x)\equiv-\int_x^{\infty}\frac{e^{-u}}{u}du =\gamma+\ln
x+\sum^\infty_{n=1}\frac{(-1)^n}{n!n}x^{n}\hspace{3cm}|x|<\infty\;.
\end{eqnarray}
The second method is first finishing the integration and then
vanishing $m$,
 \begin{eqnarray}
 I&=&\int_{\frac{1}{\Lambda^2}}^\infty\frac{d\tau}{\tau}\int_0^\infty k^2 dk^2 e^{-\tau (k^2+m^2+ bm+ C)}
 =\int_{\frac{1}{\Lambda^2}}^\infty\frac{d\tau}{\tau^3} e^{-\tau(m^2+ bm+C)}\nonumber\\
 &=&\frac{\Lambda^4}{2}e^{-\frac{(m^2\!+bm+C)^2}{\Lambda^2}}\!
 -\frac{\Lambda^2(m^2\!+\!bm\!+\!C)}{2}e^{-\frac{(m^2\!+bm+C)^2}{\Lambda^2}}\!\!
 -\frac{1}{2}(m^2\!+\!bm\!+\!C)^2\mathbf{Ei}(-\frac{(m^2\!+\!bm\!+\!C)}{\Lambda^2})\nonumber\\
 &\stackrel{m\rightarrow 0}{====}&\frac{\Lambda^4}{2}e^{-\frac{C}{\Lambda^2}}
 -\frac{\Lambda^2}{2}Ce^{-\frac{C}{\Lambda^2}}-\frac{1}{2}C^2\mathbf{Ei}(-\frac{C}{\Lambda^2})\;.
 \end{eqnarray}
 We obtain the same result as that obtained in the first method, therefore interchange the order of integration and
 $m\rightarrow 0$ limit does not change the result.

 The third method is first taking Taylor expansion in terms the power of
 $b$ and $C$ which corresponds performing the low energy expansion and then finishing integration, finally vanishing $m$,
 \begin{eqnarray}
  I&=&\int_{\frac{1}{\Lambda^2}}^\infty\frac{d\tau}{\tau}\int_0^\infty k^2 dk^2 e^{-\tau k^2-\tau m^2}
 \sum^{\infty}_{n=0}\frac{1}{n!}(-\tau b m-\tau C)^n
 =\int_{\frac{1}{\Lambda^2}}^\infty d\tau e^{-\tau m^2}
 \sum^{\infty}_{n=0}\frac{1}{n!}\tau^{n-3} (-bm-C)^n\nonumber\\
  &=&m^4\left(\frac{\Lambda^4}{2m^4}e^{-\frac{m^2}{\Lambda^2}}
 -\frac{\Lambda^2}{2m^2}e^{-\frac{m^2}{\Lambda^2}}-\frac{1}{2}\mathbf{Ei}(-\frac{m^2}{\Lambda^2})\right)
 -(bm^3+Cm^2)\bigg(\frac{\Lambda^2}{m^2}e^{-\frac{m^2}{\Lambda^2}}+\mathbf{Ei}(-\frac{m^2}{\Lambda^2})\bigg)\notag\\
 && -\frac{1}{2}(bm+C)^2\mathbf{Ei}(-\frac{m^2}{\Lambda^2})+
 \sum^{\infty}_{n=0}\frac{(-\frac{b}{m}-\frac{C}{m^2})^{n+3}}{(n+3)!}  m^4\Gamma(n+1,\frac{m^2}{\Lambda^2})\notag\\
 &\stackrel{m\rightarrow 0}{====}&\frac{\Lambda^4}{2}-C\Lambda^2-\frac{1}{2}C^2\mathbf{Ei}(-\frac{m^2}{\Lambda^2})
 +\sum^{\infty}_{n=0}\frac{n!}{(n+3)!}(-\frac{b}{m}-\frac{C}{m^2})^{n+3} m^4e^{-\frac{m^2}{\Lambda^2}}\sum_{k=0}^n\frac{1}{k!}
 \left(\frac{m^2}{\Lambda^2}\right)^k\bigg|_{m\rightarrow 0}\;.\label{I1}
 \end{eqnarray}
 We see that there are negative power of $m$ terms which will cause
 divergence when we take limit $m\rightarrow0$. This is just what has happened for the
  high order terms in the anomaly part contributions. So if we
  calculate term by terms in above expansion, we will meet
  infinities
  which seems contradict with results obtained in first two methods.
  The only way left to escape this contradiction is to sum all these
  divergences together, to see that what will happen after the summation, we introduce a
  series
  \begin{eqnarray}
  g(x,c)&\equiv&\sum^{\infty}_{n=0}\frac{n!}{(n+3)!}x^{n+3}
  \sum_{k=0}^n\frac{c^k}{k!}\;,
\end{eqnarray}
in which $c=m^2/\Lambda^2$ and $x=-\frac{b}{m}-\frac{C}{m^2}$ which
will go to negative infinity when $m\rightarrow0$.  With the help of
relation $\frac{d}{dx}\mathbf{Ei}(-x)=\frac{e^{-x}}{x}$ and boundary
condition $g''(0,c)=g'(0,c)=g(0,c)=0$, we find
\begin{eqnarray}
&&g'''(x,c)=\sum^{\infty}_{n=0}x^{n}\sum_{k=0}^n\frac{c^k}{k!}=\frac{e^{cx}}{1-x}\;,
\hspace{3cm}g''(x,c)=e^c[-\mathbf{Ei}(cx-c)+\mathbf{Ei}(-c)]\;,\notag\\
&&g'(x,c)=(x-1)e^c[-\mathbf{Ei}(cx-c)+\mathbf{Ei}(-c)]+\frac{1}{c}(e^{cx}-1)\;,\notag\\
&&g(x,c)=\frac{1}{2}(x-1)^2e^c[-\mathbf{Ei}(cx-c)+\mathbf{Ei}(-c)]+\frac{x-1}{2c}e^{cx}+\frac{1}{2c}
 +\frac{1}{2c^2}(e^{cx}-1)-\frac{x}{c}\;.
\end{eqnarray}
Then (\ref{I1}) becomes
 \begin{eqnarray}
  I\bigg|_{m=0}&=&\lim_{m\rightarrow0}\bigg[\frac{\Lambda^4}{2}-C\Lambda^2-\frac{1}{2}C^2\mathbf{Ei}(-\frac{m^2}{\Lambda^2})
 +m^4e^{-\frac{m^2}{\Lambda^2}}g(-\frac{b}{m}-\frac{C}{m^2},\frac{m^2}{\Lambda^2})\bigg]\nonumber\\
 &=&\lim_{m\rightarrow0}\bigg[\frac{\Lambda^4}{2}-C\Lambda^2-\frac{1}{2}C^2\mathbf{Ei}(-\frac{m^2}{\Lambda^2})
 +\frac{1}{2}(bm+C+m^2)^2[-\mathbf{Ei}(\frac{-bm-C-m^2}{\Lambda^2})+\mathbf{Ei}(-\frac{m^2}{\Lambda^2})]
\nonumber\\
&&+\frac{1}{2}\Lambda^2(-bm-C-m^2)e^{\frac{-bm-C-m^2}{\Lambda^2}}
+\frac{1}{2}\Lambda^2m^2e^{-\frac{m^2}{\Lambda^2}}+\frac{1}{2}\Lambda^4(
e^{\frac{-bm-C-m^2}{\Lambda^2}}-e^{-\frac{m^2}{\Lambda^2}})+\Lambda^2(bm+C)\bigg]\nonumber\\
&=&-\frac{1}{2}C^2\mathbf{Ei}(-\frac{C}{\Lambda^2})
-\frac{1}{2}\Lambda^2Ce^{-\frac{C}{\Lambda^2}}+\frac{1}{2}\Lambda^4e^{-\frac{C}{\Lambda^2}}\;.
 \end{eqnarray}
It is the same as the results obtained from first two methods, i.e.
summing all those infinities together, we obtain correct finite
result.

With above discussion, our result now is that {\it total anomaly
part contributions to the normal part of the chiral Lagrangian
vanish} ! Just take several individual terms can not reflect the
true result of the full action. In fact,
 finite result of the $p^4$ order plays a role to cancel that summations
 of all higher order terms which results in the final total null result. In
 this sense, in order to make sense for the $p^6$ and more higher order
 divergent terms,  we must sum them together and then we get $p^4$ order
 result with an extra minus sign. To avoid the appearance of divergences in $p^6$ and higher orders
 terms, what we need to do is to drop out all anomaly part contributions, since divergences from high order terms are
 finally canceled by $p^4$ order terms. In this view, our general result (\ref{Seff1}) must be changed to
\begin{eqnarray}
S_\mathrm{eff}\bigg|_\mathrm{normal~part}\approx-iN_c\mathrm{ReTr}\ln[i\slashed{\partial}+J_{\Omega}-\Sigma(\overline{\nabla}^2)]\;.
~~~~~~ \label{Seff3}
\end{eqnarray}
In fact in Ref.\cite{WQma}, we already show that including in the
anomalous part, the total effective action takes the form (see
Eq.(21) in Ref.\cite{WQma}),
\begin{eqnarray}
S_\mathrm{GND}=-iN_c\mathrm{Tr}\ln[i\slashed{\partial}+J_{\Omega}-\Sigma(\overline{\nabla}^2)]+
\mbox{Wess-Zumino terms}\;.
\end{eqnarray}

With this new viewpoint on all anomaly part contributions, we need
to modify our original numerical results on $p^4$ order LECs, since
it takes into account of the finite values of anomaly part
contributions and now we know that these nontrivial values must be
used to cancel the infinities come from all higher order terms. In
Table I, we list our modified $p^4$ order LECs for cutoff
$\Lambda=$1000$^{+100}_{-100}$MeV. The $10\%$ variation of the
cutoff is considered in our calculation to examine the effects of
cutoff dependence and the result change can be treated as the error
of our calculations. The result LECs are taken the values at
$\Lambda=1$GeV with superscript the difference caused at
$\Lambda=1.1$GeV and subscript the difference caused at
$\Lambda=0.9$GeV, i.e.,
 \begin{eqnarray}
 L_{\Lambda=1\mathrm{GeV}}\bigg|^{L_{\Lambda=1.1\mathrm{GeV}}-L_{\Lambda=1\mathrm{GeV}}}_{L_{\Lambda=0.9\mathrm{GeV}}-L_{\Lambda=1\mathrm{GeV}}}\hspace{2cm}
\bar{l}_{\Lambda=1\mathrm{GeV}}\bigg|^{\bar{l}_{\Lambda=1.1\mathrm{GeV}}-\bar{l}_{\Lambda=1\mathrm{GeV}}}_{\bar{l}_{\Lambda=0.9\mathrm{GeV}}-\bar{l}_{\Lambda=1\mathrm{GeV}}}
\hspace{0.5cm}\mbox{or}\hspace{0.5cm}
l_{\Lambda=1\mathrm{GeV}}\bigg|^{l_{\Lambda=1.1\mathrm{GeV}}-l_{\Lambda=1\mathrm{GeV}}}_{l_{\Lambda=0.9\mathrm{GeV}}-l_{\Lambda=1\mathrm{GeV}}}\;.
 \end{eqnarray}
\begin{table}[h]
%\caption
\null\noindent {\small{\bf TABLE I}. The obtained values of the
$p^4$ coefficients $L_1\cdots,L_{10}$ for three flavor quarks and
$\bar{l}_1\cdots,\bar{l}_6,l_7$ for two flavor quarks\\
\hspace*{0cm}where
$l_i\!=\!\frac{1}{32\pi^2}\gamma_i(\bar{l}_i\!+\!\ln\frac{M^2_\pi}{\mu^2})$
 for $i=1,\ldots,7$, $\mu\!=\!$770MeV and $\gamma_i$ are given in Ref.\cite{GS}. Since $\gamma_7\!=\!0$, we calculate $l_7$ instead of $\bar{l}_7$. \\
 \hspace*{0.8cm} Together with the
experimental values given in Ref.\cite{GS} and
our old result given in Ref.\cite{WQ1} for comparisons.\\
$\Lambda_\mathrm{QCD}$, $\Lambda$
 and $-\langle\bar{\psi}\psi\rangle^{\frac{1}{3}}$ are in units of MeV,
  and $L_1\cdots,L_{10},l_7$  are in units of $10^{-3}$.}
\hspace*{-1cm}\begin{tabular}{c c c c c c c c c c c c c}\hline\hline
&$\Lambda_\mathrm{QCD}$&-$\langle\bar{\psi}\psi\rangle^{\frac{1}{3}}$&$L_1$
&$L_2$ &$L_3$ &$L_4$ & $L_5$ & $L_6$ & $L_7$ & $L_8$ & $L_9$ &
$L_{10}$\\\hline $\Lambda=$1000$^{+100}_{-100}$
&453$^{-6}_{+12}$&260$^{-8}_{+9}$& 1.23$^{+0.03}_{-0.04}$ &
2.46$^{+0.05}_{-0.08}$ & -6.85$^{-0.14}_{+0.21}$ &
0.0$^{+0.0}_{-0.0}$ & 1.48$^{-0.01}_{-0.03}$ & 0.0$^{+0.0}_{-0.0}$ &
-0.51$^{+0.05}_{-0.06}$ &
1.02$^{-0.06}_{+0.06}$&8.86$^{+0.24}_{-0.37}$
&-7.40$^{-0.29}_{+0.44}$\\
Ref.\cite{WQ1}:&484&296&0.403&0.805&-3.47&0&1.47&0&-0.792&1.83&2.28&-4.08\\
 Expt:& &250&$0.9\pm 0.3$&$1.7\pm 0.7$&-$4.4\pm 2.5$&$0\pm
0.5$&$2.2\pm 0.5$&$0\pm 0.3$&-$0.4\pm 0.15$&$1.1\pm 0.3$&$7.4\pm
0.7$&-$6.0\pm 0.7$\\ \hline
&$\Lambda_\mathrm{QCD}$&-$\langle\bar{\psi}\psi\rangle^{\frac{1}{3}}$&$\bar{l}_1$
&$\bar{l}_2$ &$\bar{l}_3$ &$\bar{l}_4$ & $\bar{l}_5$ & $\bar{l}_6$ &
$l_7$ & & &
\\\hline $\Lambda=$1000$^{+100}_{-100}$ &465$^{-6}_{+12}$&227$^{-6}_{+8}$&
-4.77$^{-0.17}_{+0.24}$ & 8.01$^{+0.09}_{-0.14}$ &
1.97$^{+0.29}_{-0.35}$ & 4.34$^{-0.01}_{-0.02}$ &
17.35$^{+0.53}_{-0.80}$ & 19.98$^{+0.44}_{-0.67}$ &
-8.18$^{+0.50}_{-0.43}$ &&&\\
 Expt:& &250&$-2.3\pm 3.7$&$6.0\pm 1.3$&$2.9\pm 2.4$&$4.3\pm 0.9$&$13.9\pm 1.3$&$16.5\pm 1.1$& $O(5)$&&&\\
\hline\hline
\end{tabular}
\end{table}

In obtaining the result, we have taken the running coupling constant
as model A given by (40) of Ref.\cite{WQ1} and the low energy value
of this $\alpha_s$ is already chosen well above the critical value
to trigger the S$\chi$SB of the theory. It should be noted that
$\alpha_s$ depends on the number of quark flavors, so does for
$\Sigma$ from SDE. In fixing the $\Lambda_\mathrm{QCD}$ we have
taken input $F_0=87$MeV. The reason of taking this value is that if
the final $F_\pi$ is around value of 93MeV, then our formula shows
$F_0$ must be located around 87MeV. In Sec.VI, we will exhibit this
phenomena explicitly.
%%%%%%%%%%%%%%%%%%%%%%%%%%%%%%%%%%%%%%%%%%%%%%%%%%%%%%%%
\section{$p^6$ order of chiral Lagrangian: normal part}

The general form of $p^6$ order chiral Lagrangian was first
introduced in Ref.\cite{p6-0} and then discussed in Ref.\cite{p6-1}.
Now we can express the normal part of it in terms of our rotated
sources as what we have done in (\ref{p4}) for the $p^4$ order
terms. Considering that our computation is done under large $N_c$
limit, within this approximation, terms in the chiral Lagrangian
with two and more traces vanish when we not apply the equation of
motion. To avoid unnecessary complicities, in this paper we only
write down those terms with one trace
\begin{eqnarray}
S_\mathrm{eff}\bigg|_{p^6,~\mathrm{normal}}=\int
d^4x~\bigg[~{\displaystyle\sum_{n=1}^{94}}~\mathcal{Z}_n\mathrm{tr}_f[\bar{O}_n]~+~O(\frac{1}{N_c})\bigg]
\label{Sp6ours}
\end{eqnarray}
with $\bar{O}_n$ being $p^6$ order operator we could get in our
calculation and $\mathcal{Z}_n$ being corresponding coefficient,
$O(\frac{1}{N_c})$ are consist of the most multi-traces terms. Our
computations then give the explicit expressions of $\mathcal{Z}_n$
in terms of quark self energy. The detail expressions are given in
(\ref{ZSigma}). And the definitions of operators $\bar{O}_n$ for
$n=1,2,\ldots,94$ are given in Table.II,
\begin{eqnarray}
 &&\hspace{6.5cm}\mbox{\small{\bf TABLE II}.~~$p^6$ order operators}\nonumber\\
 &&\hspace*{-1cm}\begin{array}{|r|c|r|c|r|c|}
\hline n& \bar{O}_n & n& \bar{O}_n &n& \bar{O}_n \\ \hline
  1 &  (a_\Omega^2)^3
  & 33 &  a_\Omega^{\mu}a_\Omega^{\nu}a_{\Omega\mu}d_{\nu}p_\Omega
  & 65 &  d^2a_\Omega^{\nu}d_{\nu}p_\Omega
 \\
 2 &  a_\Omega^2a_\Omega^{\nu}a_\Omega^{\lambda}a_{\Omega\nu}a_{\Omega\lambda}
  & 34 &  a_\Omega^{\mu}a_\Omega^{\nu}(d_{\mu}a_{\Omega\nu}p_\Omega
 + p_\Omega d_{\nu}a_{\Omega\mu})
  & 66 &  d^{\mu}d^{\nu}a_{\Omega\nu}d_{\mu}p_\Omega
 \\
 3 &  a_\Omega^2a_\Omega^{\nu}a_\Omega^2a_{\Omega\nu}
  & 35 &  a_\Omega^{\mu}a_\Omega^{\nu}(d_{\nu}a_{\Omega\mu}p_\Omega
 + p_\Omega d_{\mu}a_{\Omega\nu})
  & 67 &  d^{\mu}s_{\Omega}d_{\mu}s_\Omega
 \\
 4 &  a_\Omega^{\mu}a_\Omega^{\nu}a_{\Omega\mu}a_\Omega^{\lambda}a_{\Omega\nu}a_{\Omega\lambda}
  & 36 &  a_\Omega^{\mu}p_\Omega a_{\Omega\mu}d^{\nu}a_{\Omega\nu}
  & 68 &  d^{\mu}p_{\Omega}d_{\mu}p_\Omega
 \\
 5 &  a_\Omega^{\mu}a_\Omega^{\nu}a_\Omega^{\lambda}a_{\Omega\mu}a_{\Omega\nu}a_{\Omega\lambda}
  & 37 &  a_\Omega^{\mu}p_{\Omega}a_\Omega^{\nu}(d_{\mu}a_{\Omega\nu}
 + d_{\nu}a_{\Omega\mu})
  & 69 &  iV_\Omega^{\mu\nu}V_{\Omega\mu}^{~~\lambda}V_{\Omega\nu\lambda}
 \\
 6 &  a_\Omega^2(d^{\nu}a_{\Omega\nu})^2
  & 38 &  a_\Omega^{\mu}(d_{\mu}a_\Omega^{\nu}d_{\nu}s_\Omega
 + d^{\nu}s_{\Omega}d_{\mu}a_{\Omega\nu})
  & 70 &  V_\Omega^{\mu\nu}V_{\Omega\mu\nu}a_\Omega^2
 \\
 7 &  a_\Omega^2d^{\nu}a_\Omega^{\lambda}d_{\nu}a_{\Omega\lambda}
  & 39 &  a_\Omega^{\mu}(d^{\nu}a_{\Omega\mu}d_{\nu}s_\Omega
 + d^{\nu}s_{\Omega}d_{\nu}a_{\Omega\mu})
  & 71 &  V_\Omega^{\mu\nu}V_{\Omega\mu}^{~~\lambda}a_{\Omega\nu}a_{\Omega\lambda}
 \\
 8 &  a_\Omega^2d_{\nu}a_\Omega^{\lambda}d_{\lambda}a_{\Omega\nu}
  & 40 &  a_\Omega^{\mu}(d^{\nu}a_{\Omega\nu}d_{\mu}s_\Omega
 + d_{\mu}s_{\Omega}d^{\nu}a_{\Omega\nu})
  & 72 &  V_\Omega^{\mu\nu}V_{\Omega\mu}^{~~\lambda}a_{\Omega\lambda}a_{\Omega\nu}
 \\
 9 &  a_\Omega^{\mu}a_\Omega^{\nu}(d_{\mu}a_{\Omega\nu}d^{\lambda}a_{\Omega\lambda}
 + d^{\lambda}a_{\Omega\lambda}d_{\nu}a_{\Omega\mu})
  & 41 &  (a_\Omega^2)^2s_\Omega
  & 73 &  V_\Omega^{\mu\nu}(a_{\Omega\mu}V_{\Omega\nu}^{~~\lambda}a_{\Omega\lambda}
 - a_\Omega^{\lambda}V_{\Omega\mu\lambda}a_{\Omega\nu})
 \\
 10 &  a_\Omega^{\mu}a_\Omega^{\nu}d_{\mu}a_\Omega^{\lambda}d_{\nu}a_{\Omega\lambda}
  & 42 &
  a_\Omega^{\mu}a_\Omega^{\nu}a_{\Omega\mu}a_{\Omega\nu}s_\Omega
  & 74 &  V_\Omega^{\mu\nu}a_\Omega^{\lambda}V_{\Omega\mu\nu}a_{\Omega\lambda}
 \\
 11 &  a_\Omega^{\mu}a_\Omega^{\nu}(d_{\mu}a_\Omega^{\lambda}d_{\lambda}a_{\Omega\nu}
 + d^{\lambda}a_{\Omega\mu}d_{\nu}a_{\Omega\lambda})
  & 43 &  a_\Omega^{\mu}a_\Omega^2a_{\Omega\mu}s_\Omega
  & 75 &  V_\Omega^{\mu\nu}V_{\Omega\mu\nu}s_\Omega
 \\
 12 &  a_\Omega^{\mu}a_\Omega^{\nu}(d_{\nu}a_{\Omega\mu}d^{\lambda}a_{\Omega\lambda}
 + d^{\lambda}a_{\Omega\lambda}d_{\mu}a_{\Omega\nu})
  & 44 &  ia_\Omega^{\mu}(d_{\mu}a_\Omega^{\nu}d^{\lambda}V_{\Omega\nu\lambda}
 + d^{\nu}V_{\Omega\nu}^{~~\lambda}d_{\mu}a_{\Omega\lambda})
  & 76 &  iV_\Omega^{\mu\nu}(a_{\Omega\mu}d_{\nu}p_\Omega
 + d_{\mu}p_{\Omega}a_{\Omega\nu})
 \\
 13 &  a_\Omega^{\mu}a_\Omega^{\nu}d_{\nu}a_\Omega^{\lambda}d_{\mu}a_{\Omega\lambda}
  & 45 &  ia_\Omega^{\mu}(d^{\nu}a_{\Omega\mu}d^{\lambda}V_{\Omega\nu\lambda}
 + d^{\nu}V_{\Omega\nu}^{~~\lambda}d_{\lambda}a_{\Omega\mu})
  & 77 &  iV^{\mu\nu}(p_\Omega d_{\mu}a_{\Omega\nu}
 - d_{\mu}a_{\Omega\nu}p_\Omega)
 \\
 14 &  a_\Omega^{\mu}a_\Omega^{\nu}(d_{\nu}a_\Omega^{\lambda}d_{\lambda}a_{\Omega\mu}
 + d^{\lambda}a_{\Omega\nu}d_{\mu}a_{\Omega\lambda})
  & 46 &  ia_\Omega^{\mu}(d^{\nu}a_{\Omega\nu}d^{\lambda}V_{\Omega\mu\lambda}
 - d^{\nu}V_{\Omega\mu\nu}d^{\lambda}a_{\Omega\lambda})
  & 78 &  iV_\Omega^{\mu\nu}(d_{\mu}a_{\Omega\nu}d^{\lambda}a_{\Omega\lambda}
 - d^{\lambda}a_{\Omega\lambda}d_{\mu}a_{\Omega\nu})
 \\
 15 &  a_\Omega^{\mu}a_\Omega^{\nu}d^{\lambda}a_{\Omega\mu}d_{\lambda}a_{\Omega\nu}
  & 47 &  ia_\Omega^{\mu}(d^{\nu}a_\Omega^{\lambda}d_{\mu}V_{\Omega\nu\lambda}
 - d_{\mu}V_\Omega^{\nu\lambda}d_{\nu}a_{\Omega\lambda})
  & 79 &  iV_\Omega^{\mu\nu}d_{\mu}a_\Omega^{\lambda}d_{\nu}a_{\Omega\lambda}
 \\
 16 &  a_\Omega^{\mu}a_\Omega^{\nu}d^{\lambda}a_{\Omega\nu}d_{\lambda}a_{\Omega\mu}
  & 48 &  ia_\Omega^{\mu}(d^{\nu}a_\Omega^{\lambda}d_{\nu}V_{\Omega\mu\lambda}
 - d^{\nu}V_{\Omega\mu}^{~~\lambda}d_{\nu}a_{\Omega\lambda})
  & 80 &  iV_\Omega^{\mu\nu}(d_{\mu}a_\Omega^{\lambda}d_{\lambda}a_{\Omega\nu}
 + d^{\lambda}a_{\Omega\mu}d_{\nu}a_{\Omega\lambda})
 \\
 17 &  a_\Omega^{\mu}(d_{\mu}a_\Omega^{\nu}a_{\Omega\nu} + d^{\nu}a_{\Omega\mu}a_{\Omega\nu})
 d^{\lambda}a_{\Omega\lambda}
  & 49 &  ia_\Omega^{\mu}(d^{\nu}a_\Omega^{\lambda}d_{\lambda}V_{\Omega\mu\nu}
 - d^{\nu}V_{\Omega\mu}^{~~\lambda}d_{\lambda}a_{\Omega\nu})
  & 81 &  iV_\Omega^{\mu\nu}d^{\lambda}a_{\Omega\mu}d_{\lambda}a_{\Omega\nu}
 \\
 18 &  a_\Omega^{\mu}(d_{\mu}a_\Omega^{\nu}a_\Omega^{\lambda}d_{\nu}a_{\Omega\lambda}
 + d^{\nu}a_\Omega^{\lambda}a_{\Omega\nu}d_{\lambda}a_{\Omega\mu})
  & 50 &  d^{\mu}V_{\Omega\mu}^{~\nu}d^{\lambda}V_{\Omega\nu\lambda}
  & 82 &  iV_\Omega^{\mu\nu}(a_{\Omega\mu}a_{\Omega\nu}s_\Omega
 + s_{\Omega} a_{\Omega\mu}a_{\Omega\nu})
 \\
 19 &  a_\Omega^{\mu}(d_{\mu}a_\Omega^{\nu}a_\Omega^{\lambda}d_{\lambda}a_{\Omega\nu}
 + d^{\nu}a_\Omega^{\lambda}a_{\Omega\nu}d_{\mu}a_{\Omega\lambda})
  & 51 &  d^{\mu}V_\Omega^{\nu\lambda}d_{\mu}V_{\Omega\nu\lambda}
  & 83 &  i V_\Omega^{\mu\nu}a_{\Omega\mu}s_{\Omega} a_{\Omega\nu}
 \\
 20 &  a_\Omega^{\mu}(d^{\nu}a_{\Omega\mu}a_\Omega^{\lambda}d_{\nu}a_{\Omega\lambda}
 + d^{\nu}a_\Omega^{\lambda}a_{\Omega\lambda}d_{\nu}a_{\Omega\mu})
  & 52 &  d^{\mu}V_\Omega^{\nu\lambda}d_{\nu}V_{\Omega\mu\lambda}
  & 84 &  i V_\Omega^{\mu\nu}(a_{\Omega\mu}a_{\Omega\nu}a_\Omega^2
 + a_\Omega^2a_{\Omega\mu}a_{\Omega\nu})
 \\
 21 &  a_\Omega^{\mu}d^{\nu}a_{\Omega\nu}a_{\Omega\mu}d^{\lambda}a_{\Omega\lambda}
  & 53 &  d^2a_\Omega^{\nu}d_{\nu}d^{\lambda}a_{\Omega\lambda}
  & 85 &  i V_\Omega^{\mu\nu}(a_{\Omega\mu}a_\Omega^{\lambda}a_{\Omega\nu}a_{\Omega\lambda}
 + a_\Omega^{\lambda}a_{\Omega\mu}a_{\Omega\lambda}a_{\Omega\nu})
 \\
 22 &  a_\Omega^{\mu}d^{\nu}a_\Omega^{\lambda}a_{\Omega\mu}d_{\nu}a_{\Omega\lambda}
  & 54 &  d^2a_\Omega^{\nu}d^{\lambda}d_{\nu}a_{\Omega\lambda}
  & 86 &  iV_\Omega^{\mu\nu}a_{\Omega\mu}a_\Omega^2a_{\Omega\nu}
 \\
 23 &  a_\Omega^{\mu}d^{\nu}a_\Omega^{\lambda}a_{\Omega\mu}d_{\lambda}a_{\Omega\nu}
  & 55 &  d^2a_\Omega^{\nu}d^2a_{\Omega\nu}
  & 87 &  iV_\Omega^{\mu\nu}a_{\Omega}^{\lambda}a_{\Omega\mu}a_{\Omega\nu}a_{\Omega\lambda}
 \\
 24 &  a_\Omega^2s_\Omega^2
  & 56 &  d^{\mu}d^{\nu}a_{\Omega\mu}d_{\nu}d^{\lambda}a_{\Omega\lambda}
  & 88 &  s_\Omega^3
 \\
 25 &  a_\Omega^2p_\Omega^2
  & 57 &  d^{\mu}d^{\nu}a_{\Omega\mu}d^{\lambda}d_{\nu}a_{\Omega\lambda}
  & 89 &  s_{\Omega} p_\Omega^2
 \\
 26 &  a_\Omega^{\mu}s_{\Omega}a_{\Omega\mu}s_\Omega
  & 58 &  d^{\mu}d^{\nu}a_{\Omega\nu}d_{\mu}d^{\lambda}a_{\Omega\lambda}
  & 90 &  s_{\Omega} p_{\Omega} d^{\mu}a_{\Omega\mu}
 \\
 27 &  a_\Omega^{\mu}p_{\Omega} a_{\Omega\mu}p_\Omega
  & 59 &  d^{\mu}d^{\nu}a_\Omega^{\lambda}d_{\mu}d_{\nu}a_{\Omega\lambda}
  & 91 &  s_{\Omega} d^{\mu}a_{\Omega\mu}p_\Omega
 \\
 28 &  a_\Omega^{\mu}(s_{\Omega} d_{\mu}p_\Omega
 + d_{\mu}p_{\Omega}s_\Omega)
  & 60 &  d^{\mu}d^{\nu}a_\Omega^{\lambda}d_{\mu}d_{\lambda}a_{\Omega\nu}
  & 92 &  s_{\Omega} d^{\mu}a_\Omega^{\nu}d_{\mu}a_{\Omega\nu}
 \\
 29 &  a_\Omega^{\mu}(p_{\Omega} d_{\mu}s_\Omega
 + d_{\mu}s_{\Omega}p_\Omega)
  & 61 &  d^{\mu}d^{\nu}a_\Omega^{\lambda}d_{\nu}d_{\mu}a_{\Omega\lambda}
  & 93 &  s_{\Omega} d^{\mu}a_\Omega^{\nu}d_{\nu}a_{\Omega\mu}
 \\
 30 &  a_\Omega^2d^{\nu}a_{\Omega\nu}p_\Omega
  & 62 &  d^{\mu}d^{\nu}a_\Omega^{\lambda}d_{\nu}d_{\lambda}a_{\Omega\mu}
  & 94 &  s_{\Omega} (d^{\mu}a_{\Omega\mu})^2
 \\
 31 &  a_\Omega^2p_{\Omega}d^{\nu}a_{\Omega\nu}
  & 63 &  d^{\mu}d^{\nu}a_\Omega^{\lambda}d_{\lambda}d_{\nu}a_{\Omega\mu}
  & &
 \\
 32 &  a_\Omega^2a_\Omega^{\nu}d_{\nu}p_\Omega
 + a_\Omega^{\mu}a_\Omega^2d_{\mu}p_\Omega
  & 64 &  d^{\mu}d^{\nu}a_{\Omega\mu}d_{\nu}p_\Omega
  & &\\
 \hline
 \end{array}\nonumber
 \end{eqnarray}
where some operators have $i$ in front of them to insure their
coefficients being real. In Ref.\cite{p6-1}, $p^6$ order operator
was denoted by $Y_i$ for the case of n flavor with coefficient $K_i$
\cite{p6-2}, $O_i$ for the case of three flavor with coefficient
$C_i$ and $P_i$ for the case of two flavors with coefficient $c_i$,
\begin{eqnarray}
S_\mathrm{eff}\bigg|_{p^6,~\mathrm{normal}}=\int
d^4x~\left\{\begin{array}{lll}
{\displaystyle\sum_{i=1}^{112}}~K_iY_i+\mbox{3 contact terms}&\hspace{2cm}&\mbox{n flavors}\\
{\displaystyle\sum_{i=1}^{90}}~C_iO_i+\mbox{4 contact terms}&\hspace{2cm}&\mbox{three flavors}\\
{\displaystyle\sum_{i=1}^{53}}~c_iP_i+\mbox{4 contact
terms}&\hspace{2cm}&\mbox{two flavors}
\end{array}\right.\;.\label{Sp6Ref}
\end{eqnarray}
Consider that our parametrization of the $p^6$ order chiral
Lagrangian (\ref{Sp6ours}) is general to the case of n flavor
quarks, there exist some relations among our coefficients and n
flavor coefficients given in (\ref{Sp6Ref}). With the help of
computer derivations, we have worked out these relations and list
them in Apendix\ref{KZ}. As a check, we vanish the quark self energy
$\Sigma$ in the codes which corresponds to take $m=0$ before other
further calculations and find null $p^6$ result. This verify the
analytical result discussed in Sec.IV that the anomaly part
contributions do not the contribute to normal part of chiral
Lagrangian. Another consistency check is done for those operators
which have two terms combined together by $C$ and $P$ symmetry
requirements. For $n$ flavor case, such operators are $\bar{O}_9$,
$\bar{O}_{11}$, $\bar{O}_{12}$, $\bar{O}_{14}$, $\bar{O}_{17}$,
$\bar{O}_{18}$, $\bar{O}_{19}$, $\bar{O}_{20}$, $\bar{O}_{28}$,
$\bar{O}_{29}$, $\bar{O}_{32}$, $\bar{O}_{34}$, $\bar{O}_{35}$,
$\bar{O}_{37}$, $\bar{O}_{38}$, $\bar{O}_{39}$, $\bar{O}_{40}$,
$\bar{O}_{44}$, $\bar{O}_{45}$, $\bar{O}_{46}$, $\bar{O}_{47}$,
$\bar{O}_{48}$, $\bar{O}_{49}$, $\bar{O}_{73}$, $\bar{O}_{76}$,
$\bar{O}_{77}$, $\bar{O}_{78}$, $\bar{O}_{80}$, $\bar{O}_{82}$,
$\bar{O}_{84}$, $\bar{O}_{85}$. Since in each of these operators,
there are two terms, we can compute the coefficients in front of
each terms and check if they are same. We have done the checks for
all these operators and all obtain the same analytical expressions
for the two terms in the same operator. This partly verifies the
correctness of our result given in (\ref{ZSigma}). From n flavors to
three flavors, there are some extra constraints (see (B1) in
Ref.\cite{p6-1})which make some operators depending on others.
Further from three flavors to two flavors, there are also some extra
constraints (see (B3) in Ref.\cite{p6-1})which make some more
operators depending on others. The sequence number for n flavors,
three flavors and two flavors are different, their comparisons are
list in Table 2 in Ref.\cite{p6-1}.

%%%%%%%%%%%%%%%%%%%%%%%%%%%%%%%%%%%%%%%%%%%%%%%%%%%%%%
\section{Numerical Values of $p^6$ order LECs: Normal Part}

With all above preparations in previous sections, we now come to the
stage of giving numerical values to the $p^6$ order LECs in the
normal part of the chiral Lagrangian. Note the necessary input and
process of the present computations for the $p^6$ order LECs are the
same as those for the $p^2$ and $p^4$ order LECs given in the end of
Sec.IV, we list the numerical result in Table.III, as done in
Table.I, the result LECs are taken for the values at $\Lambda=1$GeV
with superscript the difference caused at $\Lambda=1.1$GeV and
subscript the difference caused at $\Lambda=0.9$GeV,
 \begin{eqnarray}
&&\hspace{1cm}
C_{\Lambda=1\mathrm{GeV}}\bigg|^{C_{\Lambda=1.1\mathrm{GeV}}-C_{\Lambda=1\mathrm{GeV}}}_{C_{\Lambda=0.9\mathrm{GeV}}-C_{\Lambda=1\mathrm{GeV}}}\hspace{4cm}
c_{\Lambda=1\mathrm{GeV}}\bigg|^{c_{\Lambda=1.1\mathrm{GeV}}-c_{\Lambda=1\mathrm{GeV}}}_{c_{\Lambda=0.9\mathrm{GeV}}-c_{\Lambda=1\mathrm{GeV}}}\;.\notag
\end{eqnarray}
We further list result of the $p^6$ order LECs at $\Lambda=\infty$
in Table.IV. Consider that in the limit of $\Lambda=\infty$,
dropping out momentum total derivative terms in Eq.(14) is
problematic, we only take result LECs at $\Lambda=\infty$ as a
reference. Since the terms of three flavors and two flavors may have
different sequence numbers, as done in Ref.\cite{p6-1}, we put them
in the same line in our table. Since the number of independent
operators in the two flavors is smaller than that in the three
flavors, there are some operators in three flavors being independent
operators, but being dependent in two flavors, then these operators
will not have their two flavor counter parts in our table, these
leave the r.h.s. some empty blanks in the corresponding two flavor
columns. For two flavor case, Ref.\cite{NewRelation} further propose
a new relation among operators,
 \begin{eqnarray}
 0&=&8P_1-2P_2+6P_3-12P_{13}+8P_{14}-3P_{15}-2P_{16}
 -20P_{24}+8P_{25}+12P_{26}-12P_{27}-28P_{28}+8P_{36}-8P_{37}\notag\\
 &&-8P_{39}+2P_{40}+8P_{41}-8P_{42}-6P_{43}+4P_{48}\;,
 \end{eqnarray}
 which implies that one of the operators appears in above formula should be
 further dependent operator. Due to ignorance of the values of the coefficients in front of these operators,
 Ref.\cite{NewRelation} arbitrarily chooses this operator being
 $P_{27}$. Now our computations show that $c_{27}\neq 0$, so
 original choice is not suitable. Considering that $c_{37}=0$ in our computation, we instead now take $P_{37}$ as
 that dependent operator. $P_{37}$ now is a
 dependent operator, its name then is deleted in our Table.III.

To verify our choice of $F_0=87$MeV will really results in
experimental value of $F_\pi$, we exploit the relation between $F_0$
and $F_\pi$ given in Ref.\cite{Fpi}
\begin{eqnarray}
&&\hspace{-0.5cm}\frac{F_\pi}{F_0}=1+x_2(l_4^r-L)+x_2^2\bigg[\frac{1}{16\pi^2}(-\frac{1}{2}l_1^r-l_2^r+\frac{29}{12}L)
-\frac{13}{192}\frac{1}{(16\pi^2)^2}
+\frac{7}{4}k_1+k_2-2l_3^rl_4^r+2(l_4^r)^2-\frac{5}{4}k_4+r_F^r\bigg]+O(x_2^3)\;,~~~~~\label{FpiF0}\\
&&\hspace{-0.5cm}x_2=\frac{M_\pi^2}{F_\pi^2}\hspace{1cm}L=\frac{1}{16\pi^2}\ln\frac{M_\pi^2}{\mu^2}\hspace{1.5cm}k_i=(4l^r_i-\gamma_iL)L
\hspace{1.5cm}r_F^r=(8c_7+16c_8+8c_9)F_0^2~~~~\mbox{from
Ref.\cite{p6-2}}\;,
\end{eqnarray}
in which $l^r_i$ and $\gamma_i$ for $i=1,2,\ldots,7$ are defined in
Ref.\cite{GS}. Scale $\mu$ is taken to be $\rho$ mass
$\mu=M_\rho=$770MeV. Numerical calculations show that for
$\Lambda=1000^{+100}_{-100}$MeV, the contributions up to order of
$p^4$ (ignoring $x_2^2$ terms in (\ref{FpiF0})) give result
$F_\pi=92.99^{+.00}_{-.03}$MeV  and the contributions up to order of
$p^6$ (ignoring $x_2^3$ terms in (\ref{FpiF0})) give result
$F_\pi=92.97^{+.00}_{-.04}$MeV with
$r_F^r=-5.036^{+0.730}_{-1.290}\times 10^{-5}$. We see that the
$p^6$ order contributions to $F_\pi$ are very small and $F_0=87$MeV
is directly relate to $F_\pi=93$MeV.
\begin{eqnarray}
 &&\hspace{-0.5cm}\mbox{\small{\bf TABLE III.}~
The obtained values of the $p^6$ order LECs $C_1\cdots,C_{90}$ for
three flavor and $c_1\cdots,c_{53}$ for two flavors. }\notag\\
&&\hspace{-0.5cm}\mbox{\small The LECs are in units of
$10^{-3}\mathrm{GeV}^{-2}$.
The result LECs are taken the values at $\Lambda=1$GeV with superscript the difference}\notag\\
&&\hspace{-0.5cm}\mbox{\small caused at $\Lambda=1.1$GeV and
subscript the difference caused at $\Lambda=0.9$GeV. The value
$\equiv 0$ means that the constants}\notag\\
&&\hspace{-0.5cm}\mbox{\small   vanish at large $N_c$ limit.}\notag\\
 &&\hspace{0.5cm}\begin{array}{|c|c|c|c||c|c|c|c||c|c|c|c|}
 \hline i & C_i & j & c_j &
        i & C_i & j & c_j &
        i & C_i & j & c_j \\
 \hline 1 & 3.79^{+0.10}_{-0.17} & 1 & 3.58^{+0.09}_{-0.15} & 31 & -0.63^{+0.05}_{-0.09} & 17 & -1.10^{+0.12}_{-0.19} & 61 & 2.88^{-0.22}_{+0.26} & 34 & 2.84^{-0.22}_{+0.26} \\
  2 & \equiv 0 & & & 32 & 0.18^{-0.03}_{+0.04} & 18 & 0.43^{-0.07}_{+0.08} & 62 & \equiv 0  & & \\
  3 & -0.05^{+0.01}_{-0.01} & 2 & -0.03^{+0.01}_{-0.01} & 33 & 0.09^{-0.00}_{+0.03} & 19 & 0.41^{-0.06}_{+0.10} & 63 & 2.99^{-0.24}_{+0.30} & & \\
  4 & 3.10^{+0.09}_{-0.15} & 3 & 2.89^{+0.08}_{-0.13} & 34 & 1.59^{-0.10}_{+0.16} & 20 & 1.56^{-0.10}_{+0.17} & 64 & \equiv 0  & & \\
  5 & -1.01^{+0.08}_{-0.11} & 4 & 1.21^{-0.07}_{+0.06} & 35 & 0.17^{-0.12}_{+0.17} & 21 & 0.29^{-0.18}_{+0.24} & 65 & -2.43^{+0.15}_{-0.16} & 35 & 3.39^{-0.32}_{+0.41} \\
  6 & \equiv 0 & & & 36 & \equiv 0 & & & 66 & 1.71^{+0.07}_{-0.12} & 36 & 1.57^{+0.06}_{-0.10} \\
  7 & \equiv 0  & & & 37 & -0.56^{+0.09}_{-0.11} & & & 67 & \equiv 0  &  &  \\
  8 & 2.31^{-0.16}_{+0.18} & & & 38 & 0.41^{-0.08}_{+0.07} & 22 & -1.32^{+0.18}_{-0.25} & 68 & \equiv 0  & & \\
  9 & \equiv 0  & & & 39 & \equiv 0  & 23 & 0.86^{-0.12}_{+0.15} & 69 & -0.86^{-0.04}_{+0.06} & 38 & -0.68^{-0.03}_{+0.05} \\
  10 & -1.05^{+0.08}_{-0.09} & 5 & -0.98^{+0.07}_{-0.09} & 40 & -6.35^{-0.18}_{+0.32} & 24 & -4.84^{-0.14}_{+0.25} & 70 & 1.73^{-0.08}_{+0.07} & 39 & 1.81^{-0.08}_{+0.07} \\
  11 & \equiv 0  & & & 41 & \equiv 0  & & & 71 & \equiv 0  & & \\
  12 & -0.34^{+0.02}_{-0.01} & 6 & -0.33^{+0.01}_{-0.01} & 42 & 0.60^{+0.00}_{-0.00} & & & 72 & -3.30^{+0.05}_{-0.00} & 40 & -3.17^{+0.05}_{-0.02} \\
  13 & \equiv 0  & & & 43 & \equiv 0  & & & 73 & 0.50^{+0.43}_{-0.56} & 41 & 0.30^{+0.42}_{-0.54} \\
  14 & -0.83^{+0.12}_{-0.19} & 7 & -1.72^{+0.25}_{-0.35} & 44 & 6.32^{+0.20}_{-0.36} & 25 & 6.03^{+0.19}_{-0.33} & 74 & -5.07^{-0.16}_{+0.27} & 42 & -4.74^{-0.14}_{+0.24} \\
  15 & \equiv 0  & 8 & 0.86^{-0.12}_{+0.15} & 45 & \equiv 0  & & & 75 & \equiv 0  & & \\
  16 & \equiv 0  & & & 46 & -0.60^{-0.02}_{+0.04} & 26 & -1.14^{-0.05}_{+0.07} & 76 & -1.44^{-0.23}_{+0.31} & 43 & -1.29^{-0.23}_{+0.30} \\
  17 & 0.01^{-0.01}_{-0.01} & 9 & -0.84^{+0.12}_{-0.17} & 47 & 0.08^{+0.01}_{-0.00} & & & 77 & \equiv 0  & & \\
  18 & -0.56^{+0.09}_{-0.11} & & & 48 & 3.41^{+0.06}_{-0.10} & & & 78 & 17.51^{+1.02}_{-1.59} & 44 & 16.16^{+0.94}_{-1.45} \\
  19 & -0.48^{+0.09}_{-0.13} & 10 & -0.37^{+0.07}_{-0.10} & 49 & \equiv 0  & & & 79 & -0.56^{-0.30}_{+0.40} & 45 & 0.26^{-0.26}_{+0.34} \\
  20 & 0.18^{-0.03}_{+0.04} & 11 & \equiv 0  & 50 & 8.71^{+0.78}_{-1.12} & 27 & 13.57^{+1.41}_{-2.00} & 80 & 0.87^{-0.04}_{+0.03} & 46 & 0.85^{-0.04}_{+0.02} \\
  21 & -0.06^{+0.01}_{-0.01} & & & 51 & -11.49^{+0.18}_{-0.09} & 28 & 0.93^{+0.98}_{-1.25} & 81 & \equiv 0  & & \\
  22 & 0.27^{+0.19}_{-0.25} & 12 & 0.15^{+0.18}_{-0.24} & 52 & -5.04^{-0.67}_{+0.93} & & & 82 & -7.13^{-0.32}_{+0.51} & 47 & -6.73^{-0.29}_{+0.47} \\
  23 & \equiv 0  & & & 53 & -11.99^{-0.87}_{+1.33} & 29 & -11.01^{-0.81}_{+1.23} & 83 & 0.07^{+0.20}_{-0.27} & 48 & -0.22^{+0.18}_{-0.25} \\
  24 & 1.62^{+0.04}_{-0.07} & & & 54 & \equiv 0  & & & 84 & \equiv 0  & & \\
  25 & -5.98^{-0.49}_{+0.72} & 13 & -5.39^{-0.45}_{+0.66} & 55 & 16.79^{+0.96}_{-1.49} & 30 & 15.72^{+0.89}_{-1.38} & 85 & -0.82^{+0.03}_{-0.02} & 49 & -0.78^{+0.03}_{-0.01} \\
  26 & 3.35^{+0.29}_{-0.47} & 14 & 4.17^{+0.30}_{-0.49} & 56 & 19.34^{+0.52}_{-0.98} & 31 & 17.57^{+0.42}_{-0.82} & 86 & \equiv 0  & & \\
  27 & -1.54^{+0.15}_{-0.18} & 15 & -2.71^{+0.21}_{-0.25} & 57 & 7.92^{+1.34}_{-1.85} & 32 & 7.18^{+1.28}_{-1.76} & 87 & 7.57^{+0.37}_{-0.60} & 50 & 7.18^{+0.34}_{-0.55} \\
  28 & 0.30^{+0.01}_{-0.01} & & & 58 & \equiv 0  & & & 88 & -5.47^{-0.73}_{+1.03} & 51 & -4.85^{-0.69}_{+0.97} \\
  29 & -3.08^{-0.26}_{+0.32} & 16 & -2.22^{-0.22}_{+0.27} & 59 & -22.49^{-1.21}_{+1.89} & 33 & -21.19^{-1.12}_{+1.76} & 89 & 34.74^{+1.61}_{-2.62} & 52 & 32.19^{+1.46}_{-2.37} \\
  30 & 0.60^{+0.02}_{-0.03} & & & 60 & \equiv 0 & & & 90 & 2.44^{-0.38}_{+0.46} & 53 & 2.51^{-0.37}_{+0.46} \\
 \hline
 \end{array}\notag
 \end{eqnarray}

\begin{eqnarray}
 &&\hspace{-0.5cm}\mbox{\small{\bf TABLE IV.}~
The obtained values of the $p^6$ order LECs $C_1\cdots,C_{90}$ for
three flavor and $c_1\cdots,c_{53}$ for two flavors. }\notag\\
&&\hspace{1.6cm}\mbox{\small The LECs are in units of
$10^{-3}\mathrm{GeV}^{-2}$ and are taken the values at
$\Lambda=\infty$.
The value }\notag\\
&&\hspace{1.6cm}\mbox{\small $\equiv 0$ means that the constants vanish at large $N_c$.}\notag\\
 &&\hspace{2.5cm}\begin{array}{|c|c|c|c||c|c|c|c||c|c|c|c|}
 \hline i & C_i & j & c_j &
        i & C_i & j & c_j &
        i & C_i & j & c_j \\
        \hline 1 & 3.61 & 1 & 3.39 & 31 & -0.22 & 17 & -0.22 & 61 & 1.36 & 34 & 1.45 \\
 \hline 2 & \equiv 0 & & & 32 & 0.02 & 18 & 0.09 & 62 & \equiv 0 & & \\
 \hline 3 & -0.01 & 2 & 0.00 & 33 & 0.08 & 19 & 0.09 & 63 & 1.41 & & \\
 \hline 4 & 2.98 & 3 & 2.77 & 34 & 1.03 & 20 & 0.97 & 64 & \equiv 0 & & \\
 \hline 5 & -0.51 & 4 & 0.66 & 35 & -0.40 & 21 & -0.46 & 65 & -1.28 & 35 & 1.56 \\
 \hline 6 & \equiv 0 & & & 36 & \equiv 0 & & & 66 & 1.73 & 36 & 1.58 \\
 \hline 7 & \equiv 0 & & & 37 & -0.06 & & & 67 & \equiv 0 & &  \\
 \hline 8 & 1.16 & & & 38 & -0.01 & 22 & -0.25 & 68 & \equiv 0 & & \\
 \hline 9 & \equiv 0 & & & 39 & \equiv 0 & 23 & 0.20 & 69 & -0.90 & 38 & -0.71 \\
 \hline 10 & -0.49 & 5 & -0.49 & 40 & -6.10 & 24 & -4.70 & 70 & 0.91 & 39 & 1.08 \\
 \hline 11 & \equiv 0 & & & 41 & \equiv 0 & & & 71 & \equiv 0 & & \\
 \hline 12 & -0.19 & 6 & -0.20 & 42 & 0.49 & & & 72 & -2.43 & 40 & -2.37 \\
 \hline 13 & \equiv 0 & & & 43 & \equiv 0 & & & 73 & 2.47 & 41 & 2.08 \\
 \hline 14 & -0.26 & 7 & -0.42 & 44 & 6.17 & 25 & 5.86 & 74 & -4.96 & 42 & -4.61 \\
 \hline 15 & \equiv 0 & 8 & 0.20 & 45 & \equiv 0 & & & 75 & \equiv 0 & & \\
 \hline 16 & \equiv 0 & & & 46 & -0.58 & 26 & -1.11 & 76 & -2.33 & 43 & -2.08 \\
 \hline 17 & -0.15 & 9 & -0.29 & 47 & 0.08 & & & 77 & \equiv 0 & & \\
 \hline 18 & -0.06 & & & 48 & 3.13 & & & 78 & 18.97 & 44 & 17.41 \\
 \hline 19 & -0.08 & 10 & -0.09 & 49 & \equiv 0 & & & 79 & -1.81 & 45 & -0.89 \\
 \hline 20 & 0.02 & 11 & \equiv 0 & 50 & 10.73 & 27 & 17.28 & 80 & 0.49 & 46 & 0.52 \\
 \hline 21 & -0.01 & & & 51 & -8.65 & 28 & 4.93 & 81 & \equiv 0 & & \\
 \hline 22 & 1.11 & 12 & 0.91 & 52 & -7.24 & & & 82 & -7.27 & 47 & -6.83 \\
 \hline 23 & \equiv 0 & & & 53 & -13.65 & 29 & -12.49 & 83 & 0.96 & 48 & 0.59 \\
 \hline 24 & 1.55 & & & 54 & \equiv 0 & & & 84 & \equiv 0 & & \\
 \hline 25 & -7.21 & 13 & -6.46 & 55 & 18.10 & 30 & 16.83 & 85 & -0.47 & 49 & -0.49 \\
 \hline 26 & 3.93 & 14 & 4.68 & 56 & 17.99 & 31 & 16.33 & 86 & \equiv 0 & & \\
 \hline 27 & -0.60 & 15 & -1.42 & 57 & 12.69 & 32 & 11.45 & 87 & 7.83 & 50 & 7.39 \\
 \hline 28 & 0.29 & & & 58 & \equiv 0 & & & 88 & -7.83 & 51 & -6.96 \\
 \hline 29 & -3.81 & 16 & -2.78 & 59 & -23.88 & 33 & -22.35 & 89 & 35.69 & 52 & 32.93 \\
 \hline 30 & 0.58 & & & 60 & \equiv 0 & & & 90 & 0.25 & 53 & 0.51 \\
 \hline
 \end{array}\notag
 \end{eqnarray}
%%%%%%%%%%%%%%%%%%%%%%%%%%%%%%%%%%%%%%%%%%%%%%%%%%%%%%%%%%%%%%%%%
\section{Comparisons with Experiment and model results}

As we have mentioned in the introduction of this paper, present
experiment data is far enough to fix the $p^6$ order LECs. But there
do exist some combinations of the LECs which already have their
experiment or model calculation values. Usually, these LECs are
labeled by dimensionless parameters with convention \footnote{An
alternative convention is that $C_i^r$ and $c^r_i$ are used to
denote the renormalized LECs in some literatures.}of $C_i^r\equiv
C_iF_0^2$ or $c_i^r\equiv c_iF_0^2$. In this section, we collect
those combinations of LECs in the literature which have their
experiment or model calculation values and compare them with our
numerical results obtained in the last section with finite cutoff
\footnote{If the LECs at finite cutoff are replaced with those at
$\Lambda=\infty$, we have checked that qualitative feature of the
comparisons  results of this section will not change.} as the check
of our computations.
%%%%%%%%%%%%%%%%%%%%%%%%%%%%%%%%%%%%%%%%%%%%%%%%%%%%%%%%%%%%%%%%%%
\subsection{$\pi\pi$ and $\pi K$ scattering}

From the investigation of $\pi\pi$ scattering amplitudes, one can
work out the values of some combinations of $p^6$ order LECs.
Ref.\cite{p6-2} introduces following combinations,
 \begin{eqnarray}
 r^r_1&=&64c^r_1-64c^r_2+32c^r_4-32c^r_5+32c^r_6-64c^r_7-128c^r_8-64c^r_9
 +96c^r_{10}+192c^r_{11}-64c^r_{14}+64c^r_{16}+96c^r_{17}+192c^r_{18}\notag\\
 r^r_2&=&-96c^r_1+96c^r_2+32c^r_3-32c^r_4+32c^r_5-64c^r_6+32c^r_7+64c^r_8
 +32c^r_9-32c^r_{13}+32c^r_{14}-64c^r_{16}\notag\\
 r^r_3&=&48c^r_1-48c^r_2-40c^r_3+8c^r_4-4c^r_5+8c^r_6-8c^r_{12}+20c^r_{13}\notag\\
 r^r_4&=&-8c^r_3+4c^r_5-8c^r_6+8c^r_{12}-4c^r_{13}\notag\\
 r^r_5&=&-8c^r_1+10c^r_2+14c^r_3\notag\\
 r^r_6&=&6c^r_2+2c^r_3
 \end{eqnarray}
and gives the values of them by two theoretical methods of the
resonance-saturation (RS)\cite{pipi} and pure dimensional analysis
(ND) which only accounts for the order of magnitude and in Table.V.
 \begin{eqnarray}
  &&\hspace{-1cm}\mbox{\small{\bf TABLE V.}~
The obtained values for the combinations of the $p^6$ order LECs
from $\pi\pi$ scattering and our work.}\notag\\
&&\hspace{4cm}\mbox{\small The coefficients in the table are in units of $10^{-4}$}\notag\\
&&\hspace{2cm}\begin{array}{lcccccc}
 \hline  & r^r_1 & r^r_2 & r^r_3&r^r_4&r^r_5 & r^r_6\\
 \hline \mbox{RS in Ref.\cite{p6-2}} & -0.6&1.3&-1.7&-1.0&1.1&0.3\\
\mbox{ND in Ref.\cite{p6-2}} & 80&40&20&3&6&2\\
 \hline \mbox{ours}  & -9.32^{-2.62}_{+3.51} & 8.93^{+3.12}_{-4.27} & -3.06^{-0.81}_{+1.11} &
 -0.12^{+.022}_{-0.29}&0.87^{+0.04}_{-0.06}&0.42^{+0.02}_{-0.03}\\
 \hline
 \end{array}\notag
 \end{eqnarray}
 We see that all coefficients obtained from our calculations are
 consistent with those more precise RS results given in
 Ref.\cite{p6-2}. With our predictions for $p^4$ and $p^6$ order
 LECs, we can directly calculate the scattering lengths $a_l^I$ and
 slope parameters $b_l^I$ which relate to $p^4$ and $p^6$ order
 LECs through formulae given in Appendix C and D. of Ref.\cite{pipi}. We
 list experimental and our results in Table.VI. In our results, as done in Table.III, we
 take $\mu=770$MeV, but to match the result given in
 Ref.\cite{pipi} where $\mu$ is taken at $\mu=1$GeV, we also take
 $\mu=1000$MeV for comparison. We take two options, one  only includes $p^4$
 order contributions and the other combines in $p^6$ order contributions. For
 $p^6$ order contributions, for comparisons, we consider the cases
 of without and with $r_i^r$ coefficients.
 \begin{eqnarray}
 &&\hspace{-0.5cm}\mbox{\small{\bf TABLE VI.}~The obtained values for $a_l^I$ and
  $b_l^I$ in $\pi\pi$ scattering from experimental values given by Ref.\cite{pipi1} and our work.}\notag\\
 &&\hspace{-0.2cm}\begin{array}{ccccccccc}
 \hline & a_0^0 & b_0^0 & -10a_0^2 & -10b_0^2 & 10a_1^1 & 10^2b_1^1 & 10^2a_2^0 & 10^3a_2^2\\
 \hline \mbox{Ref.\cite{pipi1}} & .26\pm.05 & .25\pm.03 & .28\pm.12 & .82\pm.08 & .38\pm.02 &  & .17\pm.03 & .13\pm.30\\
 \hline p^4~\mu=10^3\mathrm{MeV} & .210^{-.000}_{+.000} & .260^{-.000}_{+.000} & .406^{+.001}_{-.001} & .662^{-.002}_{+.003} & .405^{+.001}_{-.002} & .772^{+.015}_{-.020} & .264^{+.002}_{-.003} & .195^{-.009}_{+.012}\\
  p^4~\mu=770\mathrm{MeV} & .204^{-.000}_{+.000} & .248^{-.000}_{+.000} & .411^{+.001}_{-.001} & .685^{-.002}_{+.003} & .401^{+.001}_{-.002} & .772^{+.015}_{-.020} & .235^{+.002}_{-.003} & .076^{-.009}_{+.012}\\
 \hline p^6~\mu=10^3\mathrm{MeV}~r^r_i\neq0 & .237^{-.000}_{+.000} & .307^{-.000}_{+.000} & .394^{+.001}_{-.001} & .637^{-.004}_{+.005} & .447^{+.002}_{-.003} & 1.255^{+.029}_{-.037} & .421^{+.005}_{-.008} & .339^{-.011}_{+.011}\\
  p^6~\mu=10^3\mathrm{MeV}~r^r_i=0 & .237^{-.000}_{+.000} & .305^{-.000}_{-.000} & .392^{+.001}_{-.001} & .629^{-.003}_{+.004} & .445^{+.002}_{-.002} & 1.217^{+.019}_{-.024} & .409^{+.004}_{-.006} & .337^{-.005}_{+.003}\\
  p^6~\mu=770\mathrm{MeV}~r^r_i\neq0 & .228^{-.000}_{+.000} & .287^{-.000}_{+.000} & .402^{+.001}_{-.001} & .665^{-.003}_{+.005} & .435^{+.002}_{-.003} & 1.164^{+.028}_{-.037} & .363^{+.005}_{-.008} & .212^{-.012}_{+.012}\\
  p^6~\mu=770\mathrm{MeV}~r^r_i=0 & .227^{-.000}_{+.000} & .285^{-.000}_{-.000} & .400^{+.001}_{-.001} & .657^{-.003}_{+.004} & .433^{+.002}_{-.002} & 1.125^{+.018}_{-.023} & .352^{+.004}_{-.006} & .210^{-.006}_{+.004}\\
 \hline
 \end{array}\notag
 \end{eqnarray}
 We see that the contributions from $p^6$ order LECs are rather
 small and only change the third digit of the result.

 Further, Ref.\cite{piK0} introduce coefficients in $\pi K$
 scattering
 \begin{eqnarray}
c_{01}^-&=&32m_{K^+}^3(-C_1^r+2C_3^r+2C_4^r)\;,\hspace{3cm}c_{20}^-=6m_{K^+}(-C_1^r+2C_3^r+2C_4^r)\;,\nonumber\\
c_{11}^+&=&8m_{K^+}^2(3C_1^r+6C_3^r-2C_4^r)\;,\hspace{3.3cm}c_{30}^+=\frac{1}{2}(-7C_1^r-32C_2^r+2C_3^r+10C_4^r)\;,\nonumber\\
c_{01}^+&=&16m^2_{K^+}m^2_{\pi^+}(C_6^r+C_8^r+C^r_{10}+2C^r_{11}-2C^r_{12}-2C^r_{13}+2C^r_{22}+4C^r_{23})+
16m^4_{K^+}(C^r_5+2C^r_6+C^r_{10}\nonumber\\
&&+4C^r_{11}-2C^r_{12}-4C^r_{13}+2C^r_{22}+4C^r_{23})\;,\nonumber\\
c_{10}^-&=&8m_{K^+}m^2_{\pi^+}(-4C_4^r-C_6^r-C_8^r+C^r_{10}+2C_{11}^r-2C_{12}^r-6C_{13}^r+2C^r_{22}-2C^r_{25})
+8m_{K^+}^3(-4C_4^r-C_5^r\nonumber\\
&&-2C_6^r+C_{10}^r+4C_{11}^r-2C_{12}^r-12C_{13}^r+2C_{22}^r-2C_{25}^r)\;,\nonumber\\
c_{20}^+&=&m_{K^+}^2(12C_1^r+48C_2^r-8C_4^r+C_5^r+10C_6^r+8C_7^r+4C_8^r+C_{10}^r+4C_{11}^r-2C_{12}^r
-4C_{13}^r+2C_{22}^r\nonumber\\
&&-4C_{23}^r+4C_{25}^r)
+m_{\pi^+}^2(12C_1^r+48C_2^r-8C_4^r+4C_5^r+5C_6^r+8C_7^r+C_8^r+C_{10}^r+2C_{11}^r-2C_{12}^r\nonumber\\
&&-10C_{13}^r+2C_{22}^r-4C_{23}^r+4C_{25}^r)\;.
\end{eqnarray}
In the table 1. of Ref.\cite{piK}, in terms of $c^+_{30}, c^+_{11},
c^-_{20}, c^-_{01}$, three constraints of  $p^6$ order LECs are
fixed from $\pi K$ subthreshold parameters, $\pi\pi$ amplitude and a
resonance model. And in the table 2. of Ref.\cite{piK}, in terms of
$c^+_{20}, c^+_{01}, c^-_{10}$, another three constraints of $p^6$
order LECs are fixed from the dispersive calculations and a
resonance model,
 \begin{eqnarray}
  &&\hspace{-0.5cm}\mbox{\small{\bf TABLE VII.}~
The obtained values for the combinations of the $p^6$ order LECs
from $\pi K$, $\pi\pi$ scattering and our work.}\notag\\
&&\hspace{4cm}\mbox{\small The coefficients in the l.h.s. of the
table are in units of $10^{-4}\mathrm{GeV}^{-2}$}\notag\\
&&\hspace{-1.3cm}\begin{array}{lcccc|lccc}
 \hline  & C_1+4C_3 & C_2 & C_4+3C_3 & C_1+4C_3+2C_2&&
 c_{20}^+\frac{m_\pi^4}{F_\pi^4}&c_{01}^+\frac{m_\pi^2}{F_\pi^4}&c_{10}^-\frac{m_\pi^3}{F_\pi^4}\\
 \hline \mbox{input~}c^+_{30},c^+_{11},c^-_{20} & 20.7\pm4.9 & -9.2\pm4.9 & 9.9\pm2.5  & 2.3\pm10.8&&&\\
 \mbox{input~}c^+_{30},c^+_{11},c^-_{01} & 28.1\pm4.9 & -7.4\pm4.9 & 21.0\pm2.5 & 13.4\pm10.8
 &\mbox{Dispersive}&\hspace*{-1cm}0.024\pm0.006&2.07\pm0.10&0.31\pm0.01\\
 \pi \pi \mbox{~amplitude}                           &            &            & 23.5\pm2.3 & 18.8\pm7.2&&&\\
 \mbox{Resonance model}                  & 7.2        & -0.5       & 10.0       & 6.2&\mbox{Resonance~model}
 &0.003&3.8&0.09\\
 \hline \mbox{ours}               & 35.9^{+1.3}_{-2.1} & 0.0^{+0.0}_{-0.0} & 29.5^{+1.1}_{-1.9}
 & 35.9^{+1.3}_{-2.1}& \mbox{ours}&\hspace*{-0.5cm}0.006^{-0.002}_{+0.003}&-0.159^{+0.133}_{-0.178}&0.020^{+0.037}_{-0.050}\\
 \hline
 \end{array}\notag
 \end{eqnarray}
In which for l.h.s. of the Table VII., except $C_2$, all other LECs
or combinations of LECs obtained by us have the same signs and
orders of magnitudes as those from Ref.\cite{piK}. While for r.h.s.
of the table, our results are not consistent with those obtained
from the dispersive calculations.
%%%%%%%%%%%%%%%%%%%%%%%%%%%%%%%%%%%%%%%%%%%%%%%%%%%%%%%%%%%%%%%
\subsection{Form factors}

\begin{figure}[h]
\caption{~~The space like and time like data for the vector form
factor.
\newline The red solid curve corresponds to predictions from chiral
perturbation up to $p^6$ order with LECs obtained in Table.III of
this paper. The red dashed line is the result by vanishing $p^6$
order LECs in corresponding red solid curve. The blue dot-dashed
curve corresponds to predictions from chiral perturbation up to
$p^4$ order with LECs obtained in Table.I of this paper. The blue
dotted line is the result by vanishing $p^4$ order LECs in
corresponding blue dot-dashed curve.The black x-axis of with
$|F_\pi^V|^2=1.0$ corresponds to predictions from $p^2$ order chiral
perturbation.} \label{Vectorformfactor}
\hspace*{-3cm}\begin{minipage}[b]{\textwidth}
\includegraphics[scale=0.73]{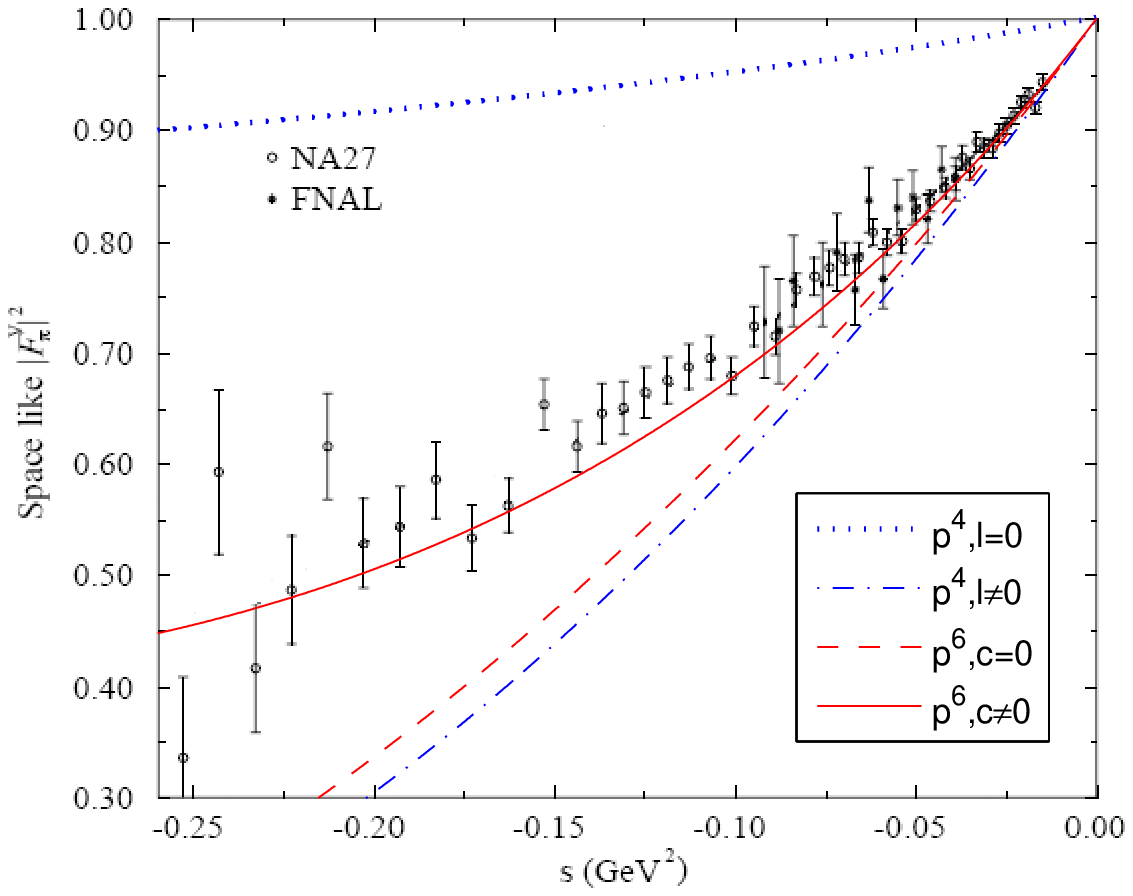}\hspace*{-1.2cm}\includegraphics[scale=0.73]{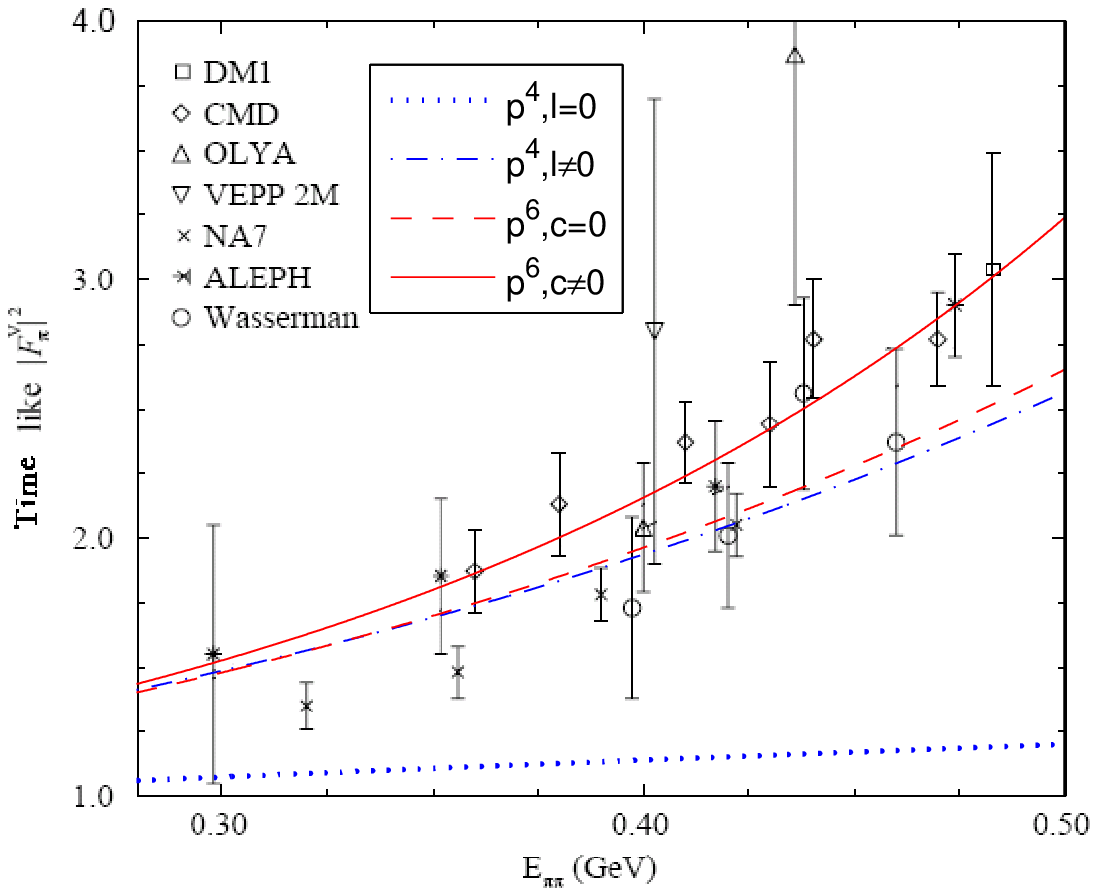}
\end{minipage}
\end{figure}

In Ref.\cite{p6-2}, in dealing with the vector form factor of the
pion, $r_{V1}^r$ and $r_{V2}^r$ are introduced into theory which
relate to $p^6$ order LECs through
\begin{eqnarray}
r_{V1}^r=-16c_6^r-4c_{35}^r-8c_{53}^r\;,\hspace{3cm}r_{V2}^r=-4c_{51}^r-4c_{53}^r\;.\label{rVdef}
\end{eqnarray}
 While for the scalar form factor, people
introduce $r_{S2}^r$ and $r_{S3}^r$ relate to $p^6$ order LECs by
\begin{eqnarray}
r_{S2}^r=32c_6^r+16c_7^r+32c_8^r+16c_9^r+16c_{20}^r\;,\hspace{3cm}r_{S3}^r=-8c_6^r\;.
\end{eqnarray}
In Ref.\cite{p6-2}, discussion of the decay of $\pi(p)\rightarrow
e\nu\gamma(q)$ further introduces $r_{A1}^r$ and $r_{A2}^r$ relate
to $p^6$ order LECs by
\begin{eqnarray}
 r^r_{A1}=48c^r_6-16c^r_{34}+8c^r_{35}-8c^r_{44}+16c^r_{46}-16c^r_{47}+8c^r_{50}\;,\hspace{2cm}
 r^r_{A2}=8c^r_{44}-16c^r_{50}+4c^r_{51}\;.
 \end{eqnarray}
In Ref.\cite{Kl3}, a naive estimation of $C_{12}^r$ is made from
scalar meson dominance (SMD) of the pion scalar form-factor and
$2C_{12}^r+C_{34}^r$ is estimated through $\lambda_0$ in $K_{l3}$
measurements (see Eq.(8.11) in Ref.\cite{Kl3}). While in
Ref.\cite{Kpi}, $C^r_{12}$ and $C^r_{12}+C^r_{34}$ are also
estimated from the $\pi K$ form factors. In Table.VIII, we list the
numerical results for above combinations of LECs given by our
calculations based on Table.III in last section and by
Ref.\cite{p6-2},\cite{Kl3},\cite{Kpi}.
 \begin{eqnarray}
 &&\hspace{-1cm}\mbox{\small{\bf TABLE VIII.}~
The obtained values for the combinations of the $p^6$ order LECs
appear in
vector and scalar form factor of pion.}\notag\\
&&\hspace{4.5cm}\mbox{\small the coefficients in the table are in units of $10^{-4}$}\notag\\
 &&\hspace{2.5cm}\begin{array}{ccc|ccc|ccc}
 \hline & \mbox{ours} & \mbox{Ref.\cite{p6-2}} & &\mbox{ours} & \mbox{Ref.\cite{p6-2}}& &\mbox{ours}
 & \mbox{Ref.\cite{p6-2}}\\
 \hline r^r_{V1} & -2.13^{+0.30}_{-0.39} & -2.5 & r^r_{S2}& 0.07^{+0.05}_{-0.08}&-0.3&r_{A1}^r&1.14^{+0.07}_{-0.09}&-0.5\\
  r^r_{V2} & 2.23^{+0.10}_{-0.16} & 2.6  & r^r_{S3}&0.20^{-0.01}_{+0.01}&0.6&r_{A2}^r&-0.38^{-0.06}_{+0.08}&1.1\\
 \hline
 \end{array}\notag\\
&&\hspace{1.8cm}\begin{array}{ccc|ccc}
 \hline  &\mbox{ours}&\mbox{Ref.\cite{Kl3}}&& \mbox{ours}&\mbox{Ref.\cite{Kpi}}\\
 \hline C_{12}^r&-0.026^{+0.001}_{-0.001}&-0.1&C^r_{12}&
 -0.026^{+0.001}_{-0.001}&(0.3\pm 5.4)\times 10^{-3}\\
  2C_{12}^r\!\!+C_{34}^r&
 0.068^{-0.006}_{+0.010}&-0.10\pm 0.17&C_{12}^r\!\!+C_{34}^r&0.094^{-0.007}_{+0.011}&(3.2\pm 1.5)\times 10^{-2}\\
 \hline
 \end{array}\notag
 \end{eqnarray}
 From which we see that among ten parameters between our predictions and values given in the literature, four of them have the same orders of magnitudes and signs
 ($r^r_{V1}$, $r^r_{V2}$, $r^r_{S3}$ and $C_{12}^r\!\!+C_{34}^r$), another one of them has different
 orders
 of magnitudes but the same signs
 ($C_{12}^r$ in Ref.\cite{Kl3}) , the left five of them have opposite signs ($r^r_{S2}$, $r^r_{A1}$, $r^r_{A2}$, $2C_{12}^r\!\!+C_{34}^r$
  and $C_{12}^r$ in Ref.\cite{Kpi}).\\
\indent Further in Fig.\ref{Vectorformfactor}, we compare the
experimental data  for vector form factors collected in Figure 4.
and Figure 5. of Ref.\cite{Fpi} with our results. In obtaining our
numerical predictions, we have exploited the formula given by
Eq.(3.16) in Ref.\cite{Fpi} which especially depends on  $p^6$ order
LECs through $r_{V1}^r, r_{V2}^r$ defined in (\ref{rVdef}) and we
input the formula $p^4$ and $p^6$ LECs obtained in Table.III of the
last section.

From Fig.\ref{Vectorformfactor}, we see that $p^6$ order LECs
explicitly improve the $p^4$ and $p^2$ order chiral perturbation
predictions and making them being more consistent with experimental
data.

%%%%%%%%%%%%%%%%%%%%%%%%%%%%%%%%%%%%%%%%%%%%%%%%%%%%%%%%%%%%%%%
\subsection{Photon-Photon Collisions}

In Ref.\cite{pigamma}, discussion of the photon-photon collision
$\gamma\gamma\rightarrow\pi^0\pi^0$ introduces $a^r_1,a_2^r$ and
$b^r$ relate to $p^6$ order LECs by
\begin{eqnarray}
 a^r_1=4096\pi^4(-c^r_{29}-c^r_{30}+c^r_{34})\hspace{0.6cm}
 a^r_2=256\pi^4(8c^r_{29}+8c^r_{30}+c^r_{31}+c^r_{32}+2c^r_{33})\hspace{0.6cm}
 b^r=-128\pi^4(c^r_{31}+c^r_{32}+2c^r_{33})\;.~~~~~~
 \end{eqnarray}
 While in Ref.\cite{pigamma1}, calculation of the photon-photon collision
$\gamma\gamma\rightarrow\pi^+\pi^-$ introduces another type of
$a^r_1,a_2^r$ and $b^r$ relate to $p^6$ order LECs by
\begin{eqnarray}
 a^r_1&=&-4096\pi^4(6c^r_6+c^r_{29}-c^r_{30}-3c^r_{34}+c^r_{35}+2c^r_{46}-4c^r_{47}+c^r_{50})\;,\notag\\
 a^r_2&=&256\pi^4(8c^r_{29}-8c^r_{30}+c^r_{31}+c^r_{32}-2c^r_{33}+4c^r_{44}+8c^r_{50}-4c^r_{51})\;,\\
 b^r&=&-128\pi^4(c^r_{31}+c^r_{32}-2c^r_{33}-4c^r_{44})\;.\notag
 \end{eqnarray}
In Table.IX, we list the numerical results for above combinations of
LECs given by our calculations based on Table.III in last section
and by Ref.\cite{pigamma} and \cite{pigamma1}.
 \begin{eqnarray}
 &&\hspace{-1cm}\mbox{\small{\bf TABLE IX.}~
The obtained values for the combinations of the $p^6$ order LECs
appear in
photon-photon collisions.}\notag\\
 &&\hspace{4cm}\begin{array}{ccc|ccc}
 \hline & \mbox{ours} & \mbox{Ref.\cite{pigamma}} & &\mbox{ours} & \mbox{Ref.\cite{pigamma1}}\\
 \hline a^r_1 & -5.65^{-0.91}_{+1.23} & -14\pm 5 & a^r_1& -5.86^{-0.49}_{+0.58}&-3.2\\
  a^r_2 & 3.79^{+0.02}_{-0.05} & 7\pm 3  & a^r_2&-0.98^{-0.07}_{+0.12}&0.7\\
  b^r & 1.66^{+0.05}_{-0.09} & 3\pm 1  & b^r&-0.23^{-0.01}_{+0.02}&0.4\\
 \hline
 \end{array}\notag
 \end{eqnarray}
 For which we see that among six parameters between our predictions and values given in the literature,
 except two have  opposite signs, other four all have the same orders of magnitudes and signs.
 %%%%%%%%%%%%%%%%%%%%%%%%%%%%%%%%%%%%%%%%%%%%%%%%%%%%%%%%%%%%%%%
 \subsection{Radiative pion decay}

 In Ref.\cite{PiDecay}, through reanalysis of the radiative pion
 decay, a group of $p^6$ order LECs are fixed.
  \begin{eqnarray}
  &&\hspace{-1cm}\mbox{\small{\bf TABLE X.}~
The obtained values for the combinations of the $p^6$ order LECs
from pion radiative decay and our work.}\notag\\
&&\hspace{4cm}\mbox{\small The coefficients in the table are in units of $10^{-5}$}\notag\\
&&\hspace{1.5cm}\begin{array}{lcccccc}
 \hline  &C^r_{12}&C^r_{13}&C^r_{61}&C^r_{62}&2C^r_{63}-C^r_{65}&C^r_{64}\\
 \hline \mbox{Ref.\cite{PiDecay}} & -0.6\pm 0.3&0\pm 0.2&1.0\pm 0.3& 0\pm 0.2&1.8\pm 0.7 &0\pm 0.2\\
 \hline
 \mbox{ours}&-0.26^{+0.01}_{-0.01}&0.0^{+0.0}_{-0.0}&2.18^{-0.17}_{+0.20}&0.0^{+0.0}_{-0.0}&6.36^{-0.42}_{+0.56}
 &0.0^{+0.0}_{-0.0}\\
  \hline\hline
&C^r_{78}&C^r_{80}&C^r_{81}&C^r_{82}& C^r_{87}&C^r_{88}\\
\hline  \mbox{Ref.\cite{PiDecay}}&10.0\pm 3.0&1.8\pm 0.4&
 0\pm 0.2& -3.5\pm 1.0&3.6\pm 1.0 &-3.5\pm 1.0\\
 \hline
 \mbox{ours}&13.26^{+0.77}_{-1.20}&0.66^{-0.03}_{+0.02}&0.0^{+0.0}_{-0.0}&-5.39^{-0.24}_{+0.39}&5.73^{+0.28}_{-0.45}
 &-4.14^{-0.55}_{+0.78}\\
\hline
 \end{array}\notag
 \end{eqnarray}
We find that all  LECs and combination of LECs from our predictions
have the same signs and orders of magnitudes as those from
experiment values.
%%%%%%%%%%%%%%%%%%%%%%%%%%%%%%%%%%%%%%%%%%%%%%%%%%%%%%%%%%%%%%%%%%%%%%
\subsection{Model calculations}

Except above phenomenological estimations on the values of some
LECs, there are model calculations for some others of them and most
of these analysis use a (single) resonance approximation. In
contrast, our calculations do not rely on the assumption of
existence of resonances.  In this subsection, we list down these
calculation values we can collect from the literature and compare
with our results.

Ref.\cite{TwoPoint} estimates values of some LECs.
\begin{eqnarray}
 &&\hspace{1cm}\mbox{\small{\bf TABLE XI.}~
The obtained values for the $p^6$ order LEC in Ref.\cite{TwoPoint} and our works}\notag\\
&&\hspace{3cm}\mbox{\small The coefficients in the table are in units of $10^{-3}\mathrm{GeV}^{-2}$}\notag\\
 &&\hspace{2cm}\begin{array}{ccccccc}
 \hline
 &C^r_{14}&C^r_{19}&C^r_{38}&C^r_{61}&C^r_{80}&C^r_{87}\\
 \hline
 \mbox{Ref.\cite{TwoPoint}} & -4.3 &-2.8&1.2&1.9&1.9&7.6\\
 \mbox{ours}&-0.83^{+0.12}_{-0.19}&-0.48^{+0.09}_{-0.13}&0.41^{-0.08}_{+0.07}&2.88^{-0.22}_{+0.26}&0.87^{-0.04}_{+0.03}
 &7.57^{+0.37}_{-0.60}\\
 \hline
 \end{array}\hspace{3cm}\notag
 \end{eqnarray}

For $C_{63}^r$ and $C_{65}^r$, Ref.\cite{C63-65} gives the value for
their combination $2C_{63}^r-C_{65}^r=(1.8\pm 0.7)\times 10^{-5}$
which, compares to our result of $6.36^{-0.48}_{+0.56}\times
10^{-5}$, is at the same order of magnitude and has the same sign.

For $C^r_{87}$, there are several works to estimate its values, we
list them in Table.XII.
\begin{eqnarray}
 &&\hspace{1cm}\mbox{\small{\bf TABLE XII.}~
The obtained values for the $p^6$ order LEC $C^r_{87}$}\notag\\
&&\hspace{2.5cm}\mbox{\small The coefficients in the table are in units of $10^{-5}$}\notag\\
 &&\hspace{2cm}\begin{array}{ccccc}
 \hline & \mbox{ours} & \mbox{Ref.\cite{C87-1}} & \mbox{Ref.\cite{C87-2}}&\mbox{Ref.\cite{C87-3}}\\
 \hline C^r_{87} & 5.73^{+0.28}_{-0.45} & 3.1\pm 1.1 & 4.3\pm 0.4 & 3.70\pm 0.14\\
 \hline
 \end{array}\hspace{3cm}\notag
 \end{eqnarray}
 where $C_{87}^r$ given in Ref.\cite{C87-2} and \cite{C87-3} are in
 form of $C_{87}$ in unit of GeV$^{-2}$, we have transformed them
 into our expression of $C^r_{87}$ with $C_{87}^r=C_{87}F_0^2$.

 Further, Ref.\cite{RL}
 exploits
 resonance Lagrangian estimates values of LECs $C_{78}$, $C_{82}$,
 $C_{87}$, $C_{88}$, $C_{89}$, $C_{90}$.
\begin{eqnarray}
&&\hspace{0cm}\mbox{\small{\bf TABLE XIII.}~
The obtained values for the $p^6$ order LEC from resonance Lagrangian given by Ref.\cite{RL} and our work}\notag\\
&&\hspace{5cm}\mbox{\small The coefficients in the table are in units of $10^{-4}/F_0^2$}\notag\\
 &&\hspace{0cm}\begin{array}{ccccccc}
 \hline  & C_{78} & C_{82} & C_{87} & C_{88} & C_{89} & C_{90}\\
 \hline
 \mbox{Lowest Meson Dominance}   & 1.09 & -0.36 & 0.40 & -0.52 &1.97&0.0\\
 \mbox{Resonance Lagrangian I } & 1.09 & -0.29 & 0.47 & -0.16&2.29&0.33\\
 \mbox{Resonance Lagrangian II} & 1.49 & -0.39 & 0.65 & -0.14&3.22&0.51\\
 \hline
 \mbox{ours}  & 1.326^{+0.077}_{-0.120} & -0.539^{-0.024}_{+0.039} & 0.573^{+0.028}_{-0.045} & -0.414^{-0.055}_{+0.078}&
 2.630^{+0.122}_{-0.198}&0.185^{-0.029}_{+0.035}\\
 \hline
 \end{array}\notag
 \end{eqnarray}
 We find that our results are consistent with those obtained from
resonance Lagrangian.

Ref.\cite{C38} estimates the value of $C_{38}$ and gives
$C^r_{38}=(2\pm 6)\times 10^{-6}$ which is also consistent with our
result of $C^r_{38}=3.1^{-0.6}_{+0.6}\times 10^{-6}$.

In terms of resonance exchange, Ref.\cite{Relation} proposes some
relations among different $p^6$ order LECs,
 \begin{eqnarray}
 C_{20}=-3C_{21}=C_{32}=\frac{1}{6}C_{35}\hspace{3cm}C_{24}=6C_{28}=3C_{30}\;.
 \end{eqnarray}
To check the validity of these relations for our results, in Table.
XIV, we write corresponding values obtained in our calculations
\begin{eqnarray}
&&\hspace{-0.5cm}\mbox{\small{\bf TABLE XIV.}~
The obtained values for the $p^6$ order LEC from our work}\notag\\
&&\hspace{2cm}\mbox{\small The coefficients in the table are in units of $10^{-3}\mathrm{GeV}^{-2}$}\notag\\
 &&\hspace{-0.5cm}\begin{array}{cccc|ccc}
 \hline  C_{20}& -3C_{21} & C_{32}& \frac{1}{6}C_{35}&C_{24}&6C_{28}&3C_{30}\\
 \hline 0.18^{-0.03}_{+0.04}&
 0.18^{-0.03}_{+0.03}&0.18^{-0.03}_{+0.04}&0.028^{-0.020}_{+0.028}&1.62^{+0.04}_{-0.07}&1.80^{+0.06}_{-0.06}&1.80^{+0.06}_{-0.09}
 \\
 \hline
 \end{array}\notag
 \end{eqnarray}
 We see that except $C_{35}$, all the other LECs satisfy the relations.
 %%%%%%%%%%%%%%%%%%%%%%%%%%%%%%%%%%%%%%%%%%%%%%%%%%%%%%%%%
\section{Summary}

In this paper, we revise our original formulation of calculating
LECs from the first principle of QCD to a chiral covariant one
suitable to computerize. With the help of computer, we successfully
obtain the analytical expressions for all the $p^6$ order LECs in
the normal part of chiral Lagrangian for pseudo scalar mesons on the
quark self energy $\Sigma(k^2)$. The ambiguities for anomaly part
contributions to the normal part of the chiral Lagrangian are
clarified and we prove that this part totally should vanish and
therefore need not to be considered in our computations. Since our
calculation is done under large $N_c$ limit, only operators of $p^6$
order with one trace and some multi-traces from the equation of
motion survive in our formulation. We set up relations among the
coefficients in front of these operators and LECs defined in
Ref.\cite{p6-1}. Then with input of $F_0=$87MeV to fix the
$\Lambda_\mathrm{QCD}$ in the running coupling constant of
$\alpha_s(k^2)$ appear in the kernel of SDE and choose cutoff of the
theory being $\Lambda=1000^{+100}_{-100}$MeV and $\Lambda=\infty$,
we calculate all $p^6$ order LECs numerically both for two flavor
and three flavor cases. Compare our result LECs with those
combinations which we can find experimental or model calculation
values in the literature,  we find that except few of them have
wrong signs, most of our predicted combinations of $p^6$ order LECs
have the same signs and orders of magnitudes with experiment or
model calculation values. This sets the solid basis for our $p^6$
order computations. For those combinations with wrong signs or wrong
order of magnitudes with experiment values, we need further
investigations. Based on these obtained $p^6$ order LECs, we expect
a very large number of predictions for various pseudo scalar meson
physics in the near future.
 %%%%%%%%%%%%%%%%%%%%%%%%%%%%%%%%
\section*{Acknowledgments}
This work was  supported by National  Science Foundation of China
(NSFC) under Grant No.10875065. We thank Prof. Y.P.Kuang for the
helpful discussions.

%\newpage
%%%%%%%%%%%%%%%%%%%%%%%%%%%%%%%%%%%%%%%%%%%%%%%%%%%%%%%%%%%%%%%%%

%%%%%%%%%%%%%%%%%%%%%%%%%%%%%%%%%%%%%%%%%%%%%%%%%%%%
\appendix
\section{Low Energy Expansion for $e^B$, $\Sigma\big((k+\tilde{F})^2\big)$ and $B$}\label{EBexp}

In this appendix, we first list down the $p^3$, $p^4$, $p^5$ and
$p^6$ order low energy expansion result for $e^B$ used in
(\ref{EBexp0}).
 \begin{eqnarray}
\frac{d^3}{d t^3}e^{B(t)}\bigg|_{t=0}&=&\frac{d^2}{d
t^2}e^{B(t)}\bigg|_{t=0}f[Ad(-B_0)](B_1)
 +\frac{d}{d t}e^{B(t)}\bigg|_{t=0}\bigg\{
 2\frac{d f}{d t}[Ad(-B(t))]\bigg|_{t=0}(B_1)
 +2f[Ad(-B_0)](B_2)\bigg\}\notag\\
 &&+e^{B_0}\bigg\{\frac{d^2 f}{d t^2}[Ad(-B(t))]\bigg|_{t=0}(B_1)
 +2\frac{d f}{d t}[Ad(-B(t))]\bigg|_{t=0}(B_2)+f[Ad(-B_0)](B_3)\bigg\}\;,\label{d3EBt}
 \end{eqnarray}
\begin{eqnarray}
 \frac{d^4}{d t^4}e^{B(t)}\bigg|_{t=0}
 &=&
 \frac{d^3}{d t^3}e^{B(t)}\bigg|_{t=0}~f[Ad(-B_0)](B_1)
 +\frac{d^2}{d t^2}e^{B(t)}\bigg|_{t=0}
 \bigg\{3\frac{d f}{dt}[Ad(-B(t))]\bigg|_{t=0}(B_1)
 +3f[Ad(-A_0)](B_2)\bigg\}\nonumber\\
 &&+\frac{d}{d t}e^{B(t)}\bigg|_{t=0}\bigg\{
 3\frac{d^2 f}{d t^2}[Ad(-B(t))]\bigg|_{t=0}(B_1)
 +6\frac{d f}{d t}[Ad(-B(t))]\bigg|_{t=0}(B_2)
 +3f[Ad(-B_0)](B_3)\bigg\}\nonumber\\
 &&+e^{B_0}\bigg\{\frac{d^3 f}{d t^3}[Ad(-B(t))]\bigg|_{t=0}(B_1)
 +3\frac{d^2 f}{d t^2}[Ad(-B(t))]\bigg|_{t=0}(B_2)\nonumber\\
 &&+3\frac{d f}{d t}[Ad(-B(t))]\bigg|_{t=0}(B_3)
 +f[Ad(-B_0)](B_4)\bigg\}\;,\label{d4EBt}
 \end{eqnarray}
 \begin{eqnarray}
\frac{d^5}{d t^5}e^{B(t)}\bigg|_{t=0}&=&
 \frac{d^4}{d t^4}e^{B(t)}\bigg|_{t=0}~f[Ad(-B_0)](B_1)
 +\frac{d^3}{d t^3}e^{B(t)}\bigg|_{t=0}
 \bigg\{4\frac{d f}{dt}[Ad(-B(t))]\bigg|_{t=0}(B_1)
 +4f[Ad(-A_0)](B_2)\bigg\}\notag\\
 &&+\frac{d^2}{d t^2}e^{B(t)}\bigg|_{t=0}
 \bigg\{6\frac{d^2 f}{d t^2}[Ad(-B(t))]\bigg|_{t=0}(B_1)
 +12\frac{d f}{d t}[Ad(-B(t))]\bigg|_{t=0}(B_2)
 +6f[Ad(-B_0)](B_3)\bigg\}\nonumber\\
 &&+\frac{d}{d t}e^{B(t)}\bigg|_{t=0}
 \bigg\{4\frac{d^3 f}{d t^3}[Ad(-B(t))]\bigg|_{t=0}(B_1)
 +12\frac{d^2 f}{d t^2}[Ad(-B(t))]\bigg|_{t=0}(B_2)+12\frac{d f}{d t}[Ad(-B(t))]\bigg|_{t=0}(B_3)\nonumber\\
 && +4f[Ad(-B_0)](B_4)\bigg\}+e^{B_0}\bigg\{\frac{d^4 f}{d t^4}[Ad(-B(t))]\bigg|_{t=0}(B_1)
 +4\frac{d^3 f}{d t^3}[Ad(-B(t))]\bigg|_{t=0}(B_2)\notag\\
 && +6\frac{d^2 f}{d t^2}[Ad(-B(t))]\bigg|_{t=0}(B_3)
 +4\frac{d f}{d t}[Ad(-B(t))]\bigg|_{t=0}(B_4)
 +f[Ad(-B_0)](B_5)\bigg\}\;,\label{d5EBt}
\end{eqnarray}
\begin{eqnarray}
\frac{d^6}{d t^6}e^{B(t)}\bigg|_{t=0}&=& \frac{d^5}{d
t^5}e^{B(t)}\bigg|_{t=0}f[Ad(-B_0)](B_1)
 +\frac{d^4}{d t^4}e^{B(t)}\bigg|_{t=0}
 \bigg\{5\frac{d f}{dt}[Ad(-B(t))]\bigg|_{t=0}(B_1)
 +5f[Ad(-B_0)](B_2)\bigg\}\nonumber\\
 &&+\frac{d^3}{d t^3}e^{B(t)}\bigg|_{t=0}
 \bigg\{10\frac{d^2 f}{d t^2}[Ad(-B(t))]\bigg|_{t=0}(B_1)
 +20\frac{d f}{d t}[Ad(-B(t))]\bigg|_{t=0}(B_2)
 +10f[Ad(-B_0)](B_3)\bigg\}\nonumber\\
 &&+\frac{d^2}{d t^2}e^{B(t)}\bigg|_{t=0}
 \bigg\{10\frac{d^3 f}{d t^3}[Ad(-B(t))]\bigg|_{t=0}(B_1)
 +30\frac{d^2 f}{d t^2}[Ad(-B(t))]\bigg|_{t=0}(B_2)\nonumber\\
 &&+30\frac{d f}{d t}[Ad(-B(t))]\bigg|_{t=0}(B_3)
 +10f[Ad(-B_0)](B_4)\bigg\}+\frac{d}{d t}e^{B(t)}\bigg|_{t=0}
 \bigg\{5\frac{d^4 f}{d t^4}[Ad(-B(t))]\bigg|_{t=0}(B_1)\notag\\
 && +20\frac{d^3 f}{d t^3}[Ad(-B(t))]\bigg|_{t=0}(B_2)
  +30\frac{d^2 f}{d t^2}[Ad(-B(t))]\bigg|_{t=0}(B_3)
 +20\frac{d f}{d t}[Ad(-B(t))]\bigg|_{t=0}(B_4)
\notag\\
 && +5f[Ad(-B_0)](B_5)\bigg\}+e^{B_0}\bigg\{\frac{d^5 f}{d t^5}[Ad(-B(t))]\bigg|_{t=0}(B_1)
 +5\frac{d^4 f}{d t^4}[Ad(-B(t))]\bigg|_{t=0}(B_2)\notag\\
 &&+10\frac{d^3 f}{d t^3}[Ad(-B(t))]\bigg|_{t=0}(B_3)
 +10\frac{d^2 f}{d t^2}[Ad(-B(t))]\bigg|_{t=0}(B_4)+5\frac{d f}{d t}[Ad(-B(t))]\bigg|_{t=0}(B_5)\notag\\
 && +f[Ad(-B_0)](B_6)\bigg\}\;.\label{d6EBt}
\end{eqnarray}
Then, we list down the $p^3$, $p^4$, $p^5$ and $p^6$ order
 low energy expansion result for $\Sigma\big((k+\tilde{F})^2\big)$ used in
 (\ref{Sigmaexp}). Note traceless terms in $p^5$ and $p^6$ orders
 are dropped out.
 \begin{eqnarray}
\frac{1}{6}\bigg[\frac{d^3}{d t^3}\Sigma[A(t)]\bigg]_{t=0}
 &=&\frac{1}{6}e^{Ad(A_0\pps)}\bigg[e^{-A_0\pps}\frac{d^2}{d t^2}e^{A(t)\pps}\bigg|_{t=0}f[Ad(-A_0\pps)](A_1\pps)\nonumber\\
 &&+e^{-A_0\pps}\frac{d}{d t}e^{A(t)\pps}\bigg|_{t=0}\bigg\{
 2\frac{d f}{d t}[Ad(-A(t)\pps)]\bigg|_{t=0}(A_1\pps)
 +2f[Ad(-A_0\pps)](A_2\pps)\bigg\}\notag\\
 &&+\bigg\{\frac{d^2 f}{d t^2}[Ad(-A(t)\pps)]\bigg|_{t=0}(A_1\pps)
 +2\frac{d f}{d t}[Ad(-A(t)\pps)]\bigg|_{t=0}(A_2\pps) \nonumber\\
 &&+f[Ad(-A_0\pps)](A_3\pps)\bigg\}\bigg]
 \Sigma(s+A(t))\bigg|_{s=0}\nonumber\\
 &=&-\frac{i}{3} (\mu \underline{\mu} \nu )k_{\nu}\skpp
 +\frac{2i}{3} (\mu \nu \lambda )k_{\mu}k_{\nu}\skpp \frac{\partial}{\partial k^{\lambda}}
 +\frac{2i}{3} (\mu \nu \lambda )k_{\nu}\skp \frac{\partial^2}{\partial k^{\mu}\partial k^{\lambda}}
 +\frac{i}{3} (\mu \underline{\mu} \nu )\skp \frac{\partial}{\partial
 k^{\nu}}\;,
\end{eqnarray}
\begin{eqnarray}
\bigg[\frac{d^4}{d t^4}\Sigma[A(t)]\bigg]_{t=0}
 &=&e^{Ad(A_0\pps)}\bigg[e^{-A_0\pps}\frac{d^3}{d t^3}e^{A(t)\pps}\bigg|_{t=0}~f[Ad(-A_0\pps)](A_1\pps)
 \nonumber\\
 &&+e^{-A_0\pps}\frac{d^2}{d t^2}e^{A(t)\pps}\bigg|_{t=0}
 \bigg\{3\frac{d f}{dt}[Ad(-A(t)\pps)]\bigg|_{t=0}(A_1\pps)
 +3f[Ad(-A_0\pps)](A_2\pps)\bigg\}\nonumber\\
 &&+e^{-A_0\pps}\frac{d}{d t}e^{A(t)\pps}\bigg|_{t=0}\bigg\{
 3\frac{d^2 f}{d t^2}[Ad(-A(t)\pps)]\bigg|_{t=0}(A_1\pps)
 +6\frac{d f}{d t}[Ad(-A(t)\pps)]\bigg|_{t=0}(A_2\pps)
 \nonumber\\
 &&+3f[Ad(-A_0\pps)](A_3\pps)\bigg\} +\bigg\{\frac{d^3 f}{d t^3}[Ad(-A(t)\pps)]\bigg|_{t=0}(A_1\pps)
 +3\frac{d^2 f}{d t^2}[Ad(-A(t)\pps)]\bigg|_{t=0}(A_2\pps)\nonumber\\
 &&+3\frac{d f}{d t}[Ad(-A(t)\pps)]\bigg|_{t=0}(A_3\pps)
 +f[Ad(-A_0\pps)](A_4\pps)\bigg\}\bigg]_{t=0}\Sigma(s+A(t))\bigg|_{s=0}\nonumber\\
 &=&-\frac{1}{4} (\mu \nu )(\underline{\mu} \lambda )k_{\nu}\skpp \frac{\partial}{\partial k^{\lambda}}
 +\frac{1}{2} (\mu \nu )(\lambda \rho )k_{\mu}k_{\lambda}\skpp \frac{\partial^2}{\partial k^{\nu}\partial k^{\rho}}
 -\frac{1}{8} (\mu \nu \underline{\nu} \lambda )k_{\lambda}\skpp \frac{\partial}{\partial k^{\mu}}
 \notag\\
 &&-\frac{1}{8} (\mu \nu \underline{\mu} \lambda )k_{\lambda}\skpp \frac{\partial}{\partial k^{\nu}}
 +\frac{1}{4} (\mu \underline{\mu} \nu \lambda )k_{\nu}\skpp \frac{\partial}{\partial k^{\lambda}}
 +\frac{1}{4} (\mu \nu \lambda \rho )k_{\nu}k_{\lambda}\skpp \frac{\partial^2}{\partial k^{\mu}\partial k^{\rho}}
 \notag\\
 &&+\frac{1}{4} (\mu \nu \lambda \rho )k_{\mu}k_{\lambda}\skpp \frac{\partial^2}{\partial k^{\nu}\partial k^{\rho}}
 +\frac{1}{4} (\mu \nu \lambda \rho )k_{\lambda}\skp \frac{\partial^3}{\partial k^{\mu}\partial k^{\nu}\partial k^{\rho}}
 +\frac{1}{4} (\mu \nu )(\underline{\mu} \underline{\nu} )\skpp
 \notag\\
 &&+\frac{1}{4} (\mu \nu )(\underline{\mu} \lambda )k_{\lambda}\skpp \frac{\partial}{\partial k^{\nu}}
 +\frac{1}{4} (\mu \nu )(\underline{\mu} \lambda )\skp \frac{\partial^2}{\partial k^{\nu}\partial k^{\lambda}}
 +\frac{1}{8} (\mu \nu \underline{\nu} \lambda )k_{\mu}\skpp \frac{\partial}{\partial k^{\lambda}}
 \notag\\
 &&+\frac{1}{8} (\mu \nu \underline{\nu} \lambda )\skp \frac{\partial^2}{\partial k^{\mu}\partial k^{\lambda}}
 +\frac{1}{8} (\mu \nu \underline{\mu} \lambda )k_{\nu}\skpp \frac{\partial}{\partial k^{\lambda}}
 +\frac{1}{8} (\mu \nu \underline{\mu} \lambda )\skp \frac{\partial^2}{\partial k^{\nu}\partial k^{\lambda}}
 \notag\\
 &&-\frac{1}{6} (\mu \nu \underline{\nu} \lambda )k_{\mu}k_{\lambda}\skppp
 -\frac{1}{6} (\mu \nu \underline{\mu} \lambda )k_{\nu}k_{\lambda}\skppp
 +\frac{1}{3} (\mu \nu \lambda \rho )k_{\mu}k_{\nu}k_{\lambda}\skppp \frac{\partial}{\partial k^{\rho}}
 +\frac{1}{3} (\mu \nu )(\underline{\mu} \lambda
 )k_{\nu}k_{\lambda}\skppp\;,
 \end{eqnarray}
 \begin{eqnarray}
 &&\hspace{-0.4cm}\frac{1}{5!}\bigg[\frac{d^5}{d
 t^5}\Sigma[A(t)]\bigg]_{t=0}\nonumber\\
 &&\hspace{-0.4cm}=\frac{1}{5!}e^{Ad(A_0\pps)}\bigg[
 e^{-A_0\pps}\frac{d^4}{d t^4}e^{A(t)\pps}\bigg|_{t=0}~f[Ad(-A_0\pps)](A_1\pps)
 +e^{-A_0\pps}\frac{d^3}{d t^3}e^{A(t)\pps}\bigg|_{t=0}
 \bigg\{4\frac{d f}{dt}[Ad(-A(t)\pps)]\bigg|_{t=0}(A_1\pps)\nonumber\\
 && +4f[Ad(-A_0\pps)](A_2\pps)\bigg\}+e^{-A_0\pps}\frac{d^2}{d t^2}e^{A(t)\pps}\bigg|_{t=0}
 \bigg\{6\frac{d^2 f}{d t^2}[Ad(-A(t)\pps)]\bigg|_{t=0}(A_1\pps)
 \nonumber\\
 &&+12\frac{d f}{d t}[Ad(-A(t)\pps)]\bigg|_{t=0}(A_2\pps)
 +6f[Ad(-A_0\pps)](A_3\pps)\bigg\}+e^{-A_0\pps}\frac{d}{d t}e^{A(t)\pps}\bigg|_{t=0}
 \bigg\{4\frac{d^3 f}{d t^3}[Ad(-A(t)\pps)]\bigg|_{t=0}(A_1\pps)\nonumber\\
 && +12\frac{d^2 f}{d t^2}[Ad(-A(t)\pps)]\bigg|_{t=0}(A_2\pps)
 +12\frac{d f}{d t}[Ad(-A(t)\pps)]\bigg|_{t=0}(A_3\pps)
 +4f[Ad(-A_0\pps)](A_4\pps)\bigg\}\notag\\
 &&+\bigg\{\frac{d^4 f}{d t^4}[Ad(-A(t)\pps)]\bigg|_{t=0}(A_1\pps)
 +4\frac{d^3 f}{d t^3}[Ad(-A(t)\pps)]\bigg|_{t=0}(A_2\pps)
 +6\frac{d^2 f}{d t^2}[Ad(-A(t)\pps)]\bigg|_{t=0}(A_3\pps)\notag\\
 &&+4\frac{d f}{d t}[Ad(-A(t)\pps)]\bigg|_{t=0}(A_4\pps)
 +f[Ad(-A_0\pps)](A_5\pps)\bigg\}\bigg]\Sigma(s+A(t))\bigg|_{s=0}\nonumber\\
 &&\hspace{-0.4cm}=\mbox{traceless terms}\;,
 \end{eqnarray}

 Finally, we list down the $p^3$, $p^4$, $p^5$, $p^6$ order
 low energy expansion result for $B$ used in (\ref{Bexp}),
\begin{eqnarray}
 B_3&=&-6 ds_\Omega^{\mu}\gamma_{\mu}\tau
 +6i dp_\Omega^{\mu}\gamma_{\mu}\gamf \tau
 -6i a_\Omega^{\mu}s \gamma_{\mu}\gamf \tau
 -6 a_\Omega^{\mu}p \gamma_{\mu}\tau
 -6i s_\Omega a_\Omega^{\mu}\gamma_{\mu}\gamf \tau
 -6 p_\Omega a_\Omega^{\mu}\gamma_{\mu}\tau
 \notag\\
 &&-3i (\mu \nu \lambda )\gamma_{\nu}\gamma_{\lambda}\tau \frac{\partial}{\partial k^{\mu}}
 +3i a_\Omega^{\mu}(\nu a_\Omega^{\lambda} )\gamma_{\mu}\gamma_{\lambda}\tau \frac{\partial}{\partial k^{\nu}}
 +3i (\mu a_\Omega^{\nu} )a_\Omega^{\lambda}\gamma_{\nu}\gamma_{\lambda}\tau \frac{\partial}{\partial k^{\mu}}
 -3i a_\Omega^{\mu}(\nu a_\Omega^{\lambda} )\gamma_{\lambda}\gamma_{\mu}\tau \frac{\partial}{\partial k^{\nu}}
 \notag\\
 &&-3i (\mu a_\Omega^{\nu} )a_\Omega^{\lambda}\gamma_{\lambda}\gamma_{\nu}\tau \frac{\partial}{\partial k^{\mu}}
 -3 (\mu \na_\Omega^{\nu} \na_\Omega^{\lambda} )\gamma_{\nu}\gamma_{\lambda}\gamf \tau \frac{\partial}{\partial k^{\mu}}
 -2i (\mu \underline{\mu} \nu )\tau \frac{\partial}{\partial k^{\nu}}
 +6i a_\Omega^{\mu}(\nu a_{\Omega\mu} )\tau \frac{\partial}{\partial k^{\nu}}
 \notag\\
 &&+6i (\mu a_\Omega^{\nu} )a_{\Omega\nu}\tau \frac{\partial}{\partial k^{\mu}}
 -12i (\mu s_\Omega)\tau k_{\mu}\skp
 +12 (\mu p_\Omega)\gamf \tau k_{\mu}\skp
 -12i (\mu s_\Omega)\tau \sk \frac{\partial}{\partial k^{\mu}}
 \notag\\
 &&+4i (\mu \underline{\mu} \nu )\tau k_{\nu}\sk \skpp
 +4i (\mu \underline{\mu} \nu )\tau k_{\nu}\skp^2
 -4i (\mu \nu \lambda )\tau k_{\nu}\frac{\partial^2}{\partial k^{\mu}\partial k^{\lambda}}
 -4i (\mu \underline{\mu} \nu )\tau \sk \skp \frac{\partial}{\partial k^{\nu}}
 \notag\\
 &&-6i (\mu \underline{\mu} a_\Omega^{\nu} )\gamma_{\nu}\gamf \tau \skp
 -12i (\mu \nu a_\Omega^{\lambda} )\gamma_{\lambda}\gamf \tau k_{\mu}k_{\nu}\skpp
 -6i (\mu \nu a_\Omega^{\lambda} )\gamma_{\lambda}\gamf \tau k_{\mu}\skp \frac{\partial}{\partial k^{\nu}}
 -6i (\mu \nu a_\Omega^{\lambda} )\gamma_{\lambda}\gamf \tau k_{\nu}\skp \frac{\partial}{\partial k^{\mu}}
 \notag\\
 &&-6i (\mu \nu a_\Omega^{\lambda} )\gamma_{\lambda}\gamf \tau \sk \frac{\partial^2}{\partial k^{\mu}\partial k^{\nu}}
 +3 (\mu \nu )a_{\Omega\mu}\gamf \tau \frac{\partial}{\partial k^{\nu}}
 +3 a_\Omega^{\mu}(\underline{\mu} \nu )\gamf \tau \frac{\partial}{\partial k^{\nu}}
 -6i (\mu \nu )a_\Omega^{\lambda}\gamma_{\lambda}\gamf \tau k_{\mu}\skp \frac{\partial}{\partial k^{\nu}}
 \notag\\
 &&+3 (\mu \nu a_{\Omega\mu} )\gamf \tau \frac{\partial}{\partial k^{\nu}}
 +3 (\mu \nu a_{\Omega\nu} )\gamf \tau \frac{\partial}{\partial k^{\mu}}
 +6 (\mu \nu a_\Omega^{\lambda} )\gamf \tau k_{\lambda}\frac{\partial^2}{\partial k^{\mu}\partial k^{\nu}}
 -6 (\mu \nu \lambda )\gamma_{\mu}\tau k_{\nu}\skp \frac{\partial}{\partial k^{\lambda}}
 \notag\\
 &&-6 (\mu \underline{\mu} \nu )\gamma_{\nu}\tau \skp
 +12 (\mu \nu \lambda )\gamma_{\nu}\tau k_{\mu}k_{\lambda}\skpp
 +6 (\mu \nu \lambda )\gamma_{\nu}\tau k_{\mu}\skp \frac{\partial}{\partial k^{\lambda}}
 +6 (\mu \nu \lambda )\gamma_{\nu}\tau k_{\lambda}\skp \frac{\partial}{\partial k^{\mu}}
 \\
 &&-6i a_\Omega^{\mu}(\nu \lambda )\gamma_{\mu}\gamf \tau k_{\nu}\skp \frac{\partial}{\partial k^{\lambda}}
 -8i (\mu \nu \lambda )\tau k_{\mu}k_{\nu}\sk \skpp \frac{\partial}{\partial k^{\lambda}}
 -8i (\mu \nu \lambda )\tau k_{\mu}k_{\nu}\skp^2 \frac{\partial}{\partial k^{\lambda}}
 -8i (\mu \nu \lambda )\tau k_{\nu}\sk \skp \frac{\partial^2}{\partial k^{\mu}\partial
 k^{\lambda}}\;,\notag
  \end{eqnarray}
 \begin{eqnarray}
 B_4&=&-24 s_\Omega^2 \tau
 -24 p_\Omega^2 \tau
 +24i [s_\Omega, p_\Omega] \gamf \tau
 +24i (\mu d^{\nu}s_\Omega)\gamma_{\nu}\tau \frac{\partial}{\partial k^{\mu}}
 +24 (\mu d^{\nu}p_\Omega )\gamma_{\nu}\gamf \tau \frac{\partial}{\partial k^{\mu}}
 -24 a_\Omega^{\mu}(\nu s_\Omega)\gamma_{\mu}\gamf \tau \frac{\partial}{\partial k^{\nu}}\notag\\
 && -24 (\mu a_\Omega^{\nu} )s_\Omega \gamma_{\nu}\gamf \tau \frac{\partial}{\partial k^{\mu}}
 +24i a_\Omega^{\mu}(\nu p_\Omega)\gamma_{\mu}\tau \frac{\partial}{\partial k^{\nu}}
 +24i (\mu a_\Omega^{\nu} )p_\Omega \gamma_{\nu}\tau \frac{\partial}{\partial k^{\mu}}
 -24 s_\Omega (\mu a_\Omega^{\nu} )\gamma_{\nu}\gamf \tau \frac{\partial}{\partial k^{\mu}}
 \notag\\
 &&-24 (\mu s_\Omega)a_\Omega^{\nu}\gamma_{\nu}\gamf \tau \frac{\partial}{\partial k^{\mu}}
+24i p_\Omega (\mu a_\Omega^{\nu} )\gamma_{\nu}\tau
\frac{\partial}{\partial k^{\mu}}
 +24i (\mu p_\Omega)a_\Omega^{\nu}\gamma_{\nu}\tau \frac{\partial}{\partial k^{\mu}}
 -6 (\mu \nu \lambda \rho )\gamma_{\lambda}\gamma_{\rho}\tau \frac{\partial^2}{\partial k^{\mu}\partial k^{\nu}}
 \notag\\
 &&+6 a_\Omega^{\mu}(\nu \lambda a_\Omega^{\rho} )\gamma_{\mu}\gamma_{\rho}\tau \frac{\partial^2}{\partial k^{\nu}\partial k^{\lambda}}
 +12 (\mu a_\Omega^{\nu} )(\lambda a_\Omega^{\rho} )\gamma_{\nu}\gamma_{\rho}\tau \frac{\partial^2}{\partial k^{\mu}\partial k^{\lambda}}
 +6 (\mu \nu a_\Omega^{\lambda} )a_\Omega^{\rho}\gamma_{\lambda}\gamma_{\rho}\tau \frac{\partial^2}{\partial k^{\mu}\partial k^{\nu}}
 \notag\\
 &&-6 a_\Omega^{\mu}(\nu \lambda a_\Omega^{\rho} )\gamma_{\rho}\gamma_{\mu}\tau \frac{\partial^2}{\partial k^{\nu}\partial k^{\lambda}}
 -12 (\mu a_\Omega^{\nu} )(\lambda a_\Omega^{\rho} )\gamma_{\rho}\gamma_{\nu}\tau \frac{\partial^2}{\partial k^{\mu}\partial k^{\lambda}}
 -6 (\mu \nu a_\Omega^{\lambda} )a_\Omega^{\rho}\gamma_{\rho}\gamma_{\lambda}\tau \frac{\partial^2}{\partial k^{\mu}\partial k^{\nu}}
 \notag\\
 &&+6i (\mu\nu(d^{\lambda}a_\Omega^{\rho}-d^{\rho}a_\Omega^{\lambda}))\gamma_{\lambda}\gamma_{\rho}\gamf \tau \frac{\partial^2}{\partial k^{\mu}\partial k^{\nu}}
 -3 (\mu \nu \underline{\nu} \lambda )\tau \frac{\partial^2}{\partial k^{\mu}\partial k^{\lambda}}
 -3 (\mu \nu \underline{\mu} \lambda )\tau \frac{\partial^2}{\partial k^{\nu}\partial k^{\lambda}}
 +12 a_\Omega^{\mu}(\nu \lambda a_{\Omega\mu} )\tau \frac{\partial^2}{\partial k^{\nu}\partial k^{\lambda}}\notag\\
 && +24 (\mu a_\Omega^{\nu} )(\lambda a_{\Omega\nu} )\tau \frac{\partial^2}{\partial k^{\mu}\partial k^{\lambda}}
 +12 (\mu \nu a_\Omega^{\lambda} )a_{\Omega\lambda}\tau \frac{\partial^2}{\partial k^{\mu}\partial k^{\nu}}
-12 (\mu \nu )(\underline{\mu} \underline{\nu} )\tau \sk \skpp
 -12 (\mu \nu )(\underline{\mu} \underline{\nu} )\tau \skp^2
  \notag\\
 &&-6 (\mu \nu )(\underline{\mu} \lambda )\tau \frac{\partial^2}{\partial k^{\nu}\partial k^{\lambda}}
+8 (\mu \underline{\mu} \nu )a_\Omega^{\lambda}\gamma_{\lambda}\gamf
\tau k_{\nu}\skpp
 -8 (\mu \underline{\mu} \nu )a_\Omega^{\lambda}\gamma_{\lambda}\gamf \tau \skp \frac{\partial}{\partial k^{\nu}}
 -24 (\mu \underline{\mu} s_\Omega)\tau \skp-24i (\mu \underline{\mu} p)\gamf \tau \skp
 \notag\\
 && -48 (\mu \nu s_\Omega)\tau k_{\mu}k_{\nu}\skpp
 -48i (\mu \nu p_\Omega)\gamf \tau k_{\mu}k_{\nu}\skpp
 -24 (\mu \nu s_\Omega)\tau k_{\mu}\skp \frac{\partial}{\partial k^{\nu}}
 -24i (\mu \nu p_\Omega)\gamf \tau k_{\mu}\skp \frac{\partial}{\partial k^{\nu}}
  \notag\\
 &&-24 (\mu \nu s_\Omega)\tau k_{\nu}\skp \frac{\partial}{\partial k^{\mu}}
-24i (\mu \nu p_\Omega)\gamf \tau k_{\nu}\skp
\frac{\partial}{\partial k^{\mu}}
 -24 (\mu \nu s_\Omega)\tau \sk \frac{\partial^2}{\partial k^{\mu}\partial k^{\nu}}
 -8i (\mu \nu \underline{\nu} \lambda )\gamma_{\mu}\tau k_{\lambda}\skpp
 \notag\\
 &&+8i (\mu \nu \underline{\nu} \lambda )\gamma_{\mu}\tau \skp \frac{\partial}{\partial k^{\lambda}}
  +8 a_\Omega^{\mu}(\nu \underline{\nu} \lambda )\gamma_{\mu}\gamf \tau k_{\lambda}\skpp
 -8 a_\Omega^{\mu}(\nu \underline{\nu} \lambda )\gamma_{\mu}\gamf \tau \skp \frac{\partial}{\partial k^{\lambda}}
 +8 (\mu \nu \underline{\nu} \lambda )\tau k_{\mu}k_{\lambda}\sk \skppp\notag\\
 && +8 (\mu \nu \underline{\mu} \lambda )\tau k_{\nu}k_{\lambda}\sk \skppp
 +24 (\mu \nu \underline{\nu} \lambda )\tau k_{\mu}k_{\lambda}\skp \skpp
+24 (\mu \nu \underline{\mu} \lambda )\tau k_{\nu}k_{\lambda}\skp
\skpp
 +6 (\mu \nu \underline{\nu} \lambda )\tau k_{\lambda}\sk \skpp \frac{\partial}{\partial k^{\mu}}
  \notag\\
 &&+6 (\mu \nu \underline{\mu} \lambda )\tau k_{\lambda}\sk \skpp \frac{\partial}{\partial k^{\nu}}
-12 (\mu \underline{\mu} \nu \lambda )\tau k_{\nu}\sk \skpp
\frac{\partial}{\partial k^{\lambda}}
 -6 (\mu \nu \underline{\nu} \lambda )\tau k_{\mu}\sk \skpp \frac{\partial}{\partial k^{\lambda}}
 -6 (\mu \nu \underline{\mu} \lambda )\tau k_{\nu}\sk \skpp \frac{\partial}{\partial k^{\lambda}}
 \notag\\
 &&+6 (\mu \nu \underline{\nu} \lambda )\tau k_{\lambda}\skp^2 \frac{\partial}{\partial k^{\mu}}
 +6 (\mu \nu \underline{\mu} \lambda )\tau k_{\lambda}\skp^2 \frac{\partial}{\partial k^{\nu}}
 -12 (\mu \underline{\mu} \nu \lambda )\tau k_{\nu}\skp^2 \frac{\partial}{\partial k^{\lambda}}
 -6 (\mu \nu \underline{\nu} \lambda )\tau k_{\mu}\skp^2 \frac{\partial}{\partial k^{\lambda}}\notag\\
 && -6 (\mu \nu \underline{\mu} \lambda )\tau k_{\nu}\skp^2 \frac{\partial}{\partial k^{\lambda}}
 -24 (\mu \nu )s_\Omega \tau k_{\mu}\skp \frac{\partial}{\partial k^{\nu}}
 +24i (\mu \nu )p_\Omega \gamf \tau k_{\mu}\skp \frac{\partial}{\partial
k^{\nu}}
 -6 (\mu \nu \lambda \rho )\tau k_{\lambda}\frac{\partial^3}{\partial k^{\mu}\partial k^{\nu}\partial k^{\rho}}
  \notag\\
 &&-6 (\mu \nu \underline{\nu} \lambda )\tau \sk \skp \frac{\partial^2}{\partial k^{\mu}\partial k^{\lambda}}
-6 (\mu \nu \underline{\mu} \lambda )\tau \sk \skp
\frac{\partial^2}{\partial k^{\nu}\partial k^{\lambda}}
 -12i (\mu \nu )(\lambda a_{\Omega\mu} )\gamf \tau \frac{\partial^2}{\partial k^{\nu}\partial k^{\lambda}}
 -16 (\mu \nu \underline{\nu} a_\Omega^{\lambda} )\gamma_{\lambda}\gamf \tau k_{\mu}\skpp
 \notag\\
 &&-8 (\mu \nu \underline{\nu} a_\Omega^{\lambda} )\gamma_{\lambda}\gamf \tau \skp \frac{\partial}{\partial k^{\mu}}
 -16 (\mu \nu \underline{\mu} a_\Omega^{\lambda} )\gamma_{\lambda}\gamf \tau k_{\nu}\skpp
 -16 (\mu \underline{\mu} \nu a_\Omega^{\lambda} )\gamma_{\lambda}\gamf \tau k_{\nu}\skpp
 -32 (\mu \nu \lambda a^{\rho} )\gamma_{\rho}\gamf \tau k_{\mu}k_{\nu}k_{\lambda}\skppp\notag\\
 && -16 (\mu \nu \lambda a_\Omega^{\rho} )\gamma_{\rho}\gamf \tau k_{\nu}k_{\lambda}\skpp \frac{\partial}{\partial k^{\mu}}
 -8 (\mu \nu \underline{\mu} a_\Omega^{\lambda} )\gamma_{\lambda}\gamf \tau \skp \frac{\partial}{\partial k^{\nu}}
 -16 (\mu \nu \lambda a_\Omega^{\rho} )\gamma_{\rho}\gamf \tau k_{\mu}k_{\lambda}\skpp \frac{\partial}{\partial k^{\nu}}
   \notag\\
 &&-8 (\mu \nu \lambda a_\Omega^{\rho} )\gamma_{\rho}\gamf \tau k_{\lambda}\skp \frac{\partial^2}{\partial k^{\mu}\partial k^{\nu}}
-8 (\mu \underline{\mu} \nu a_\Omega^{\lambda}
)\gamma_{\lambda}\gamf \tau \skp \frac{\partial}{\partial k^{\nu}}
-16 (\mu \nu \lambda a_\Omega^{\rho} )\gamma_{\rho}\gamf \tau
k_{\mu}k_{\nu}\skpp \frac{\partial}{\partial k^{\lambda}}\notag\\
&& -8 (\mu \nu \lambda a^{\rho} )\gamma_{\rho}\gamf \tau k_{\nu}\skp
\frac{\partial^2}{\partial k^{\mu}\partial k^{\lambda}}
 -8 (\mu \nu \lambda a_\Omega^{\rho} )\gamma_{\rho}\gamf \tau k_{\mu}\skp \frac{\partial^2}{\partial k^{\nu}\partial k^{\lambda}}
 -8 (\mu \nu \lambda a_\Omega^{\rho} )\gamma_{\rho}\gamf \tau \sk \frac{\partial^3}{\partial k^{\mu}\partial k^{\nu}\partial k^{\lambda}}
\notag\\
 && -8i (\mu \nu \lambda )a_{\Omega\nu}\gamf \tau \frac{\partial^2}{\partial k^{\mu}\partial k^{\lambda}}
 -12i (\mu a_\Omega^{\nu} )(\underline{\nu} \lambda )\gamf \tau \frac{\partial^2}{\partial k^{\mu}\partial k^{\lambda}}
-8i a_\Omega^{\mu}(\nu \underline{\mu} \lambda )\gamf \tau
\frac{\partial^2}{\partial k^{\nu}\partial k^{\lambda}}
 -24 s_\Omega (\mu \nu )\tau k_{\mu}\skp \frac{\partial}{\partial k^{\nu}}
  \notag\\
 &&-24i p_\Omega (\mu \nu )\gamf \tau k_{\mu}\skp \frac{\partial}{\partial k^{\nu}}
-16 (\mu \nu )(\underline{\mu} \lambda )\tau k_{\nu}k_{\lambda}\sk
\skppp
 -48 (\mu \nu )(\underline{\mu} \lambda )\tau k_{\nu}k_{\lambda}\skp \skpp
 +12 (\mu \nu )(\underline{\mu} \lambda )\tau k_{\nu}\sk \skpp \frac{\partial}{\partial k^{\lambda}}
 \notag\\
 &&-12 (\mu \nu )(\underline{\mu} \lambda )\tau k_{\lambda}\sk \skpp \frac{\partial}{\partial k^{\nu}}
 +12 (\mu \nu )(\underline{\mu} \lambda )\tau k_{\nu}\skp^2 \frac{\partial}{\partial k^{\lambda}}
 -12 (\mu \nu )(\underline{\mu} \lambda )\tau k_{\lambda}\skp^2 \frac{\partial}{\partial k^{\nu}}
 \notag\\
 &&-12 (\mu \nu )(\underline{\mu} \lambda )\tau \sk \skp \frac{\partial^2}{\partial k^{\nu}\partial k^{\lambda}}
 -16 (\mu \nu \lambda )a_\Omega^{\rho}\gamma_{\rho}\gamf \tau k_{\mu}k_{\nu}\skpp \frac{\partial}{\partial k^{\lambda}}
 -24 (\mu \nu )(\lambda a_\Omega^{\rho} )\gamma_{\rho}\gamf \tau k_{\mu}\skp \frac{\partial^2}{\partial k^{\nu}\partial k^{\lambda}}
 \notag\\
 &&-16 (\mu \nu \lambda )a_\Omega^{\rho}\gamma_{\rho}\gamf \tau k_{\nu}\skp \frac{\partial^2}{\partial k^{\mu}\partial k^{\lambda}}
 -24 (\mu a_\Omega^{\nu} )(\underline{\mu} \lambda )\gamma_{\nu}\gamf \tau \skp \frac{\partial}{\partial k^{\lambda}}
 -48 (\mu a_\Omega^{\nu} )(\lambda \rho )\gamma_{\nu}\gamf \tau k_{\mu}k_{\lambda}\skpp \frac{\partial}{\partial k^{\rho}}
 \notag\\
 &&-24 (\mu a_\Omega^{\nu} )(\lambda \rho )\gamma_{\nu}\gamf \tau k_{\lambda}\skp \frac{\partial^2}{\partial k^{\mu}\partial k^{\rho}}
 -4i (\mu \nu \lambda a_{\Omega\mu} )\gamf \tau \frac{\partial^2}{\partial k^{\nu}\partial k^{\lambda}}
 -4i (\mu \nu \lambda a_{\Omega\nu} )\gamf \tau \frac{\partial^2}{\partial k^{\mu}\partial k^{\lambda}}
 \notag\\
 &&-4i (\mu \nu \lambda a_{\Omega\lambda} )\gamf \tau \frac{\partial^2}{\partial k^{\mu}\partial k^{\nu}}
 -8i (\mu \nu \lambda a_\Omega^{\rho} )\gamf \tau k_{\rho}\frac{\partial^3}{\partial k^{\mu}\partial k^{\nu}\partial k^{\lambda}}
 +16i (\mu \nu \lambda \rho )\gamma_{\mu}\tau k_{\nu}k_{\lambda}\skpp \frac{\partial}{\partial k^{\rho}}
 \notag\\
 &&+16i (\mu \nu \lambda \rho )\gamma_{\mu}\tau k_{\lambda}\skp \frac{\partial^2}{\partial k^{\nu}\partial k^{\rho}}
 -24i (\mu \nu )(\lambda \rho )\gamma_{\mu}\tau k_{\lambda}\skp \frac{\partial^2}{\partial k^{\nu}\partial k^{\rho}}
 +24i (\mu \nu )(\lambda \rho )\gamma_{\lambda}\tau k_{\mu}\skp \frac{\partial^2}{\partial k^{\nu}\partial k^{\rho}}
 \notag\\
 &&-16i (\mu \underline{\mu} \nu \lambda )\gamma_{\nu}\tau k_{\lambda}\skpp
 -8i (\mu \underline{\mu} \nu \lambda )\gamma_{\nu}\tau \skp \frac{\partial}{\partial k^{\lambda}}
 +16i (\mu \nu \underline{\mu} \lambda )\gamma_{\lambda}\tau k_{\nu}\skpp
 +16i (\mu \nu \underline{\nu} \lambda )\gamma_{\lambda}\tau k_{\mu}\skpp\notag\\
 && -32i (\mu \nu \lambda \rho )\gamma_{\lambda}\tau k_{\mu}k_{\nu}k_{\rho}\skppp
 -16i (\mu \nu \lambda \rho )\gamma_{\lambda}\tau k_{\mu}k_{\nu}\skpp \frac{\partial}{\partial k^{\rho}}
 +8i (\mu \nu \underline{\mu} \lambda )\gamma_{\lambda}\tau \skp \frac{\partial}{\partial k^{\nu}}
 -16i (\mu \nu \lambda \rho )\gamma_{\lambda}\tau k_{\mu}k_{\rho}\skpp \frac{\partial}{\partial k^{\nu}}
  \notag\\
 &&-8i (\mu \nu \lambda \rho )\gamma_{\lambda}\tau k_{\mu}\skp \frac{\partial^2}{\partial k^{\nu}\partial k^{\rho}}
+8i (\mu \nu \underline{\nu} \lambda )\gamma_{\lambda}\tau \skp
\frac{\partial}{\partial k^{\mu}}
 -16i (\mu \nu \lambda \rho )\gamma_{\lambda}\tau k_{\nu}k_{\rho}\skpp \frac{\partial}{\partial k^{\mu}}
 -8i (\mu \nu \lambda \rho )\gamma_{\lambda}\tau k_{\nu}\skp \frac{\partial^2}{\partial k^{\mu}\partial k^{\rho}}
 \notag\\
 &&-8i (\mu \nu \lambda \rho )\gamma_{\lambda}\tau k_{\rho}\skp \frac{\partial^2}{\partial k^{\mu}\partial k^{\nu}}
 -16 a^{\mu}(\nu \lambda \rho )\gamma_{\mu}\gamf \tau k_{\nu}k_{\lambda}\skpp \frac{\partial}{\partial k^{\rho}}
 -16 a^{\mu}(\nu \lambda \rho )\gamma_{\mu}\gamf \tau k_{\lambda}\skp \frac{\partial^2}{\partial k^{\nu}\partial k^{\rho}}
 \notag\\
 &&-16 (\mu \nu \lambda \rho )\tau k_{\mu}k_{\nu}k_{\lambda}\sk \skppp \frac{\partial}{\partial k^{\rho}}
 -48 (\mu \nu \lambda \rho )\tau k_{\mu}k_{\nu}k_{\lambda}\skp \skpp \frac{\partial}{\partial k^{\rho}}
 -12 (\mu \nu \lambda \rho )\tau k_{\nu}k_{\lambda}\sk \skpp \frac{\partial^2}{\partial k^{\mu}\partial k^{\rho}}
 \notag\\
 &&-12 (\mu \nu \lambda \rho )\tau k_{\mu}k_{\lambda}\sk \skpp \frac{\partial^2}{\partial k^{\nu}\partial k^{\rho}}
 -12 (\mu \nu \lambda \rho )\tau k_{\nu}k_{\lambda}\skp^2 \frac{\partial^2}{\partial k^{\mu}\partial k^{\rho}}
 -12 (\mu \nu \lambda \rho )\tau k_{\mu}k_{\lambda}\skp^2 \frac{\partial^2}{\partial k^{\nu}\partial k^{\rho}}
 \notag\\
 &&-12 (\mu \nu \lambda \rho )\tau k_{\lambda}\sk \skp \frac{\partial^3}{\partial k^{\mu}\partial k^{\nu}\partial k^{\rho}}
 +24i (\mu \nu )(\underline{\mu} \lambda )\gamma_{\nu}\tau \skp \frac{\partial}{\partial k^{\lambda}}
 -48i (\mu \nu )(\lambda \rho )\gamma_{\mu}\tau k_{\nu}k_{\lambda}\skpp \frac{\partial}{\partial k^{\rho}}
 \notag\\
 &&-24 (\mu \nu )(\lambda \rho )\tau k_{\mu}k_{\lambda}\sk \skpp \frac{\partial^2}{\partial k^{\nu}\partial k^{\rho}}
 -24 (\mu \nu )(\lambda \rho )\tau k_{\mu}k_{\lambda}\skp^2 \frac{\partial^2}{\partial k^{\nu}\partial
 k^{\rho}}\;,
 \end{eqnarray}

 %%%%%%%%%%%%%%%%%%%%%%%%%%%%%%%%%%%%%%%%%%%%%%%%%%%%%%%%%%%%%%%%%%%%%%%%%%%%%%%%%%%%%%%
 \section{Analytical expressions of $\mathcal{Z}_i$ on
 $\Sigma(k^2)$}\label{ZSigma}

 %
where
 \begin{eqnarray}
 &&\int dK\equiv N_c\int \frac{d^4k}{(2\pi)^4}e^{-\tau(k^2+\Sigma_k^2)}\int_{\frac{1}{\Lambda^2}}^\infty
 \frac{d\tau}{\tau}\;,
 \hspace{2cm} X=\frac{1}{k^2+\Sigma_k^2}\;,\\
  &&\int^\infty_{\frac{1}{\Lambda^2}}\frac{d\tau}{\tau}e^{-\tau(k^2+\Sigma_k^2)}\tau
    =Xe^{-\frac{(k^2+\Sigma_k^2)}{\Lambda^2}}\;,\\
 &&\int^\infty_{\frac{1}{\Lambda^2}}\frac{d\tau}{\tau}e^{-\tau(k^2+\Sigma_k^2)}\tau^2
    =\left(X^2+\frac{X}{\Lambda^2}\right)e^{-\frac{(k^2+\Sigma_k^2)}{\Lambda^2}}\;,\\
 &&\int^\infty_{\frac{1}{\Lambda^2}}\frac{d\tau}{\tau}e^{-\tau(k^2+\Sigma_k^2)}\tau^3
    =\left(2X^3+2\frac{X^2}{\Lambda^2}+\frac{X}{\Lambda^4}\right)e^{-\frac{(k^2+\Sigma_k^2)}{\Lambda^2}}\;,\\
 &&\int^\infty_{\frac{1}{\Lambda^2}}\frac{d\tau}{\tau}e^{-\tau(k^2+\Sigma_k^2)}\tau^4
    =\left(6X^4+6\frac{X^3}{\Lambda^2}+3\frac{X^2}{\Lambda^4}+\frac{X}{\Lambda^6}\right)
   e^{-\frac{(k^2+\Sigma_k^2)}{\Lambda^2}}\;,\\
 &&\int^\infty_{\frac{1}{\Lambda^2}}\frac{d\tau}{\tau}e^{-\tau(k^2+\Sigma_k^2)}\tau^5
    =\left(24X^5+24\frac{X^4}{\Lambda^2}+12\frac{X^3}{\Lambda^4}
   +4\frac{X^2}{\Lambda^6}+\frac{X}{\Lambda^8}\right)
  e^{-\frac{(k^2+\Sigma_k^2)}{\Lambda^2}}\;,\\
 &&\int^\infty_{\frac{1}{\Lambda^2}}\frac{d\tau}{\tau}e^{-\tau(k^2+\Sigma_k^2)}\tau^6
 =\left(120X^6+120\frac{X^5}{\Lambda^2}+60\frac{X^4}{\Lambda^4}
 +20\frac{X^3}{\Lambda^6}+5\frac{X^2}{\Lambda^8}+\frac{X}{\Lambda^{10}}\right)
 e^{-\frac{(k^2+\Sigma_k^2)}{\Lambda^2}}\;.
 \end{eqnarray}
 As we mentioned before, since the momentum integrations in
 (\ref{ZSigma}) are convergent due to suppression factor $e^{-\tau
 k^2}$, the integrand is only accurate up to some total derivatives
 , but if we insist on taking the limit of $\Lambda\rightarrow\infty$,
 the dropping out momentum space total derivatives becomes
 problematic.
 We choose these momentum space total derivatives to reduce the
high order self energy derivatives as much as possible. For example,
following relation with $m,m_0,m_1,m_2,m_3,m_4$ being zero or
positive integers can be used to reduce the term with a $\skppppp$
to the terms with lower order self energy derivatives,
 \begin{eqnarray}
 &&\hspace{-0.5cm}\int\frac{d^4k}{(2\pi)^4}e^{-\tau(k^2+\Sigma_k^2)}\tau^n
 (k^{2+m}\sk^{m_0}\skp^{m_1}\skpp^{m_2}\skppp^{m_3}\skpppp^{m_4}\skppppp)\notag\\
 &=&\int\frac{d^4k}{(2\pi)^4}e^{-\tau(k^2+\Sigma_k^2)}\tau^n
 \Big(-\frac{4+m}{2(m_4+1)}k^m\sk^{m_0}\skp^{m_1}\skpp^{m_2}\skppp^{m_3}\skpppp^{m_4+1}\notag\\
 &&-\frac{m_0}{m_4+1}k^{2+m}\sk^{m_0-1}\skp^{m_1+1}\skpp^{m_2}\skppp^{m_3}\skpppp^{m_4+1}
 -\frac{m_1}{m_4+1}k^{2+m}\sk^{m_0}\skp^{m_1-1}\skpp^{m_2+1}\skppp^{m_3}\skpppp^{m_4+1}\notag\\
 &&-\frac{m_2}{m_4+1}k^{2+m}\sk^{m_0}\skp^{m_1}\skpp^{m_2-1}\skppp^{m_3+1}\skpppp^{m_4+1}
 -\frac{m_3}{m_4+1}k^{2+m}\sk^{m_0}\skp^{m_1}\skpp^{m_2}\skppp^{m_3-1}\skpppp^{m_4+2}\notag\\
 &&+\frac{1}{m_4+1}\tau k^{2+m}\sk^{m_0}\skp^{m_1}\skpp^{m_2}\skppp^{m_3}\skpppp^{m_4+1}
 +\frac{2}{m_4+1}\tau k^{2+m}\sk^{m_0+1}\skp^{m_1+1}\skpp^{m_2}\skppp^{m_3}\skpppp^{m_4+1}\Big)
 \end{eqnarray}
Similarly, we have a series relations to  reduce the term with a
$\skpppp$, or a $\skppp$,  a $\skpp$,  a $\skp$  to the terms with
lower order self energy derivatives,
\begin{eqnarray}
 &&\hspace{-0.5cm}\int\frac{d^4k}{(2\pi)^4}e^{-\tau(k^2+\Sigma_k^2)}\tau^n
 (k^{2+m}\sk^{m_0}\skp^{m_1}\skpp^{m_2}\skppp^{m_3}\skpppp)\notag\\
 &=&\int\frac{d^4k}{(2\pi)^4}e^{-\tau(k^2+\Sigma_k^2)}\tau^n
 \Big(-\frac{4+m}{2(m_3+1)}k^m\sk^{m_0}\skp^{m_1}\skpp^{m_2}\skppp^{m_3+1}
 -\frac{m_0}{m_3+1}k^{2+m}\sk^{m_0-1}\skp^{m_1+1}\skpp^{m_2}\skppp^{m_3+1}\notag\\
 &&-\frac{m_1}{m_3+1}k^{2+m}\sk^{m_0}\skp^{m_1-1}\skpp^{m_2+1}\skppp^{m_3+1}
 -\frac{m_2}{m_3+1}k^{2+m}\sk^{m_0}\skp^{m_1}\skpp^{m_2-1}\skppp^{m_3+2}\notag\\
 &&+\frac{1}{m_3+1}\tau k^{2+m}\sk^{m_0}\skp^{m_1}\skpp^{m_2}\skppp^{m_3+1}
 +\frac{2}{m_3+1}\tau
 k^{2+m}\sk^{m_0+1}\skp^{m_1+1}\skpp^{m_2}\skppp^{m_3+1}\Big)\\
  &&\hspace{-0.5cm}\int\frac{d^4k}{(2\pi)^4}e^{-\tau(k^2+\Sigma_k^2)}\tau^n
 (k^{2+m}\sk^{m_0}\skp^{m_1}\skpp^{m_2}\skppp)\notag\\
 &=&\int\frac{d^4k}{(2\pi)^4}e^{-\tau(k^2+\Sigma_k^2)}\tau^n
 \Big(-\frac{4+m}{2(m_2+1)}k^m\sk^{m_0}\skp^{m_1}\skpp^{m_2+1}
 -\frac{m_0}{m_2+1}k^{2+m}\sk^{m_0-1}\skp^{m_1+1}\skpp^{m_2+1}\\
 &&-\frac{m_1}{m_2+1}k^{2+m}\sk^{m_0}\skp^{m_1-1}\skpp^{m_2+2}
 +\frac{1}{m_2+1}\tau k^{2+m}\sk^{m_0}\skp^{m_1}\skpp^{m_2+1}
 +\frac{2}{m_2+1}\tau
 k^{2+m}\sk^{m_0+1}\skp^{m_1+1}\skpp^{m_2+1}\Big)\notag\\
  &&\hspace{-0.5cm}\int\frac{d^4k}{(2\pi)^4}e^{-\tau(k^2+\Sigma_k^2)}\tau^n
 (k^{2+m}\sk^{m_0}\skp^{m_1}\skpp)\notag\\
 &=&\int\frac{d^4k}{(2\pi)^4}e^{-\tau(k^2+\Sigma_k^2)}\tau^n
 \Big(-\frac{4+m}{2(m_1+1)}k^m\sk^{m_0}\skp^{m_1+1}
 -\frac{m_0}{m_1+1}k^{2+m}\sk^{m_0-1}\skp^{m_1+2}\notag\\
 &&+\frac{1}{m_1+1}\tau k^{2+m}\sk^{m_0}\skp^{m_1+1}
 +\frac{2}{m_1+1}\tau k^{2+m}\sk^{m_0+1}\skp^{m_1+2}\Big)\\
  &&\hspace{-0.5cm}\int\frac{d^4k}{(2\pi)^4}\tau^{n+1}
  k^{2+m}\sk^{m_0+1}\skp\notag\\
 &=&\int\frac{d^4k}{(2\pi)^4}e^{-\tau(k^2+\sk^2)}
 \tau^n\Big((1+\frac{m}{4})k^m\sk^{m_0}+\frac{1}{2}m_0k^{m+2}\sk^{m_0-1}\skp-\frac{1}{2}\tau
 k^{2+m}\sk^{m_0}\Big)\;.
 \end{eqnarray}
With the help of above relations, the highest self energy
derivatives appear in the final result (\ref{ZSigma}) is $\skppp$.
 %%%%%%%%%%%%%%%%%%%%%%%%%%%%%%%%%%%%%%%%%%%%%%%%%%%%%%%%%%%%%%%%%%%%%%%%%%%%%%%%%%%%%%%%
\section{Relations among $K_i$ and $\mathcal{Z}_i$}\label{KZ}


\newpage
To help understanding mutual relation between the definition of
symbols in our formulation and those in Ref.\cite{p6-1}, in Table
XIII, we give a comparison.
\begin{eqnarray}
&&\hspace{-1cm}\mbox{\small{\bf TABLE XIV}. Comparisons between the
symbols
introduce in Ref.\cite{p6-1} (the first and third columns) }\notag\\
&&\hspace{1cm}\mbox{\small and corresponding
ones defined in present paper (the second and fourth columns). }\notag\\
 &&\hspace*{-1cm}
\begin{array}{|c|c||c|c|}
 \hline
 \mbox{Ref.~\cite{p6-1}}&\mbox{Present paper}& \mbox{Ref.~\cite{p6-1}}&\mbox{Present paper}\\
 \hline g_R&V_R&\chi_-^{\mu} & 4iB_0 d^{\mu}p_\Omega-4iB_0 s_\Omega a_\Omega^{\mu}-4iB_0a_\Omega^{\mu} s_\Omega\\
 \hline g_L&V_L&F_{L,R}^{\mu\nu} & f_{L,R}^{\mu\nu}\\
  \hline u(\varphi)&\Omega &f_+^{\mu\nu} & 2V^{\mu\nu}-2i(a_\Omega^\mu a_\Omega^{\nu}-a_\Omega^{\nu}a_\Omega^{\mu}) \\
 \hline u^\mu & 2a_\Omega^\mu& \nabla^{\lambda}f_+^{\mu\nu} & 2d^{\lambda}V^{\mu\nu}-2id^{\lambda}(a_\Omega^\mu a_\Omega^{\nu}
 -a_\Omega^{\nu}a_\Omega^{\mu})\\
 \hline u\cdot u=u^\mu u^\mu & 4a_\Omega^2 &f_-^{\mu\nu} & -2(d^\mu a_\Omega^{\nu}-d^{\nu}a_\Omega^{\mu}) \\
 \hline \chi & \chi &\nabla^{\lambda}f_-^{\mu\nu} & -2(d^{\lambda}d^\mu a_\Omega^{\nu}-d^{\lambda}d^{\nu}a_\Omega^{\mu}) \\
 \hline \chi_+ & 4B_0s_\Omega & h^{\mu\nu} & 2(d^\mu a_\Omega^{\nu}+d^{\nu}a_\Omega^{\mu})\\
 \hline \chi_+^{\mu} & 4B_0 d^{\mu}s_\Omega+4B_0 p_\Omega a_\Omega^{\mu}+4B_0a_\Omega^{\mu}p_\Omega & \nabla^\mu & d^\mu\\
 \hline \chi_- & 4iB_0p_\Omega &\Gamma^\mu & -iv_\Omega^\mu\\
 \hline
 \end{array}\notag
 \end{eqnarray}
 In obtaining (\ref{KZrelation}), some dependent terms must be
 reduced into independent terms, in Table XV, the first column is
 the dependent operators and the second column is the sum of its corresponding
 independent operators.
  \begin{eqnarray}
  &&\hspace{1cm}\mbox{\small{\bf TABLE XV}. Expand dependent operator
  in terms of independent operators}\notag\\
 &&\hspace{-0.5cm}\begin{array}{|c|c|}
 \hline \mbox{dependent operator} & \mbox{corresponding independent operators}\\
 \hline \mathrm{tr}(u_{\mu}u_{\nu}h^{\mu\lambda}h^{\nu\lambda}) &
  -\frac{1}{2}Y_{3}+\frac{1}{2}Y_{28}-\frac{1}{N_f}Y_{30}+\frac{1}{N_f}Y_{34}
  -\frac{1}{2N_f^2}Y_{36}-\frac{1}{2}Y_{37}+Y_{54}-Y_{60}\notag\\
  &+Y_{68}+\frac{1}{2}Y_{92}+Y_{94}+\frac{1}{2}Y_{105}\\
 \hline  \mathrm{tr}(u_{\mu}u_{\nu}h^{\mu\lambda}\fm^{\nu\lambda} &
  2Y_{1}-\frac{1}{2}Y_{3}+\frac{3}{2}Y_{5}
  +\frac{3}{N_f}Y_{30}-3Y_{31}+2Y_{33}-\frac{3}{N_f}Y_{34}+\frac{3}{2N_f^2}Y_{36}-\frac{1}{2}Y_{37}\notag\\
  +u_{\mu}u_{\nu}\fm^{\mu\lambda}h^{\nu\lambda})&
  -2Y_{49}-Y_{52}+4Y_{54}+Y_{58}-2Y_{60}-Y_{64}-2Y_{66}-4Y_{67}+Y_{68}\notag\\
  &+Y_{86}-2Y_{90}+\frac{1}{2}Y_{92}+2Y_{94}-2Y_{95}
  -\frac{3}{2}Y_{97}+Y_{105}\\
 \hline  \mathrm{tr}(u_{\mu}u_{\nu}h^{\nu\lambda}h^{\mu\lambda}) &
  -Y_{1}-\frac{1}{2}Y_{5}+\frac{1}{2}Y_{28}-\frac{2}{N_f}Y_{30}+Y_{31}
  -Y_{33}+\frac{2}{N_f}Y_{34}-\frac{1}{N_f^2}Y_{36}\notag\\
  &+Y_{49}+Y_{52}-2Y_{54}+Y_{64}+2Y_{67}+Y_{90}+Y_{95}+\frac{1}{2}Y_{97}-\frac{1}{2}Y_{105}\\
\hline
 \end{array}\notag
 \end{eqnarray}
\end{document}